\documentclass[preprint,aps,nofootinbib]{revtex4}
\usepackage{color}
\usepackage{graphicx}
\usepackage{subfigure}

\begin{document}
\def\oalphas{O(\alpha_{s})}
\def\met{\not\!\!E_{T}}
\def\D0{{D\O}}

\title{Single top quark production and decay in the $t$-channel at next-to-leading
order at the LHC}
\author{Reinhard Schwienhorst}
\email{schwier@pa.msu.edu}
\affiliation{Department of Physics $\&$ Astronomy, Michigan State University,
East Lansing, MI 48824, USA}

\author{Qing-Hong Cao}
\email{caoq@hep.anl.gov}
\affiliation{HEP Division, Argonne National Laboratory, Argonne, IL 60439, U.S.A}
\affiliation{Enrico Fermi Institute, University of Chicago, Chicago, Illinois 60637, U.S.A.}

\author{C.-P. Yuan}
\email{yuan@pa.msu.edu}
\affiliation{Department of Physics $\&$ Astronomy, Michigan State University,
East Lansing, MI 48824, USA}

\author{Charles Mueller}
\email{muell149@msu.edu}
\affiliation{Department of Physics $\&$ Astronomy, Michigan State University,
East Lansing, MI 48824, USA}
\date{\today}

\begin{abstract}
We present a study of single top and single antitop quark production in the $t$-channel
mode at the LHC $pp$ collider at 7~TeV, 10~TeV and 14~TeV, including
next-to-leading order QCD corrections to the production and
decay of the top quark. We discuss the effects of different $O(\alpha_{s})$
contributions on the inclusive cross section and important kinematic
distributions, after imposing the relevant cuts to select $t$-channel
single top quark events. 
\end{abstract}

\maketitle

\section{Introduction\label{sec:Introduction}}

\begin{figure}[!h!tbp]
\includegraphics[width=1.6in]{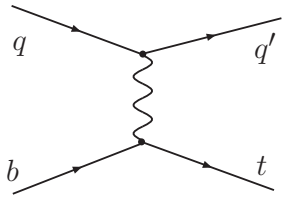}
\caption{Representative Feynman diagram for $t$-channel single top quark production.
\label{fig:tchan-tree}}
\end{figure}

The top quark and its electroweak interaction are important within the Standard
Model and provide a window to physics beyond the standard model. In particular 
the production of single top quarks through electroweak interactions
is a sensitive process at hadron colliders that is being studied at both the 
Tevatron proton-antiproton collider and the Large Hadron Collider (LHC) proton-proton collider. 
Electroweak single top quark production proceeds through the $s$-channel
decay of a virtual $W$ ($q\bar{q}'\to W^{*}\to t\bar{b}$),
the $t$-channel exchange of a virtual $W$ ($bq\to tq'$
and $b\bar{q}'\to t\bar{q}$, shown in Fig.~\ref{fig:tchan-tree}),
and the associated production of a top quark and a $W$ boson ($bg\to tW^{-}$).
The single top cross section is proportional to the Cabibbo-Kobayashi-Maskawa
(CKM) quark mixing matrix element $|V_{tb}|^{2}$, and the single
top cross section measurement provides a direct determination of $|V_{tb}|$
without assumptions about the number of quark generations. A study
of spin correlations in single top quark production can be used to
test the left-handed nature of the top quark charged-current weak
interaction and to look for anomalous top quark 
couplings~\cite{Chen:2005vr,anomCouplD0,anomCouplCombD0}.

The \D0 and CDF collaborations at the Fermilab Tevatron proton-antiproton collider
have observed single top quark production for the first 
time~\cite{singletopobsD0,singletopobsCDF,singletopobsTeV},
following the evidence for single top production from 
\D0~\cite{Abazov:2006gd,singletopevidencePRDD0},
confirmed by CDF~\cite{singletopevidenceCDF}. The Tevatron measurements combine the 
$s$-channel and $t$-channel signals
to maximize the sensitivity to the single top quark signal. The \D0 collaboration has
also reported a separate measurement of the $t$-channel cross section~\cite{tchannelD0},
independent of the $s$-channel. These measurements rely heavily on multivariate
analysis techniques~\cite{singletopevidencePRDD0,singletopsearchPRDD0,singletopobsCDFPRD}
which require accurate modeling of both the single top signals and
the various background processes. 

At the LHC, single top quark production will play an important role in searches for
new physics, in the single top quark final state and as background
in other searches. All three single top processes should be observed
at the LHC, where the $t$-channel has the largest cross section, followed
by associated production and the $s$-channel. It is important to separate
these processes from each other since they have different systematic 
uncertainties and different sensitivities to new 
physics~\cite{Tait:2000sh,Cao:2007ea,AguilarSaavedra:2010zi,Zhang:2010dr}. 

The next-to-leading order (NLO) $\oalphas$ corrections to single
top quark production have already been carried out in 
Refs.~\cite{Smith:1996ij,Bordes:1994ki},
which shows an uncertainty on the total cross section of about $5\%$
by varying the factorization and renormalization scales. The differential
cross sections for on-shell single top quark production have been
calculated~\cite{Harris:2002md,Sullivan:2004ie}, and the complete
NLO calculations including both the single top quark production and
decay have been done independently by several 
groups~\cite{Campbell:2004ch, Cao:2004ky, Cao:2004ap, Cao:2005pq,
Frixione:2008yi, Alioli:2009je, Re:2010bp, Falgari:2010sf}.
For the $t$-channel process, the NLO corrections have also been calculated
for the $2\to3$ process 
$bg\to tq'\overline{b}$~\cite{Campbell:2009ss}, and interference between
top quark production and decay have been included~\cite{Falgari:2010sf}. The total cross
section has been calculated at NLO including higher order soft gluon
corrections~\cite{Kidonakis:2007ej,Kidonakis:2010tc}.
Recently, resummation effects on the $s$- and $t$-channel single top productions 
in the framework of soft-collinear-effective-theory have been 
investigated~\cite{Zhu:2010mr,Wang:2010ue}. 
In previous studies we presented a detailed phenomenological analysis,
focusing on signal cross sections and kinematical distributions, of
$t$-channel single top quark production at the Tevatron~\cite{Cao:2005pq},
and $s$-channel single top quark production at the LHC~\cite{Heim:2009ku}.
Here we analyze $t$-channel single top quark production at the LHC
proton-proton collider at center of mass (CM) energies of 7~TeV, 10~TeV and 14~TeV.
Since the LHC is $pp$ collider, the production cross sections and
kinematic distributions are different for $b\bar{t}(j)$ and $t\bar{b}(j)$.
We therefore consider the production of a $t$-quark separately from
the production of a $\overline{t}$-quark. 

The paper is organized as follows. In Sec.~\ref{sec:Cross-Section-Inclusive},
we first present the inclusive cross section for single top quark
production at the LHC in the $t$-channel mode. We then investigate
its dependence on the top quark mass, renormalization and factorization
scales, and parton distribution functions (PDF). In Sec.~\ref{sec:Single-Top-Acceptance},
we examine the acceptance of single top quark events for various sets
of kinematic cuts. In Sec.~\ref{sec:EventDistr}, we explore the
kinematical information in the final state objects. Section~\ref{sec:Conclusions}
contains our conclusions.

\section{Inclusive cross section\label{sec:Cross-Section-Inclusive}}

The LHC will produce large samples of $t$-channel single top quark
events, already in the initial 7~TeV run and even more so in 14~TeV
running, and single top quark production can be studied in detail.
The accuracy of the cross section and $|V_{tb}|$ measurements at
the Tevatron is limited by the statistical uncertainty of the small
single top samples collected and the large backgrounds~\cite{singletopobsD0,singletopobsCDF}.
The situation will be drastically different at the LHC where systematic
uncertainties dominate. In particular the measurement of $|V_{tb}|$
requires accurate theoretical predictions of both the inclusive cross
section and kinematic distributions. In this section, we show the
inclusive production rates and discuss their dependence on the factorization
and renormalization scales as well as the top quark mass.

We present numerical results for the production of single top quark
events considering the leptonic decay of the $W$-boson from the top
quark decay at the LHC. We consider three different different CM energies
for the LHC $pp$ collider: the start-up CM energy of 7~TeV, an intermediate
energy of 10~TeV, and the design energy of 14~TeV. Unless otherwise
specified, we use the NLO PDF set CTEQ6.6M~\cite{Pumplin:2002vw,Nadolsky:2008zw},
defined in the $\overline{MS}$ scheme, and the NLO (2-loop) running
coupling $\alpha_{s}$ with $\Lambda_{\overline{MS}}$ provided by
the PDFs. For the CTEQ6.6M PDFs, $\Lambda_{\overline{MS}}^{(4)}=0.326$~GeV
for four active quark flavors. The values of the relevant electroweak
parameters are: $\alpha=1/137.0359895$, $G_{\mu}=1.16637\times10^{-5}\,\mbox{GeV}^{-2}$,
$M_{W}=80.40\,\mbox{GeV}$, $M_{Z}=91.1867\,\mbox{GeV}$,
and $\sin^{2}\theta_{W}=0.231$~\cite{Cao:2004yy}. The square of
the weak gauge coupling is $g^{2}=4\sqrt{2}M_{W}^{2}G_{\mu}$. We
extend our previous Tevatron studies~\cite{Cao:2004ap,Cao:2005pq,Cao:2004ky}
to the LHC proton-proton collider, including both top and antitop
production. We focus on the electron final state, though our analysis
procedure also applies to the muon final state. Including $\oalphas$
corrections to the decay $W\to\bar{q}q^{\prime}$, the decay branching
ratio of the $W$ boson into leptons is $Br(W\to\ell^{+}\nu)=0.108$~\cite{Cao:2004yy}.
Unless otherwise specified, we will choose the top quark mass to be
$m_{t}=173\,\mbox{GeV}$, consistent with the world 
average~\cite{Abazov:2008ds,Aaltonen:2008gj,TevatronElectroweakWorkingGroup:2010yx},
and the renormalization scale ($\mu_{R}$) as well as the factorization
scale ($\mu_{F}$) to be equal to the top quark mass.

\subsection{Theoretical cutoff dependence\label{sub:Theoretical-Cutoff-Dependence}}

The NLO QCD differential cross sections are calculated using the one-cutoff
phase space slicing (PSS) method~\cite{Giele:1991vf,Giele:1993dj}.
This procedure introduces a theoretical cutoff parameter ($s_{min}$)
in order to isolate soft and collinear singularities associated with
real gluon emission sub-processes by partitioning the phase space into
soft, collinear and hard regions such that
\begin{equation}
\left|\mathcal{M}^{{\rm r}}\right|^{2}=\left|\mathcal{M}^{{\rm r}}\right|_{{\rm soft}}^{2}+\left|\mathcal{M}^{{\rm r}}\right|_{{\rm collinear}}^{2}+\left|\mathcal{M}^{{\rm r}}\right|_{{\rm hard}}^{2}\,.\label{eq:nlomat}
\end{equation}
In the soft and collinear regions the cross section is proportional
to the Born-level cross section. Using dimensional regularization,
we evaluate the real gluon emission diagrams in n-dimensions under
the soft gluon approximation in the soft region and under the collinear
approximation in the collinear region, and integrate out the corresponding
phase space volume analytically. The resulting divergences are canceled
by virtual corrections or absorbed into the perturbative parton distribution
functions in the factorization procedure. For our study, we found
a cutoff value of $s_{min}=1\,{\rm GeV}^{2}$ to be appropriate for
studying the $t$-channel single top process. For comparison, we also
used a value of $s_{min}=1\,{\rm GeV}^{2}$ in our Tevatron $t$-channel
study~\cite{Cao:2005pq}, and a value of $s_{min}=5\,{\rm GeV}^{2}$
in our Tevatron~\cite{Cao:2004ap} and LHC~\cite{Heim:2009ku} $s$-channel
studies. A detailed discussion of the phase space slicing method can
be found in Ref.~\cite{Cao:2004ky}.

\subsection{Inclusive cross section\label{sub:Inclusive-Cross-Section}}

\begin{figure}[!h!tbp]
\includegraphics[scale=0.33]{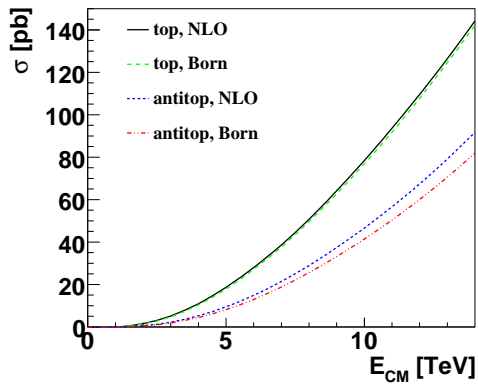}
\caption{Inclusive $t$-channel single top quark cross section at the LHC proton-proton
collider as a function of CM energy for a top quark mass of 173~GeV.\label{fig:deps-cm}}
\end{figure}

The LHC has been designed for a CM energy of 14~TeV, though during
the initial start-up of the accelerator the energy is only 7~TeV.
Figure~\ref{fig:deps-cm} shows the dependence of the inclusive
cross section on the CM energy. %

At Born-level, the production cross section for top quarks is about
3/2 larger than that for antitop quarks due to the initial state configuration
(up quark PDF vs down quark PDF). The difference between top and antitop
is smaller at NLO, and the antitop production cross section is larger
at NLO compared to Born-level. This is in contrast to the top quark
production cross section, which is almost identical between Born-level
and NLO. This difference between top and antitop quarks is due to
the different $b$ quark and gluon PDF momentum fraction regions.

We group the higher-order QCD corrections into three separate gauge
invariant sets according to their origin: corrections from the light
quark line (LIGHT), corrections from the heavy quark line (HEAVY),
and corrections from the top quark decay (TDEC). If appropriate we
further subgroup the HEAVY corrections into those with real gluon
emission (HEAVY(g)) and those with emission of a $\bar{b}$-quark
(HEAVY($\bar{b}$), also called $2\to3$), see also Fig.~\ref{fig:real_tchan}.
The explicit diagrams and definitions for these three sets can be
found in Ref.~\cite{Cao:2004ky}. The inclusive cross section as
well as the individual $\oalphas$ contributions are listed in Table~\ref{tab:inclusive}. 

\begin{table}
\begin{centering}
\begin{tabular}{l|c|c|c|c}
 & \multicolumn{2}{c|}{Top quark production} & \multicolumn{2}{c}{Antitop quark production}\\
\hline 
 & Cross section & Fraction of & Cross section & Fraction of\\
 & (pb) & NLO ($\%$) & (pb) & NLO ($\%$)\\
\hline
7~TeV CM energy: &  &  &  & \\
~~Born-level & 38.6 & 96.6 & 19.1 & 87.2\\
~~~~$O(\alpha_{s})$ HEAVY & -3.6 & -9.0 & 0.84 & 3.8\\
~~~~$O(\alpha_{s})$ LIGHT & 3.9 & 9.7 & 1.8 & 8.0\\
~~~~$O(\alpha_{s})$ TDEC & 1.1 & 2.7 & 0.22 & 1.0\\
~~$O(\alpha_{s})$ sum & 1.4 & 3.4 & 2.8 & 12.8\\
~~NLO & 40.0 & 100 & 21.9 & 100\\
\hline
14~TeV CM energy: &  &  &  & \\
~~Born-level & 146.5 & 97.2 & 84.3 & 87.7\\
~~~~$O(\alpha_{s})$ HEAVY & -9.0 & -6.0 & 6.9 & 7.2\\
~~~~$O(\alpha_{s})$ LIGHT & 9.0 & 6.0 & 3.9 & 4.1\\
~~~~$O(\alpha_{s})$ TDEC & 4.2 & 2.8 & 1.0 & 1.0\\
~~$O(\alpha_{s})$ sum & 4.2 & 2.8 & 11.8 & 12.3\\
~~NLO & 150.7 & 100 & 96.1 & 100\\
\hline
\end{tabular}
\par\end{centering}

\caption{Inclusive $t$-channel single top production cross section for different
sub-processes at the 7~TeV and 14~TeV LHC for a top quark mass of
173~GeV. \label{tab:inclusive}}

\end{table}

The effects of the finite widths of the top quark and $W$-boson have
been included. The cross section is about twice as large for top quark
production compared to antitop because there are two up quarks in
the proton and only one down quark. The ratio between top and antitop
quark production decreases at NLO due to the heavy quark correction,
which has opposite sign for top and antitop quark production. In top
quark production, the HEAVY correction is negative, whereas in antitop
quark production it is positive. This is a reflection of the different
gluon and $b$~quark momentum fractions that the two processes are
sensitive to. The TDEC contribution is small compared to the other
two.

\subsection{Top quark mass, PDF and renormalization/factorization scale dependence\label{sub:Top-Quark-Mass}}

The inclusive cross section as given in Table~\ref{tab:inclusive}
has three uncertainty components. The top quark mass is not known
exactly and a change in the mass results in a changing cross section.
The $t$-channel cross section is especially sensitive to the PDFs,
in particular of the gluon and the $b$~quark. The renormalization
and factorization scales also contribute to the uncertainty of the
theoretical prediction. The renormalization scale $\mu_{R}$ is introduced
when redefining the bare parameters in terms of the renormalized parameters
in the $O(\alpha_{s})$ corrections, while the factorization scale
$\mu_{F}$ is introduced when absorbing the collinear divergences
into parton distribution functions. Therefore, both $\mu_{R}$ and
$\mu_{F}$ are unphysical and the final predictions should not depend
on them. However, since we work at a fixed order in perturbation theory,
we indeed see a dependence of the predicted cross section on $\mu_{R}$
and $\mu_{F}$, which is formally of higher order. Here, we examine
the top quark mass, PDF, and scale dependence of the $t$-channel
inclusive cross section.

\begin{figure}[!h!tbp]
\subfigure[]{
\includegraphics[scale=0.33]{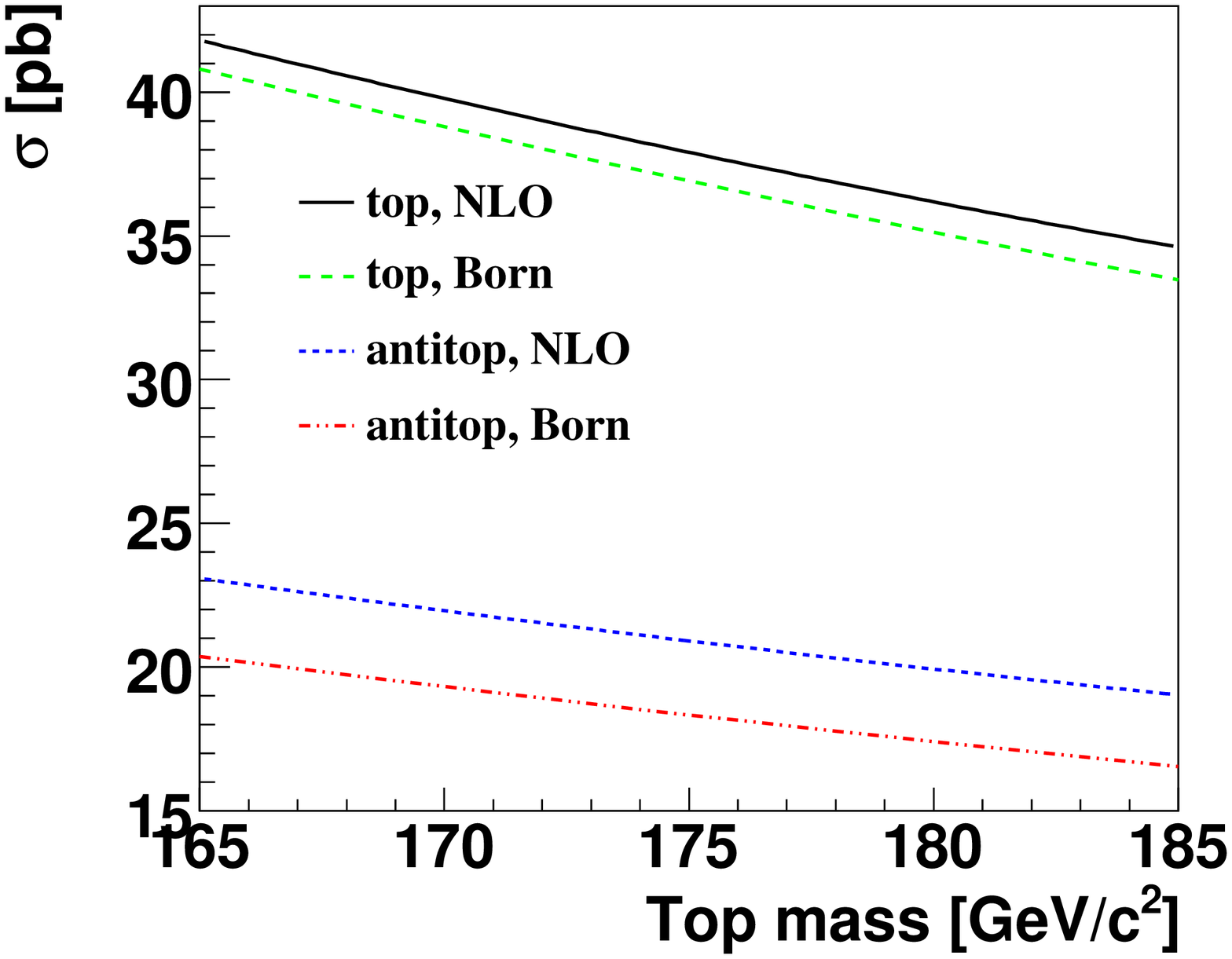}}
\subfigure[]{
\includegraphics[scale=0.33]{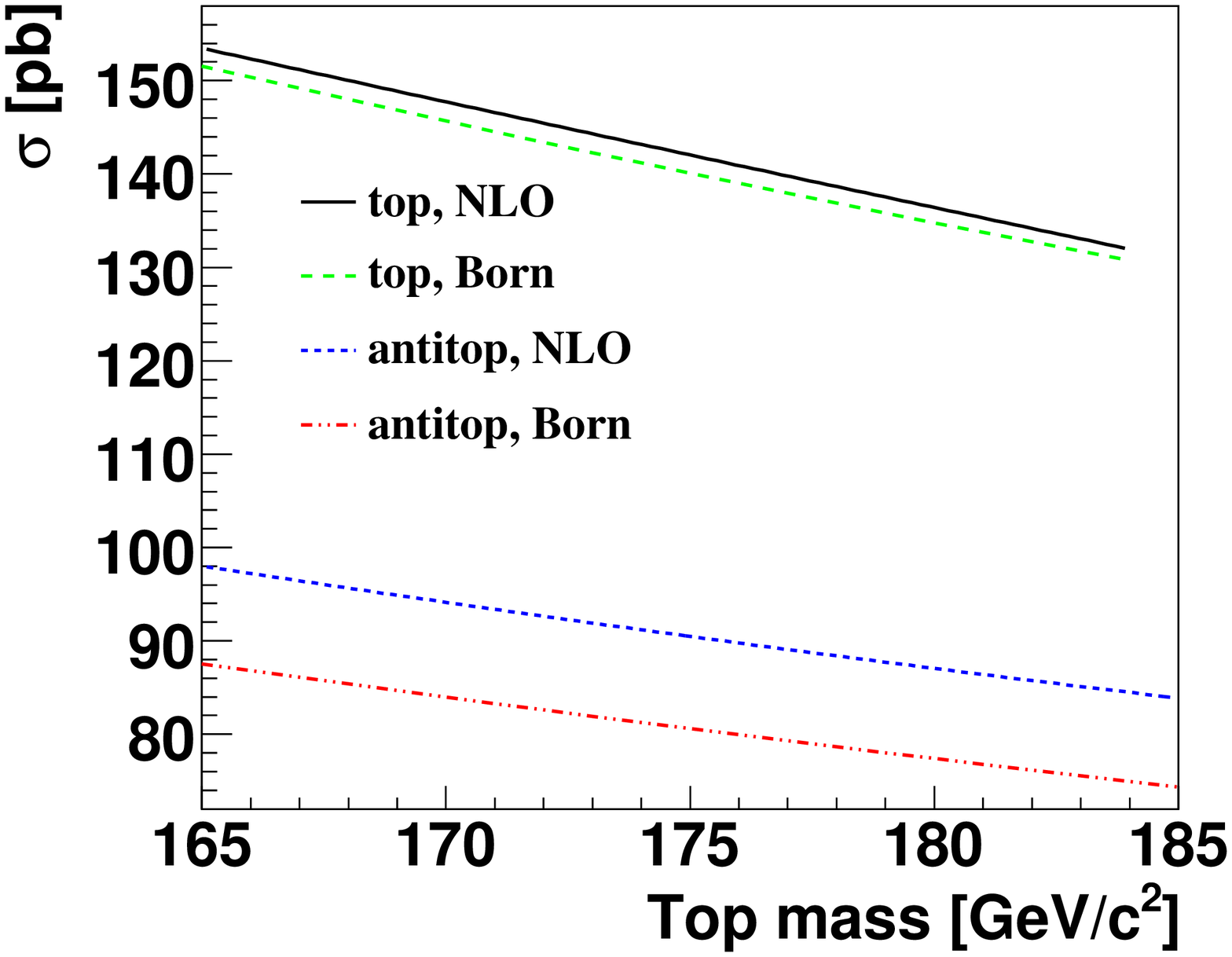}}
\caption{Top quark mass dependence of the inclusive $t$-channel single top
quark cross section (a) at 7~TeV and (b) at 14~TeV for the LHC proton-proton
collider. \label{fig:deps-mt}}

\end{figure}

Figure~\ref{fig:deps-mt} shows the top quark mass dependence of the cross
section for top and antitop quark production, at Born-level and NLO.
Each cross section changes by about $\pm1.1\%$ when the top quark
mass $m_{t}$ is varied by its current uncertainty of 1.1~GeV around
173~GeV~\cite{TevatronElectroweakWorkingGroup:2010yx}. The uncertainty
is larger for lower CM energies, as show in Table~\ref{tab:sys}. 

\begin{figure}[!h!tbp]
\subfigure[]{
\includegraphics[scale=0.33]{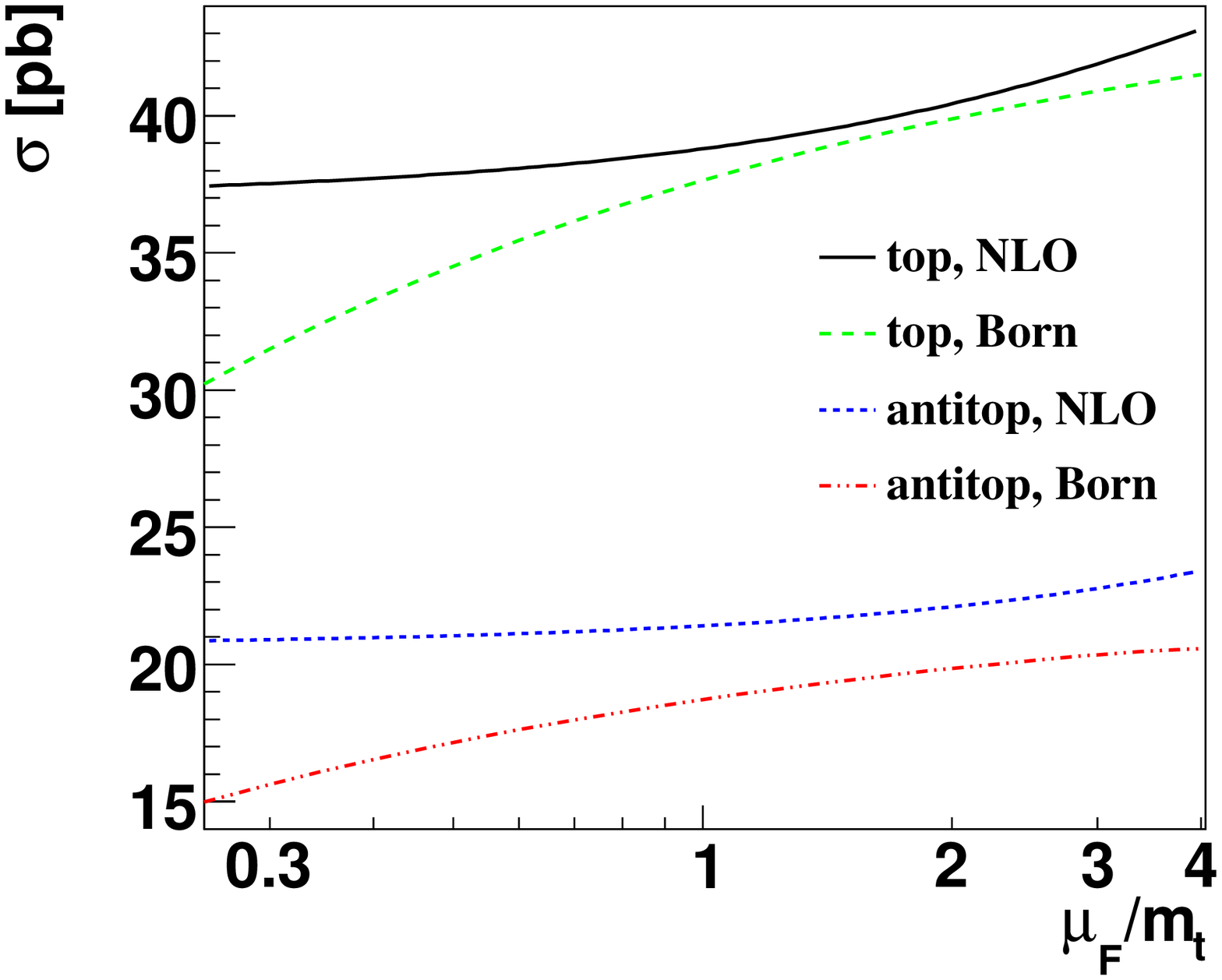}}
\subfigure[]{
\includegraphics[scale=0.33]{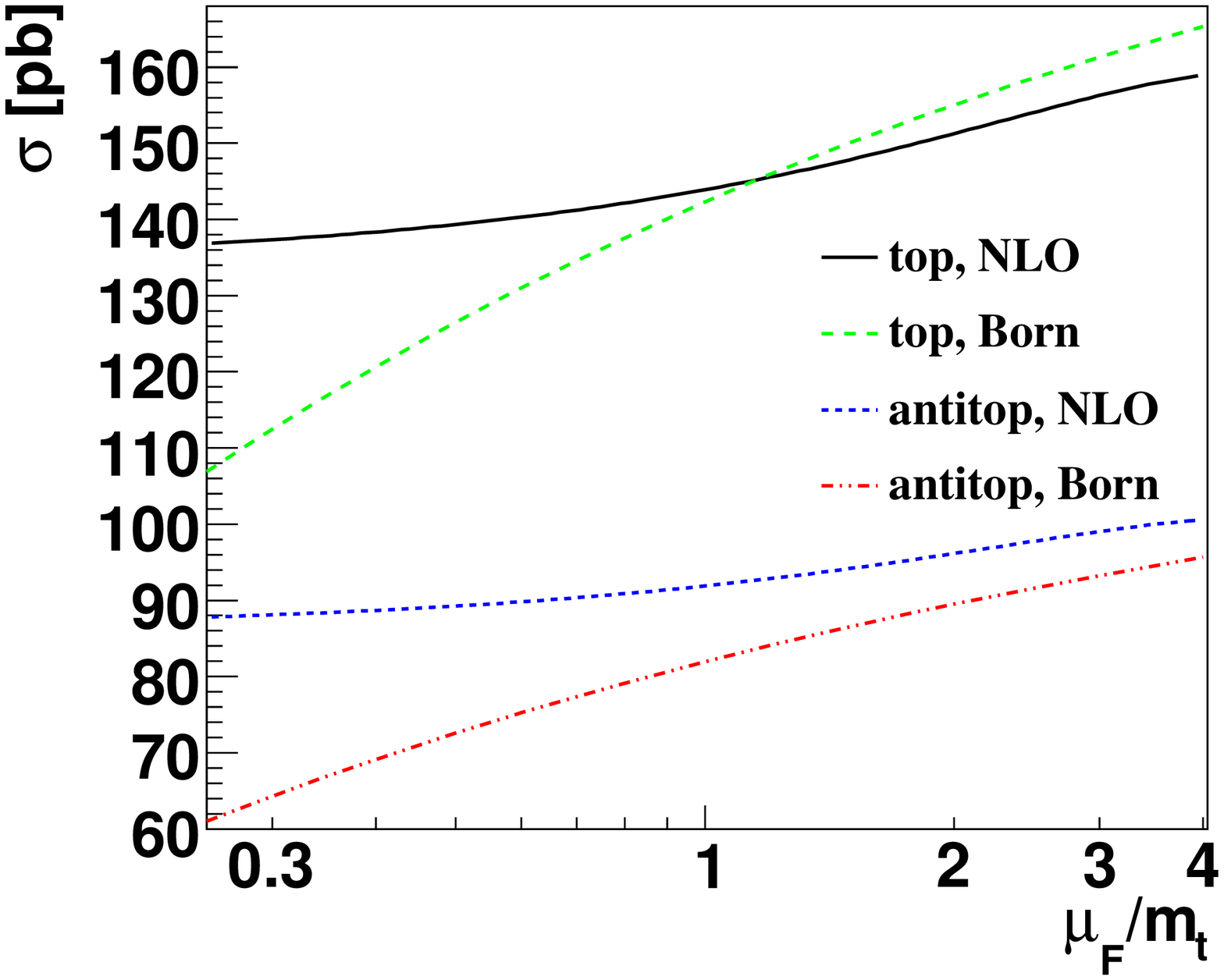}}

\caption{Inclusive $t$-channel single top quark production cross section at
a CM energy of (a) 7~TeV and (b) 14~TeV at the LHC for $m_{t}=173$~GeV,
versus the ratio of the factorization scale $\mu_{F}$ to its typical
value $\mu_{0}=m_{t}$. \label{fig:varyscale}}
\end{figure}

The usual practice for estimating the yet-to-be calculated higher
order QCD correction to a perturbative cross section is to vary the
scale by a factor of 2 up and down. In Fig.~\ref{fig:varyscale}
we show the variation of the total cross section for $t$-channel
single top production for a range of scales around the nominal $\mu_{F}=\mu_{R}=m_{t}$.
The figure shows that the NLO calculation reduces the scale dependence,
which can also be seen in Table~\ref{tab:sys}. 

Figure~\ref{fig:varyscale} also shows that the NLO cross section
is below Born-level, except for top quark production with a large
scale. 

We examine the PDF uncertainty following the standard prescription
using the 44 CTEQ Eigenvectors~\cite{Nadolsky:2008zw,Pumplin:2002vw}.
We evaluate this uncertainty for Born-level and NLO, both top and
antitop production, and three different LHC CM energies in Table~\ref{tab:sys}.
The PDF uncertainty is small and does not change much with top quark
mass or collider CM energy.

\begin{table}
\begin{centering}
\begin{tabular}{|l|c|c|c|c|}
\hline 
 & cross & Mass ($\pm1.1$~GeV) & PDF & Scale\\
 & section {[}pb{]}  & uncertainty (\%) & uncertainty (\%) & uncertainty (\%)\\
\hline
7~TeV, top, Born & 38.6 & 1.1 & 0.5 & 7.1\\ 
7~TeV, top, NLO & 40.0 & 1.1 & 0.5 & 4.4\\
7~TeV, antitop, Born & 19.1 & 1.2 & 0.5 & 7.2\\
7~TeV, antitop, NLO & 21.9 & 1.1 & 0.5 & 3.9\\
\hline 
10~TeV, top, Born & 79.1 & 0.93 & 0.5 & 8.6\\
10~TeV, top, NLO & 80.7 & 0.93 & 0.5 & 4.4\\
10~TeV, antitop, Born & 42.6 & 1.0 & 0.5 & 8.8\\
10~TeV, antitop, NLO & 48.7 & 0.93 & 0.5 & 3.9\\
\hline 
14~TeV, top, Born & 146.5 & 0.87 & 0.5 & 10.0\\
14~TeV, top, NLO & 150.7 & 0.87 & 0.5 & 4.1\\
14~TeV, antitop, Born & 84.3 & 0.91 & 0.5 & 10.0\\
14~TeV, antitop, NLO & 96.1 & 0.87 & 0.5 & 3.7\\
\hline
\end{tabular}
\par\end{centering}

\caption{The top quark mass, PDF and scale uncertainties for the $t$-channel
single top quark production cross section at the LHC for top and antitop
quark production, at Born-level and NLO, for a top quark mass of 173~GeV
and various LHC CM energies. \label{tab:sys}}

\end{table}

\section{Kinematic acceptance studies\label{sec:Single-Top-Acceptance}}

In this section we explore the final state object kinematics of $t$-channel
single top quark events. We focus on the leptonic decay mode of the
$W$-boson from the top quark decay because those are the focus of
the LHC experiments as well~\cite{Gerber:2007xk,Clement:2009zz}.
Therefore, the signature of $t$-channel single top quark events which
is accessible experimentally consists of one charged lepton, missing
transverse energy, together with two or three jets. Since we are studying
the effects of NLO QCD radiative corrections on the production rate
and the kinematic distributions of single top quark events at the
parton level in this paper, we do not include any detector effects,
such as jet energy resolution or $b$-tagging efficiency. Only an
approximation of kinematic acceptances of a generic detector are considered.
In this section, we first present the event topology of the $t$-channel
single top quark process, and then introduce a jet finding algorithm
and the various kinematical cuts used in studying event acceptance.
Unless otherwise specified, figures and notation refer to the production
of top quarks, and we show antitop quark distributions only when different
from those for top quarks. We show distributions for a CM energy
of 7~TeV unless otherwise noted.

\subsection{Event topology\label{sub:Event-Topology}}

\begin{figure}[!h!tbp]
\includegraphics[scale=0.5]{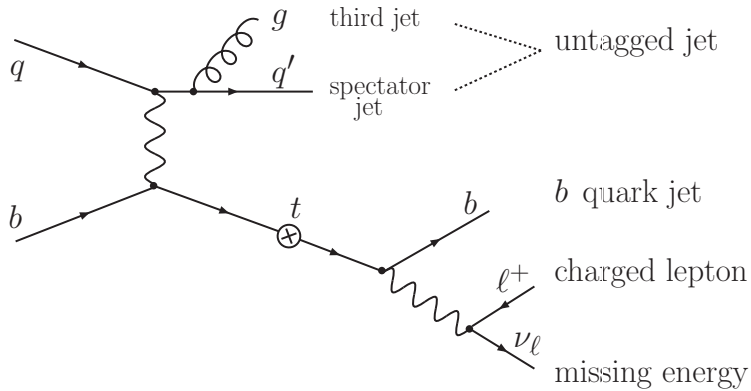}

\caption{Pictorial illustration of the $t$-channel process and the final
state notation used in this paper.\label{fig:notation}}
\end{figure}

At Born-level, the collider signature of the $t$-channel single top
quark process includes two jets (one $b$-tagged jet from the $b$~quark
from the top quark decay, and one non-$b$-tagged jet from the light
quark), one charged lepton, and missing transverse energy ($\met$)
in the final state. This signature becomes more complicated beyond
Born-level, as Figure~\ref{fig:notation} indicates. We use the same
notation as in Ref.~\cite{Cao:2005pq}: the light quark jet is also
called {}``spectator jet'', and the label {}``untagged jet'' refers
to all jets which do not contain a $b$~or $\bar{b}$~quark.

\begin{figure}[!h!tbp]
\includegraphics[width=0.7\linewidth]{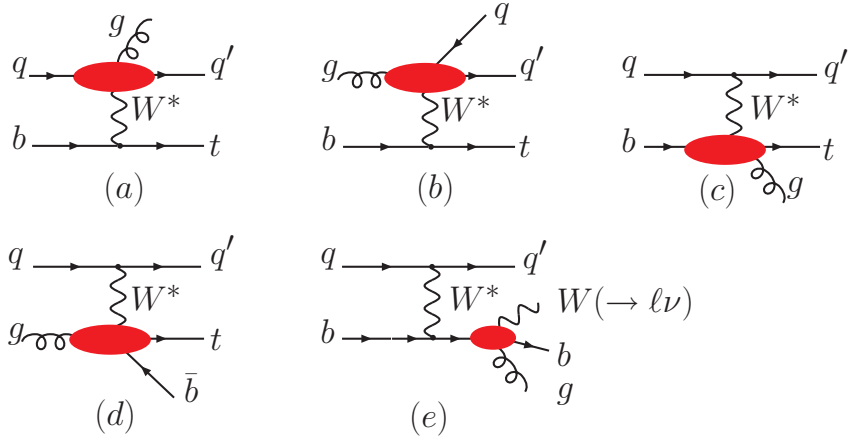}

\caption{Representative diagrams of the real emission corrections to $t$-channel
single top quark production: (a) and (b) represent the real radiative
corrections to the LIGHT quark line, while (c) and (d) represent the
real radiative corrections to the HEAVY quark line, and (e) represents
the real radiative corrections to the top quark decay. The NLO QCD
corrections are indicated by the large shaded ellipse. Detailed Feynman
diagrams can be found in Ref.~\cite{Cao:2004ky}.\label{fig:real_tchan}}
\end{figure}

At NLO, besides the charged lepton and $\met$, there may be two jets
(one $b$-tagged jet and one untagged jet) as for the Born-level,
or there may be three jets. The flavor composition of the three-jet
final state depends on the origin of the third jet. When it is a gluon
or anti-quark (cases (a-c) in Fig.~\ref{fig:real_tchan}), there
is one $b$-tagged jet and two untagged jets. When it is a $\bar{b}$~quark
(case (d) in Fig.~\ref{fig:real_tchan}, also called $W$-gluon fusion),
there are two $b$-tagged jets and one untagged jet. Therefore, prescriptions
are needed to identify the $b$~quark jet from the top quark decay
and the light quark jet produced with the top quark and to separate
them from the additional jet.

The unique signature of the $t$-channel single top quark process
is the spectator jet in the forward direction, and the kinematics
of this jet are used to suppress the copious backgrounds from $t\bar{t}$
and $Wb\bar{b}$ production. Studying the kinematics of this spectator
jet is important in order to have a better prediction of the acceptance
of $t$-channel single top quark events and of the distribution of
several important kinematic variables. In this work, we study the
impact of the NLO QCD corrections on the kinematic properties of the
spectator jet at the LHC. As pointed out in Ref.~\cite{Yuan:1989tc},
in the effective-$W$ approximation, a high-energy $t$-channel single
top quark event is dominated by the fusion diagram of a longitudinal
$W$~boson and a $b$~quark. This effective longitudinal $W$~boson
is also found in the production of a heavy Higgs boson via the $WW$
fusion process, hence understanding the effective longitudinal $W$~boson
and the kinematics of the light quark jet in $t$-channel single top
quark production is essential to better predict the kinematics of
Higgs boson events via $WW$ fusion. We will show that the spectator
jet is not uniquely identified anymore at NLO and will compare proposed
solutions to this problem. In our Tevatron study~\cite{Cao:2005pq},
we differentiated three categories of events, based on final state
jet multiplicity and flavor composition. Here we will explore the
same three cases:
\begin{enumerate}
\item Born-level-type exclusive two-jet events (containing the $b$~quark
and the light quark), which have unique quark-jet assignments because
the $b$-tagged jet is identified as the $b$~quark jet, and the
untagged jet is assigned to the spectator jet.
\item Exclusive three-jet events where the additional jet is a gluon. In
this case it is straightforward to identify the $b$~quark jet from
the top quark decay correctly, but the spectator jet is not uniquely
identified anymore.
\item Exclusive three-jet events where the additional jet is a $\bar{b}$~quark.
This is the $2\to3$ process for which NLO corrections have
been calculated~\cite{Campbell:2009ss}. If both of these jets are
tagged, then the spectator jet is uniquely identified but the $b$~quark
jet from the top quark decay is not. If only one of the jets is tagged,
then both the spectator quark and the $b$~quark from the top quark
decay are potentially mis-identified.
\end{enumerate}
We will demonstrate that the fraction of events in these three categories
shows significant variation depending on the event reconstruction
details.

\subsection{Acceptance\label{sub:Acceptance}}

In order to meaningfully discuss the effects of gluon radiation in
single top quark events, we must define a jet as an infrared-safe
observable. In this study, we adopt the cone-jet algorithm~\cite{Alitti:1990aa}
as explained in our previous work~\cite{Cao:2004ap,Cao:2005pq}.
More specifically, we adopt the $E$-scheme cone-jet approach (4-momenta
of particles in a cone are simply added to form a jet) with radius
$R=\sqrt{\Delta\eta^{2}+\Delta\phi^{2}}$ in order to define $b$,
$q$ and possibly extra $g$, $\bar{q}$, or $\bar{b}$ jets, where
$\Delta\eta$ and $\Delta\phi$ are the separation of particles in
the pseudo-rapidity $\eta$ and the azimuthal angle $\phi$, respectively.
In recent studies, the LHC collaborations ATLAS and CMS have used cone
sizes of $R=0.4$ and $R=0.5$, respectively~\cite{Gerber:2007xk,Aad:2009wy,CMS-TDR-II-0954-3899-34-6-S01}.
We will consider both in this paper, as well as a larger cone size
$R=1.0$ for comparison. The same $R$ value will also be required
for the separation between the lepton and each jet, i.e. lepton isolation.

The kinematic cuts imposed on the final state objects are:\begin{eqnarray}
P_{T}^{\ell}\ge25\,\mbox{GeV} & , & \left|\eta_{\ell}\right|\le\eta_{\ell}^{max},\nonumber \\
\met\ge\met^{min} & , & E_{T}^{j1,2}\ge E_{T}^{j1,2\, min},\nonumber \\
E_{T}^{j3}\ge15\,\mbox{GeV} & , & \left|\eta_{j}\right|\le5.0,\nonumber \\
\Delta R_{\ell j}\ge R_{cut} & , & \Delta R_{jj}\ge R_{cut},\label{eq:cuts}\end{eqnarray}
where the jet cuts are applied to both the $b$- and light quark jets
as well as any additional gluon or quark jets in the final state.
We consider two different sets of cuts in the following: a loose set
of cuts corresponding to the basic experimental event selection, and
a tight set of cuts used by the experiments to separate $t$-channel
single top events from the backgrounds. The loose cuts utilize a lepton
pseudorapidity corresponding to the full detector,$\eta_{\ell}^{max}=2.5$
and require$\met^{min}=25$~GeV and $E_{T}^{j1,2\, min}=25$~GeV
for the leading two jets. The tight cuts restrict leptons to the central
detector, $\eta_{\ell}^{max}=1.5$, and require$\met^{min}=40$~GeV
and $E_{T}^{j1,2\, min}=50$~GeV for the leading two jets in order
to suppress backgrounds. Each event is furthermore required to have
at least one lepton and two jets passing all selection criteria. The
cut on the separation in $R$ between the lepton and the jets as well
as between different jets is given by $R_{cut}$. In addition, the
$b$~quark jet from the top decay is required to be in the central
detector ($|\eta^{b-\mbox{jet}}|<2.5$), and the other high-$E_{T}$~jet
is required to be at a high pseudo-rapidity, ${|\eta}^{jet}|>2.5$.

\begin{table}
\begin{centering}
\begin{tabular}{|l|r|r|r|r|r|r|r|r|}
\hline 
$\sigma\times Br$ [pb] & \multicolumn{4}{c|}{7~TeV} & \multicolumn{4}{c|}{14~TeV}\\
\hline
 & \multicolumn{2}{c|}{top} & \multicolumn{2}{c|}{antitop} & \multicolumn{2}{c|}{top} & \multicolumn{2}{c|}{antitop}\\
 & LO & NLO & LO & NLO &  LO  &  NLO  & LO & NLO\\
\hline 
loose cuts, $R_{cut}=0.4$ & & & & & & & & \\
~~$tq+tqj$ & 1.66 & 1.75 & 0.83 & 0.96 & 6.2 & 6.6 & 3.6 & 4.2\\
~~$tqj$ &  & 1.28 &  & 0.64 &  & 5.5 &  & 3.2\\
\hline
loose cuts, $R_{cut}=0.5$  & & & & & & & & \\
~~$tq+tqj$ & 1.65 & 1.75 & 0.82 & 0.96 & 6.1 & 6.6 & 3.6 & 4.2\\
~~$tqj$ &  & 1.19 &  & 0.60 &  & 5.1 &  & 3.0\\
\hline
tight cuts, $R_{cut}=0.4$  & & & & & & & & \\
~~$tq+tqj$  & 0.12 & 0.14 & 0.06 & 0.07 & 0.61 & 0.72 & 0.29 & 0.36\\
~~$tqj$ &  & 0.089 &  & 0.05 &  & 0.60 &  & 0.29\\
\hline
tight cuts, $R_{cut}=0.5$  & & & & & & & & \\
~~$tq+tqj$  & 0.12 & 0.14 & 0.05 & 0.06 & 0.61 & 0.73 & 0.29 & 0.36\\
~~$tqj$ &  & 0.084 &  & 0.03 &  & 0.57 &  & 0.27\\
\hline
\end{tabular}
\par\end{centering}

\caption{The $t$-channel single top (antitop) quark production cross sections
times decay branching ratio $t\to bW^{+}\to be^{+}\nu$
($\bar{t}\to\bar{b}W^{-}\to\bar{b}e^{-}\bar{\nu}$)
at the 7~TeV and 14~TeV LHC under various cut scenarios listed in
the text. \label{tab:total}}
\end{table}

Table~\ref{tab:total} shows the single top quark production
cross sections in pb for the loose and tight set of cuts listed in
Eq.~(\ref{eq:cuts}) for different CM energies and jet cone sizes.
The acceptance for the loose cuts is about 4.3\%, similar for top
and antitop quarks and slightly higher at NLO than at Born-level.
Differences between LO and NLO come out when going to the tight set
of cuts. From loose to tight, the Born-level top (antitop) quark acceptance
goes down by a factor of ten (twelve). The decrease is slightly smaller
at NLO and for exclusive 3-jet events. Changing $R_{cut}$ from 0.4
to 0.5 does not change the acceptance very much, but it reduces the
number of 3-jet events, mainly because more gluon radiation is clustered
into the $b$~quark and light quark jets. For the loose cuts, the
increase from Born-level to NLO is comparable to the inclusive cross
section increase; whereas for the tight cuts, the increase from Born-level
to NLO is larger, 20\% for top quarks and 25\% for antitop quarks.
This is due to the HEAVY correction, which contributes 20\% of the
NLO cross section after tight cuts, and is positive for both top and
antitop. For the loose set of cuts, the HEAVY correction contributes
about 10\% of the NLO cross section, again with a positive sign for
both top and antitop quarks.

\begin{figure}[!h!tbp]
\subfigure[]{
\includegraphics[scale=0.33]{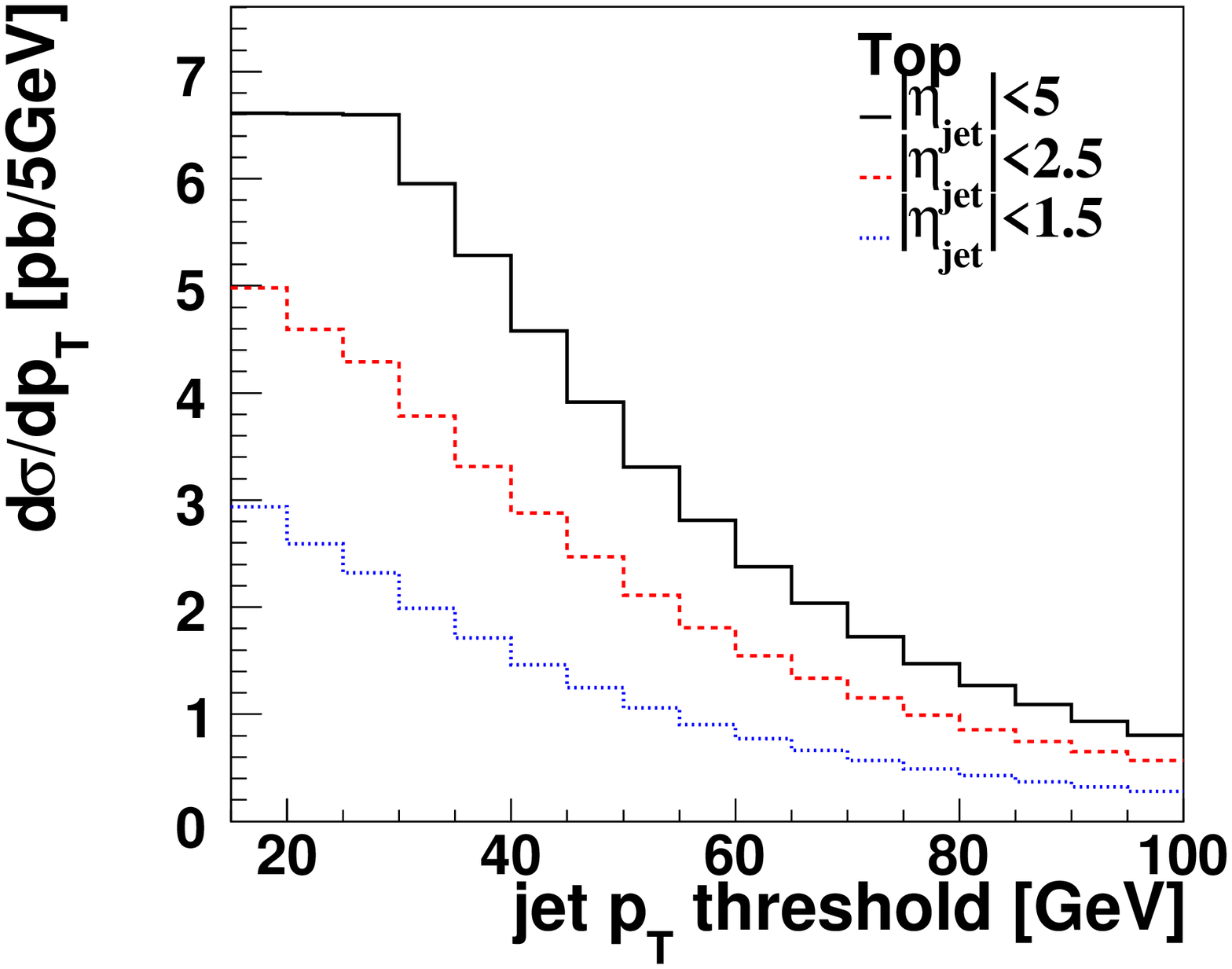}}
\subfigure[]{
\includegraphics[scale=0.33]{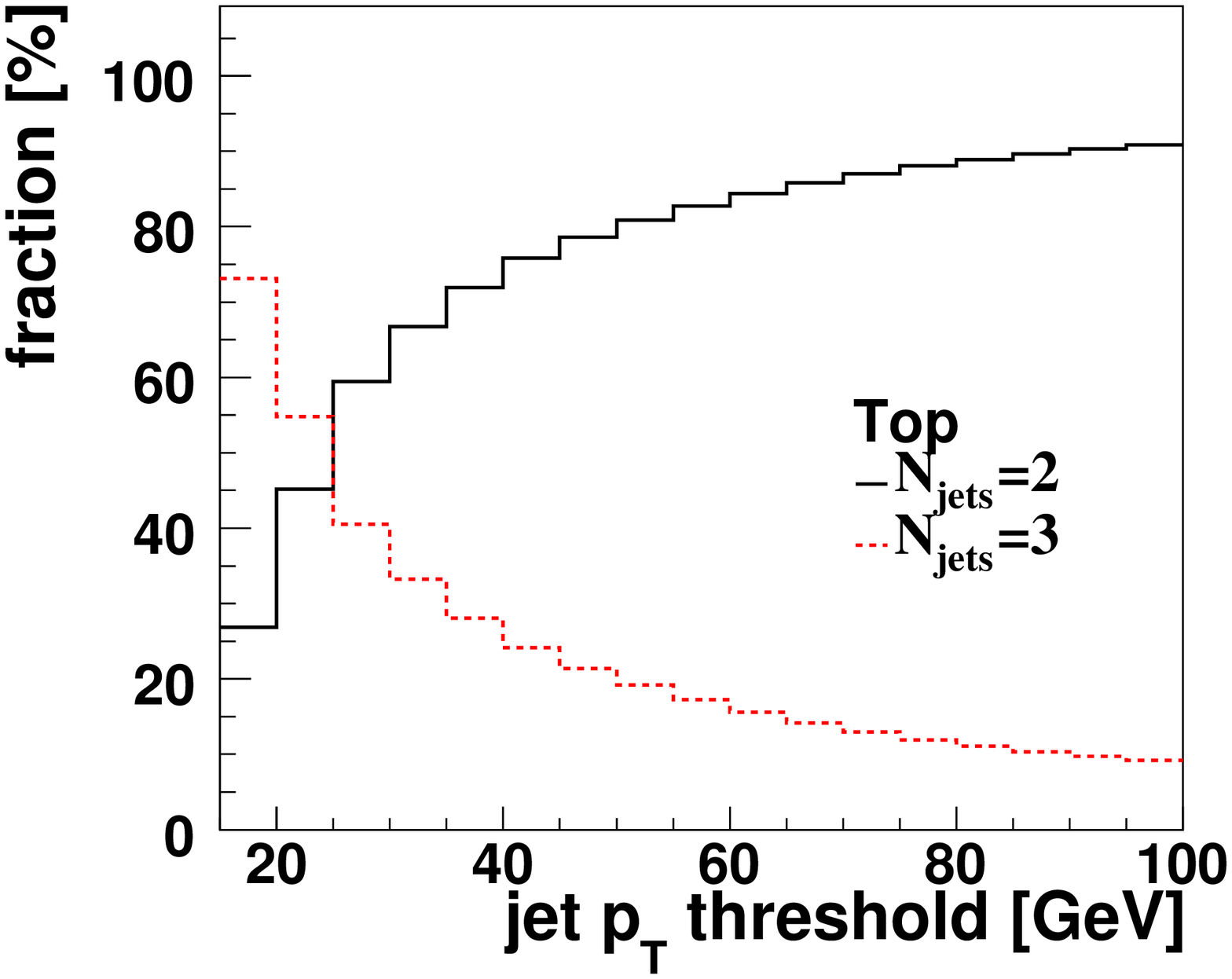}}

\caption{Cross section and fraction of three-jet events at the 7~TeV LHC at
NLO for varying jet $p_{T}$ cuts, requiring only that $n_{jets}\geq2$,
and not making any cuts on the electron or neutrino. Shown is the
total cross section for events with two or three jets as a function
of the jet $E_{T}$ cut for three different jet pseudo-rapidity cuts
(a) and the fraction of two-jet and three-jet events as a function
of jet $p_{T}$ (b). The jet cone size is 0.4 and the lowest threshold
considered is 15~GeV. \label{fig:njets_jet_pt}}

\end{figure}

Figure~\ref{fig:njets_jet_pt}(a) shows how the cross section changes
as a function of the jet $E_{T}$ cut when applying the loose set
of cuts (without lepton requirements but including a requirement of
there being at least two jets in the event). The figure also shows
the dependence of the fraction of two-jet events and three-jet events
on the jet $E_{T}$ cut. There are only two-jet events at Born-level, 
whereas $O(\alpha_{s})$ corrections can produce an additional
soft jet. The fraction of events with these additional jets is as
high as 83\% for a jet threshold of 15~GeV (and a jet clustering
cone size of 0.4), and drops off to about 10\% for a jet threshold
of 100~GeV. Since the typical jet $E_{T}$ thresholds considered
by experiments are in the range of 15~GeV to 25~GeV, the fraction
of 3-jet events will be very high, and it will be important to study
3-jet $t$-channel single top quark events at the LHC in detail.

For antitop quark events the distribution is similar, but the fraction
of 2-jet events for the lowest threshold of 15~GeV is higher, 33\%
(compared to 26\% for top quarks). At a CM energy of 14~TeV (10~TeV),
the fraction of 2-jet events in top quark production decreases to
17\% (21\%) as a result of the additional available phase space for
gluon radiation.

We will use the loose set of cut values for the following discussion:
$\eta_{l}^{max}=2.5$, $\eta_{j}^{max}=5.0$, and $R_{cut}=0.4$,
$E_{Tj}^{min}=15$~GeV, cf. Eq.~(\ref{eq:cuts}).

\section{Event distributions\label{sec:EventDistr}}

In this section we examine the kinematic properties of $t$-channel
single top quark events. The $t$-channel final state includes
one $b$-tagged jet, one untagged jet, one charged lepton, and missing
energy at Born-level, thus it is straightforward to reconstruct the
top quark from the $b$-tagged jet and the reconstructed $W$~boson,
and to identify the light quark jet as the untagged jet. At NLO, however,
an additional jet can be radiated, which will complicate the reconstruction
of the top quark final state. First, the additional jet can be either
a $b$-tagged jet or an untagged jet. When it is a $b$-tagged jet,
we need to select which of the two $b$-tagged jets corresponds to
the $b$~quark jet from the top quark decay. Here we always choose
the highest $p_{T}$ $b$-tagged jet. Similarly, when the additional
jet is an untagged jet, we need to select which of the two untagged
jets is the spectator jet. We will explore two different methods to
resolve this ambiguity: selecting the leading untagged jet and selecting
the most forward untagged jet. Second, the additional untagged jet
can come from either the production or the decay of the top quark.
Production-stage emission occurs before the top quark goes on shell,
thus the $W$~boson and $b$~quark momenta will combine to give
the top quark momentum. Decay-stage emission occurs only after the
top quark goes on shell, thus the gluon momentum must also be included
in order to reconstruct the top quark momentum properly. 

We also examine various kinematic distributions of the final state
particles and then study the effects of NLO corrections on distributions
concerning the reconstructed top quark, in particular spin correlations
between the final state particles. Finally, we explore the impact
of the radiated jet in exclusive three-jets events. We use only the
loose set of cuts to maximize the efficiency when examining the distributions
and efficiencies in detail.

\subsection{Final state object distributions\label{sub:Final-State-Object}}

\subsubsection{Charged lepton and missing transverse energy}

In this section we examine various kinematic distributions of final
state objects after event reconstruction and after applying the loose
set of cuts, cf. Table~\ref{tab:inclusive} and Eq.~(\ref{eq:cuts}).
We concentrate on inclusive two-jet events in this section because
they give more reliable infrared-safe predictions. 

\begin{figure}[!h!tbp]
\subfigure[]{
\includegraphics[width=0.33\linewidth]{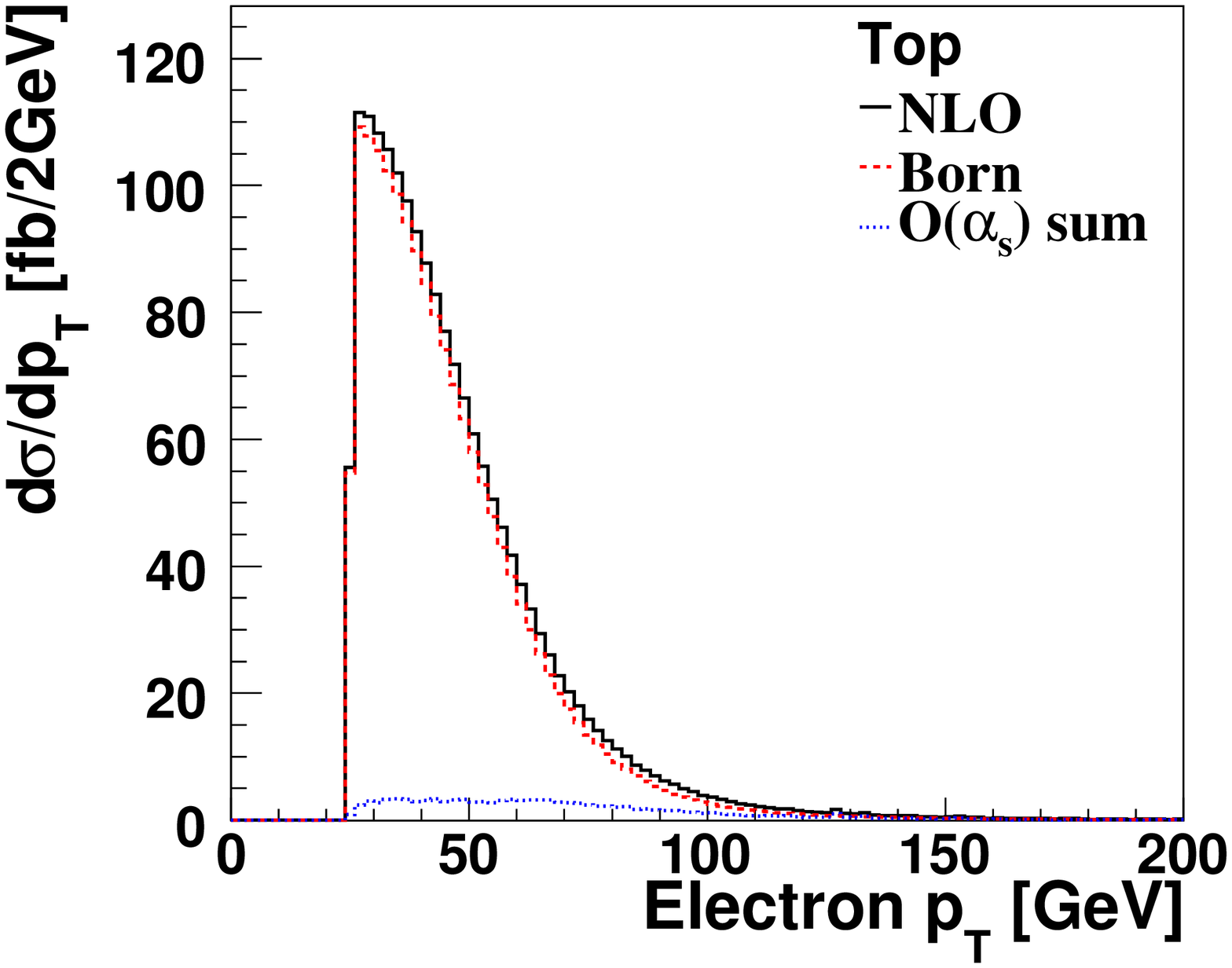}}
\subfigure[]{
\includegraphics[width=0.33\linewidth]{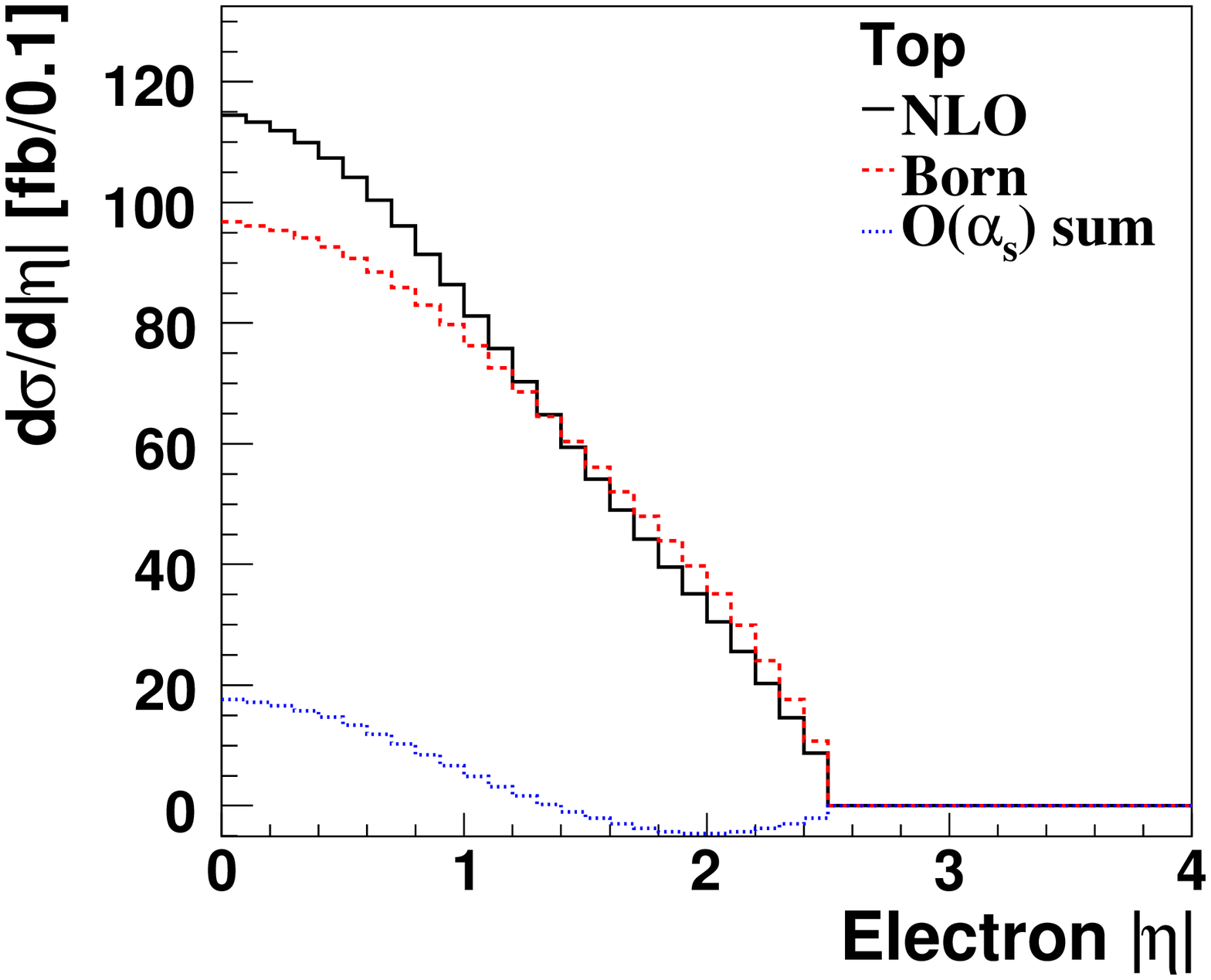}}
\subfigure[]{
\includegraphics[width=0.33\linewidth]{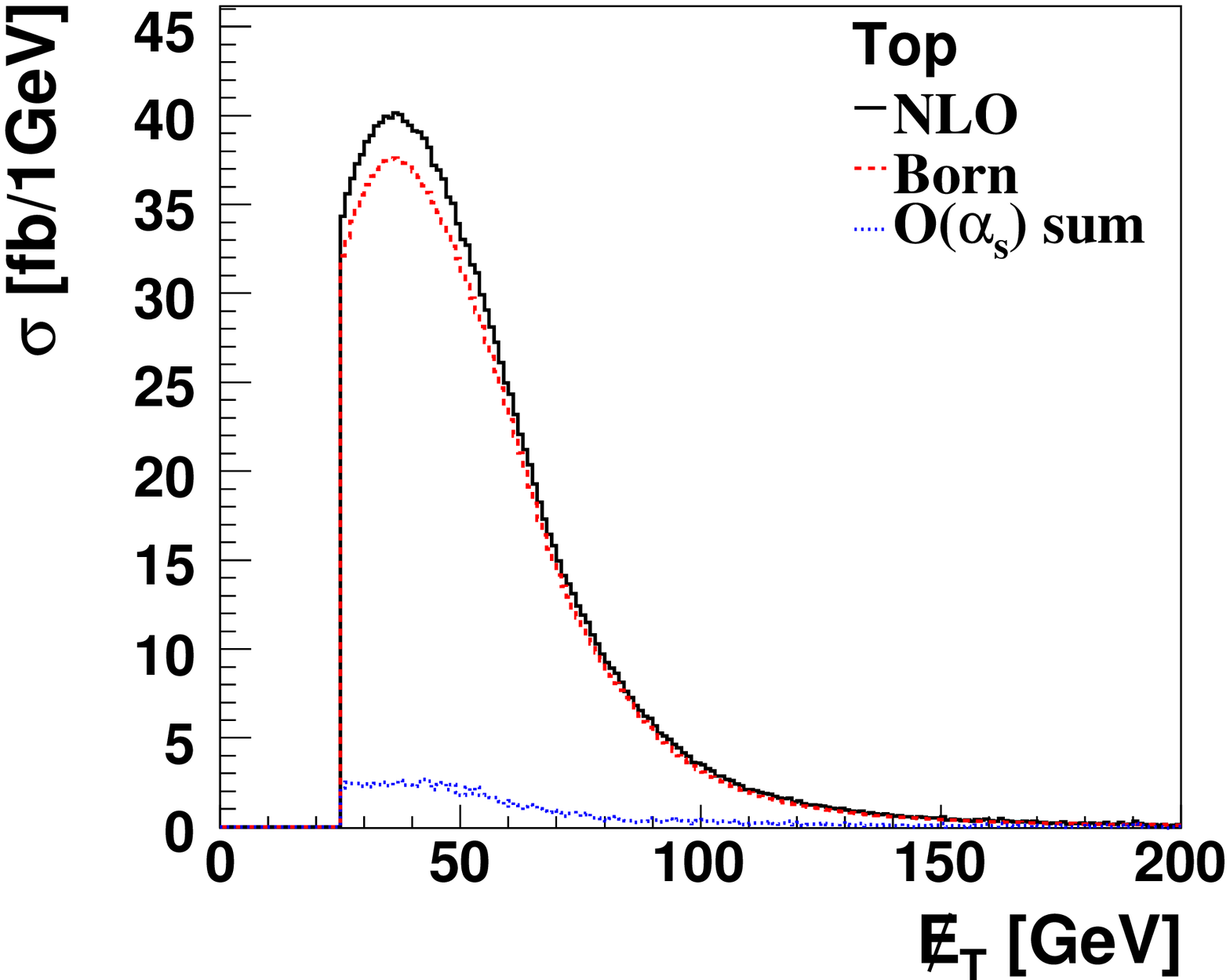}}

\caption{(a) transverse momentum of the electron and (b) its pseudo-rapidity
and (c) missing transverse energy, after selection cuts, comparing
Born-level to $\oalphas$ corrections, for top quark production at
the 7~TeV LHC.\label{fig:pte-etae}}
\end{figure}

Figure~\ref{fig:pte-etae} show the transverse momentum of the electron
and its pseudorapidity as well as the missing transverse energy for
top quark events. The antitop quark distributions are very similar.
The lepton transverse momentum is typically smaller than the missing
transverse energy because the neutrino from the $W$-boson decay moves
preferentially along the direction of the top quark, both for top
and antitop quark production. This is due to the left-handed nature
of the charged current interaction and can easily be seen when examining
the spin correlations between the charged lepton and the top quark
in the top quark rest frame. The $\oalphas$ corrections do not change
the distributions much. The pseudo-rapidity distribution of the electron
is given in Fig.~\ref{fig:pte-etae}(b). Since the LHC is a proton-proton
collider, all pseudorapidity ($\eta$) distributions will be symmetric
about zero and we thus present distributions of the absolute value
$|\eta|$. This distribution is more central for NLO events than for
Born-level events because the LIGHT and HEAVY $\oalphas$ corrections
tend to reduce the $z$-momentum imbalance of the two initial state
partons. We will comment more on the subject of angular correlations
in Sec.~\ref{sub:Object-Correlations}.

\subsubsection{Spectator jet}

\begin{figure}[!h!tbp]
\subfigure[]{
\includegraphics[width=0.33\linewidth]{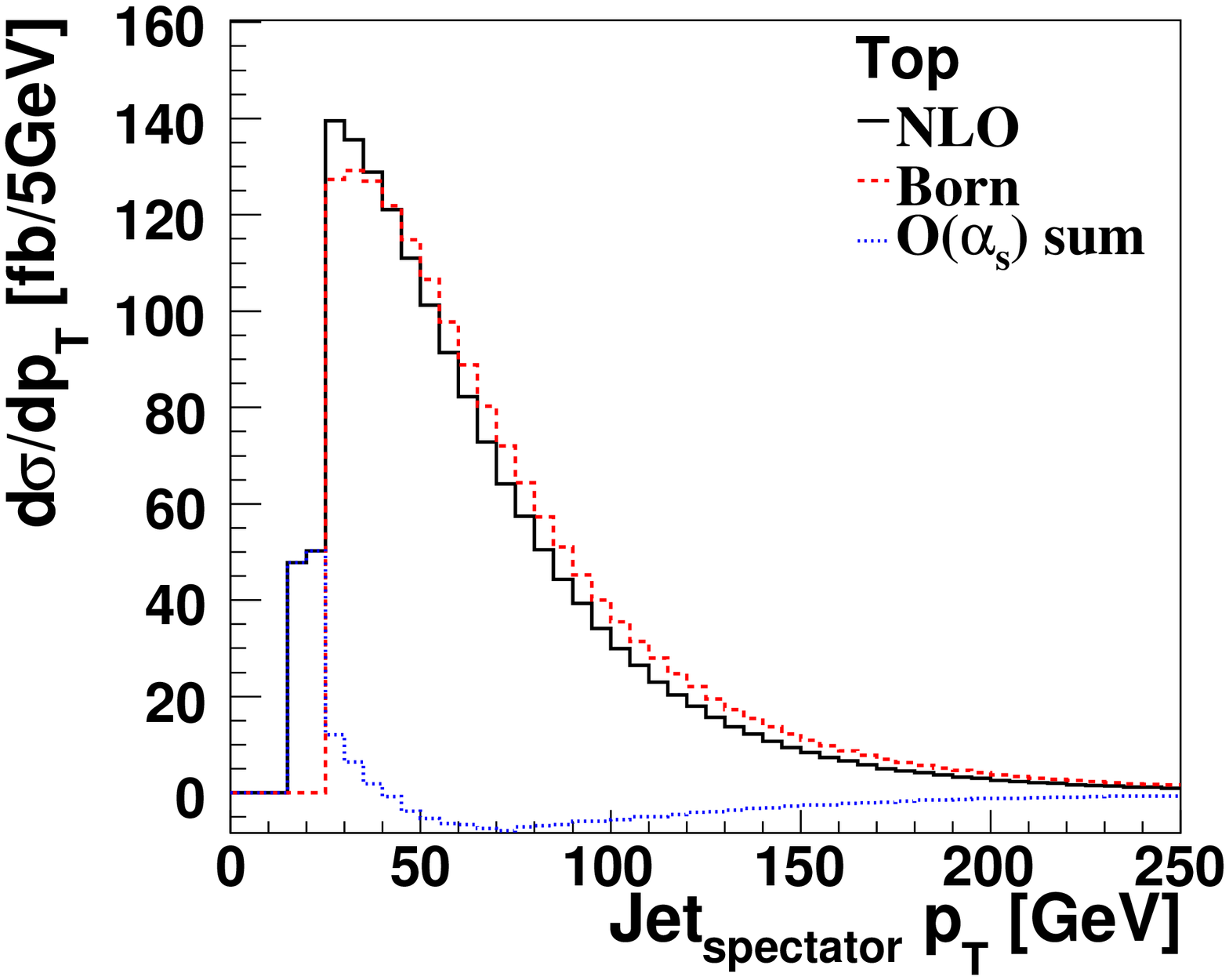}}
\subfigure[]{
\includegraphics[width=0.33\linewidth]{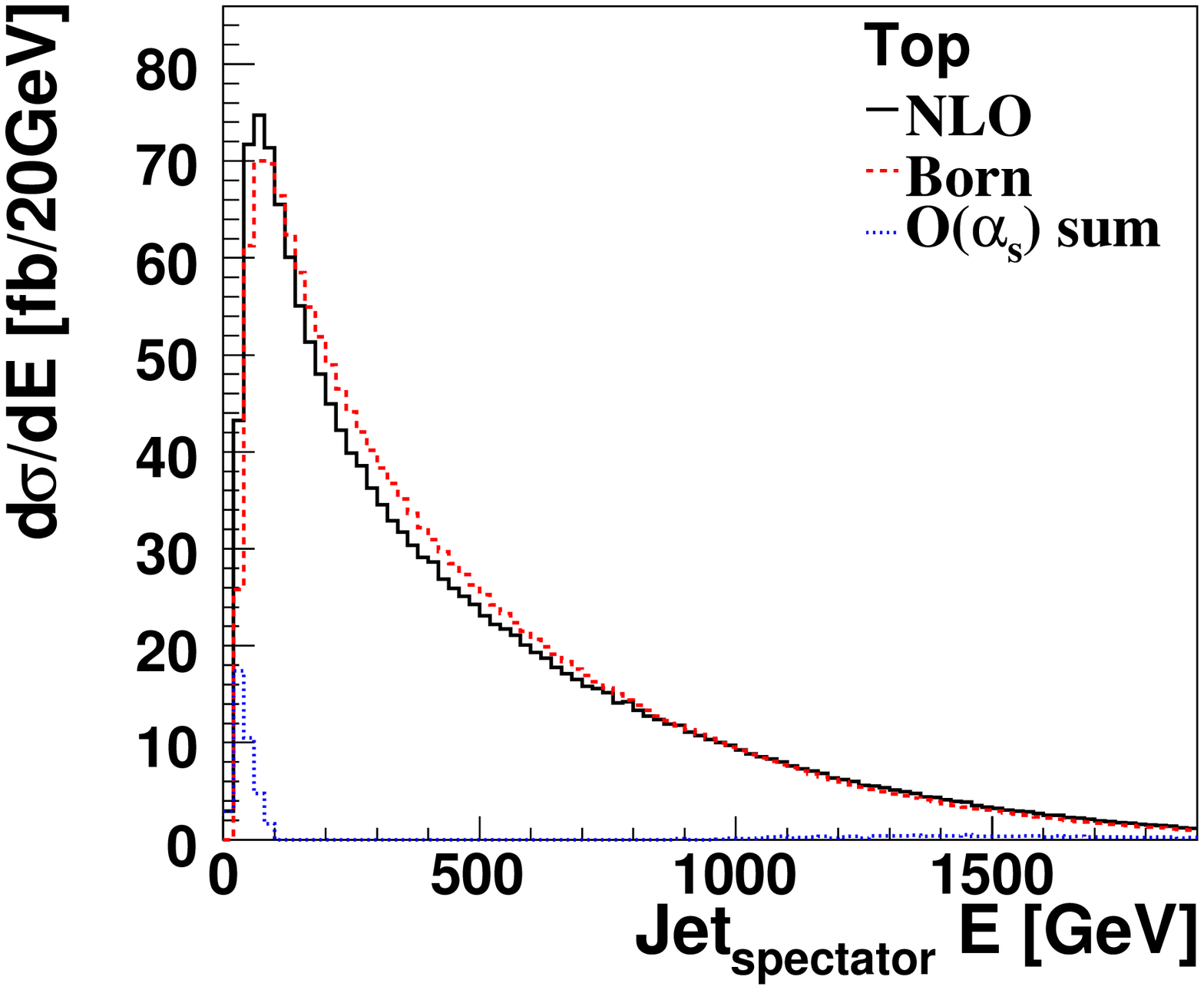}}

\subfigure[]{
\includegraphics[width=0.33\linewidth]{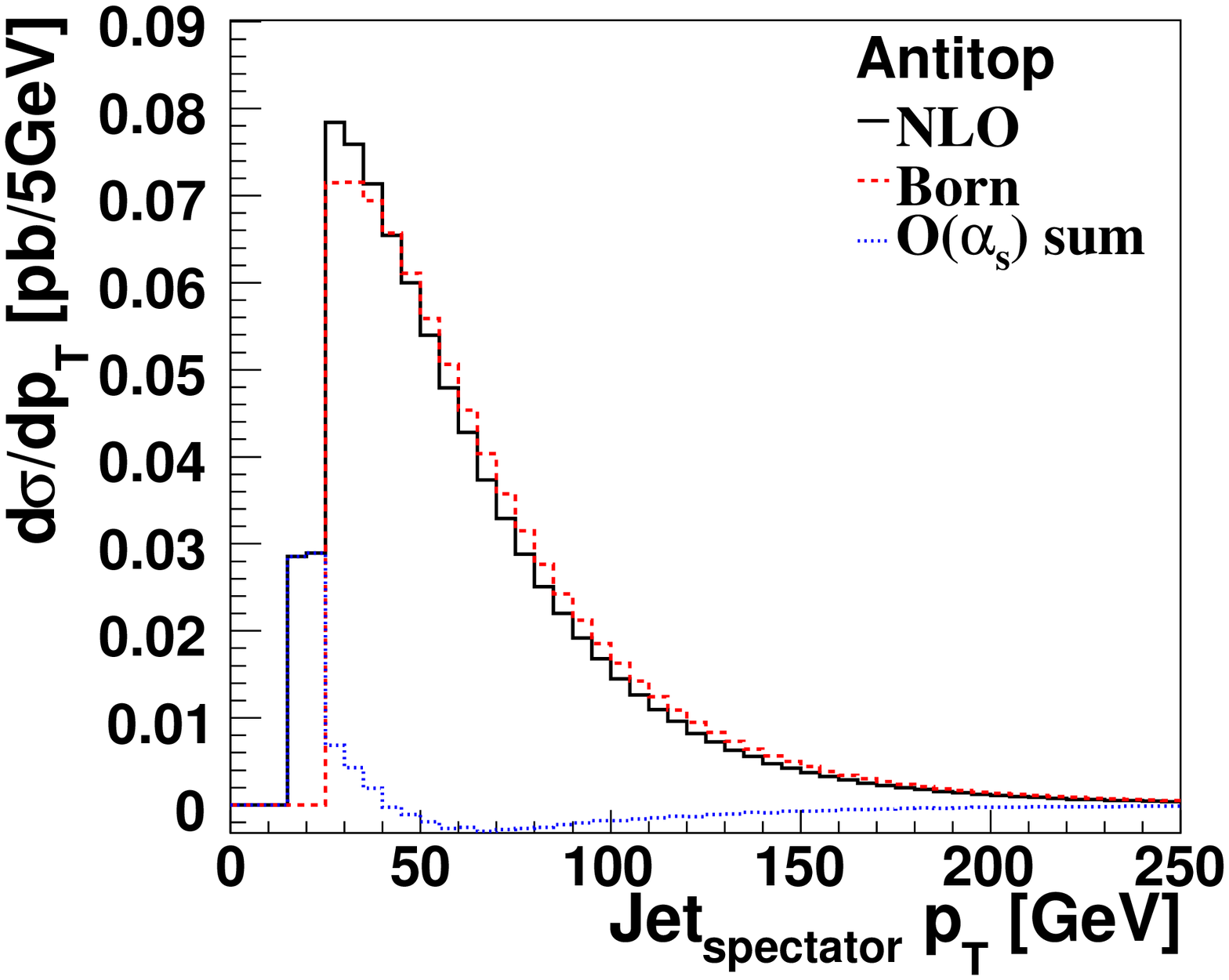}}
\subfigure[]{
\includegraphics[width=0.33\linewidth]{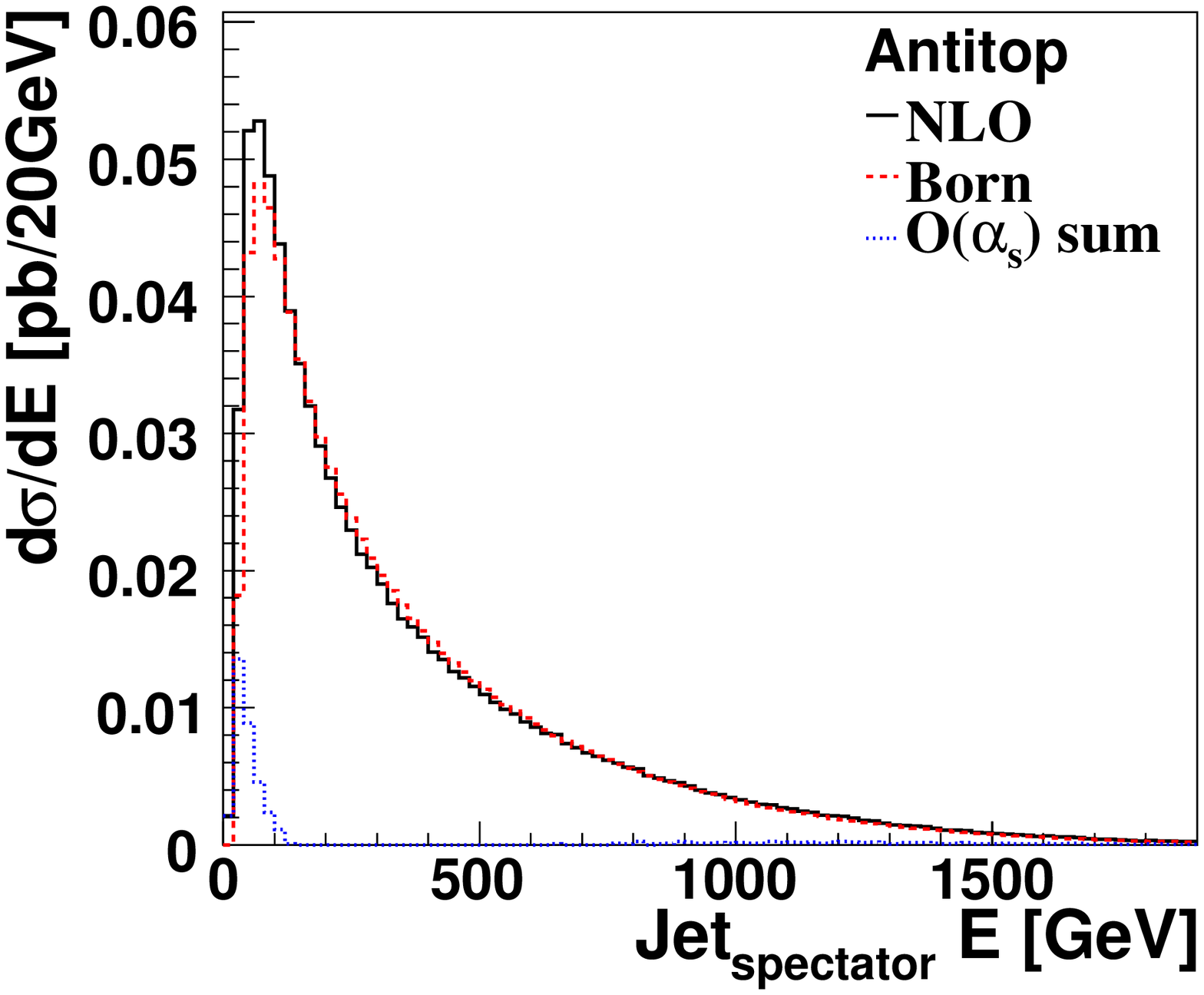}}

\caption{(a, c) transverse momentum and (b, d) energy of the spectator jet
after selection cuts, comparing Born-level to $\oalphas$ corrections,
(a, b) for top quarks and (c, d) antitop quarks at the 7~TeV LHC.
\label{fig:spectator}}
\end{figure}

One of the unique features of $t$-channel single top quark production
is the light quark jet. This spectator jet can be used to disentangle
$t$-channel single top quark events from the copious backgrounds.
Therefore, its kinematic distributions need to be well studied, especially
the impact of $O(\alpha_{s})$ corrections. The transverse momentum
and energy of the spectator jet are shown in Figure~\ref{fig:spectator}.

Since the spectator jet comes from the initial state quark after emitting
the effective $W$~boson, its transverse momentum peaks around $\sim M_{W}/2$
as can be seen in Figure~\ref{fig:spectator}. The spectator jet
$p_{T}$ distribution in top quark events is slightly higher than
in antitop quark events due to the typically harder PDF for up quarks
versus down quarks. The $O(\alpha_{s})$ corrections lower the transverse
momentum and energy distributions of the spectator jet in both cases
due to the LIGHT correction. The spectator jet $p_{T}$ distributions
also have a jump at 25~GeV due to events with a spectator jet $p_{T}<25$~GeV
which fail the loose selection cuts unless a third jet from real emission
has $p_{T}>25$~GeV. The spectator jet distributions for a CM energy
of 14~TeV are similar, except that the tail at high spectator jet
energies extends farther.

\begin{figure}[!h!tbp]
\subfigure[]{
\includegraphics[width=0.33\linewidth]{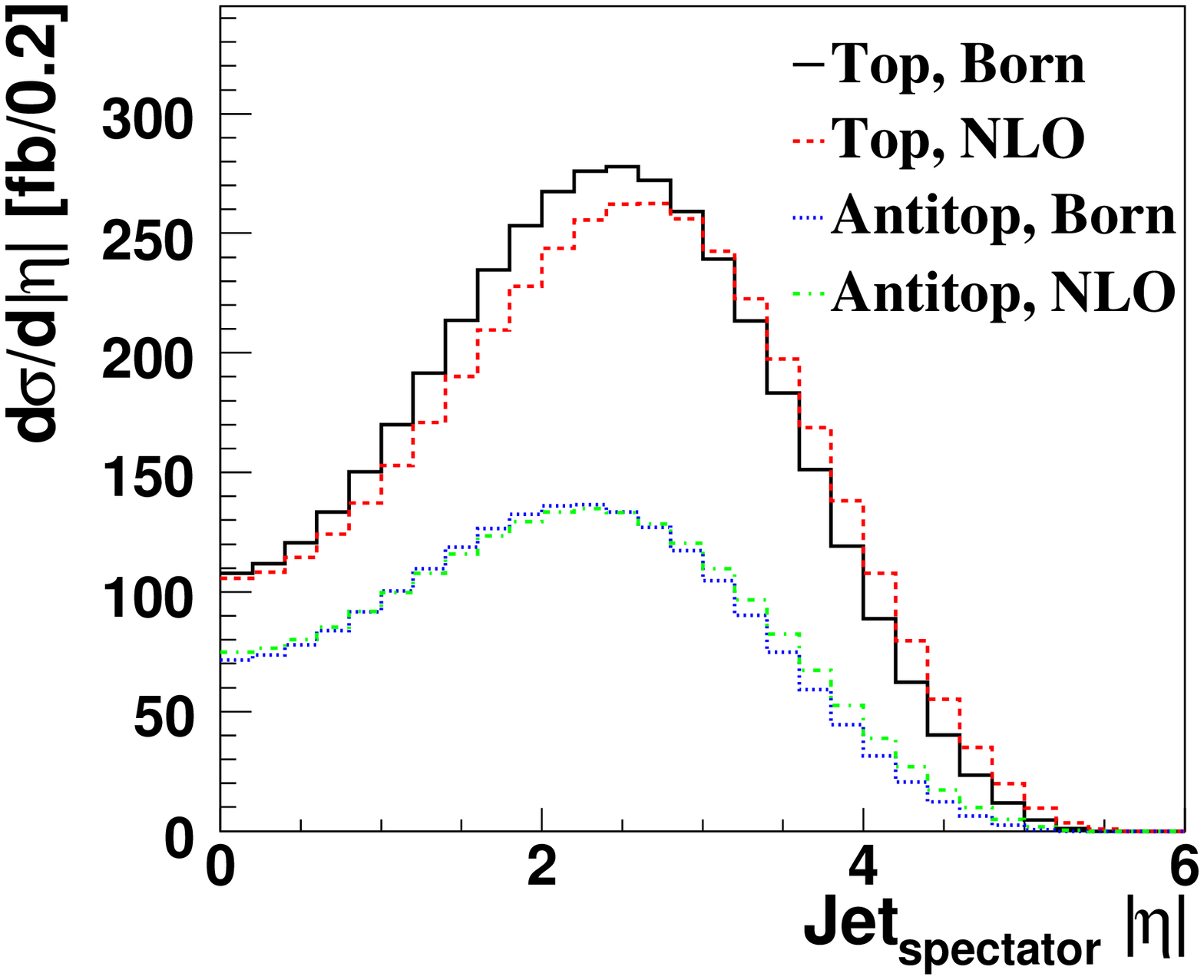}}
\subfigure[]{
\includegraphics[width=0.33\linewidth]{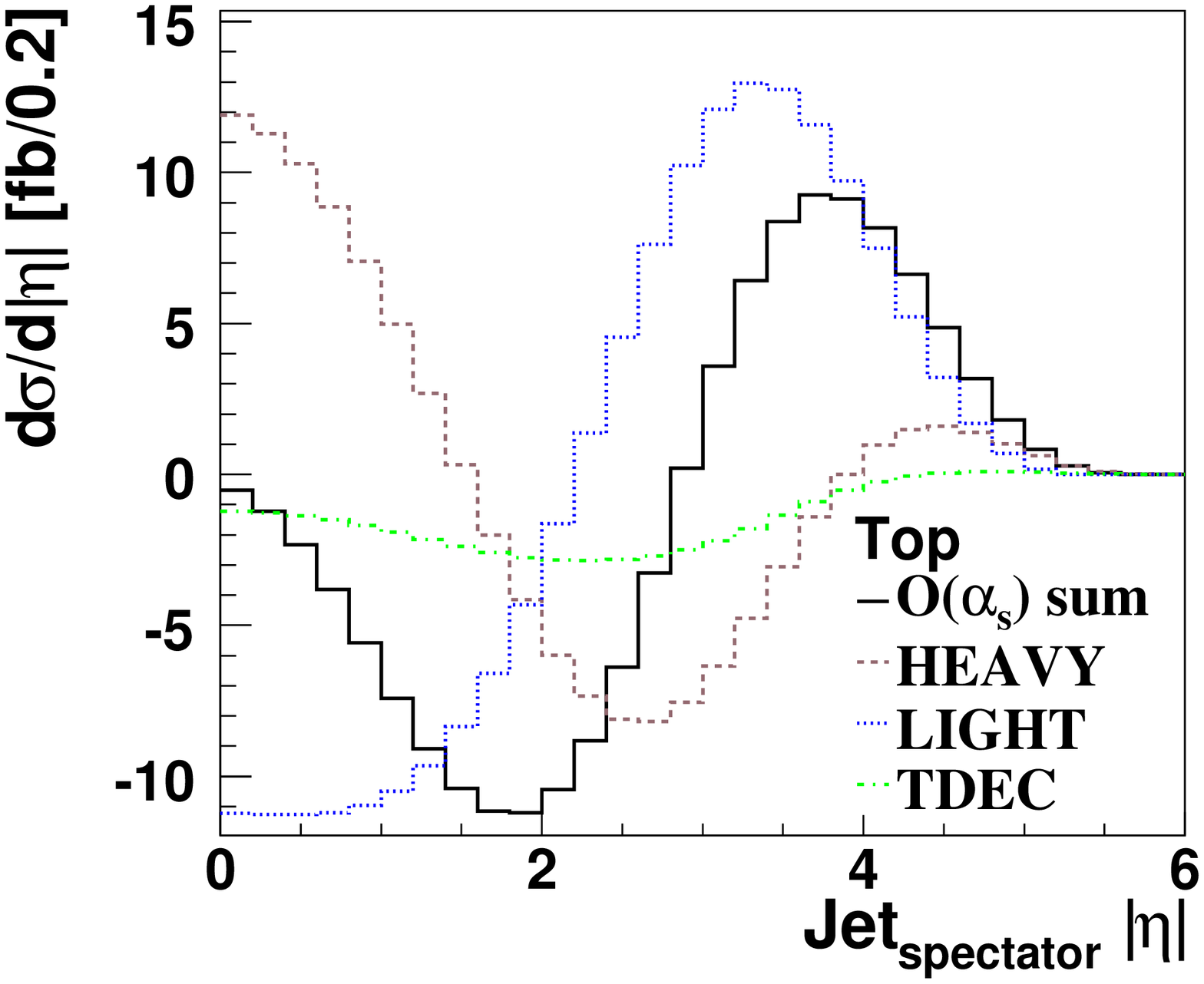}}

\subfigure[]{
\includegraphics[width=0.33\linewidth]{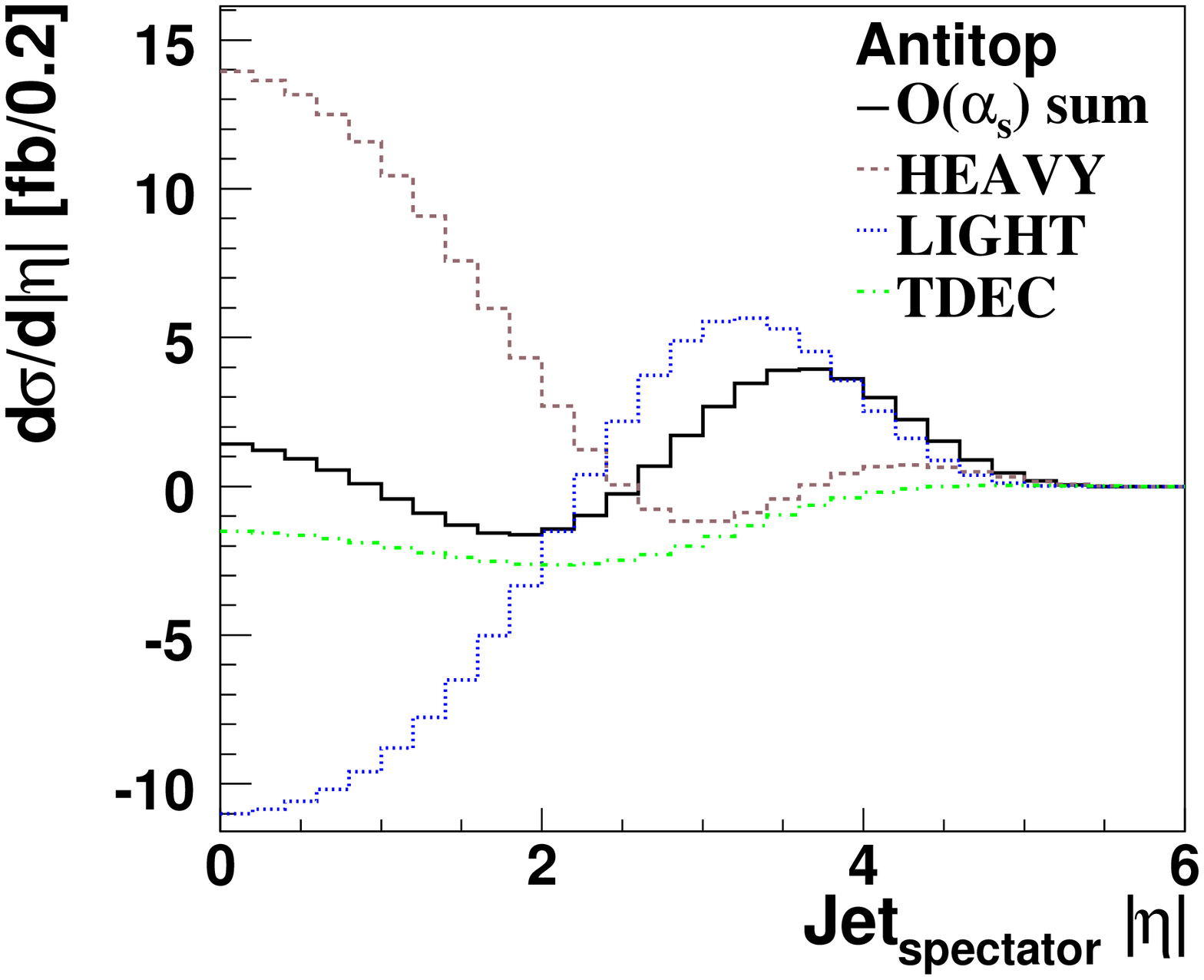}}
\subfigure[]{
\includegraphics[width=0.33\linewidth]{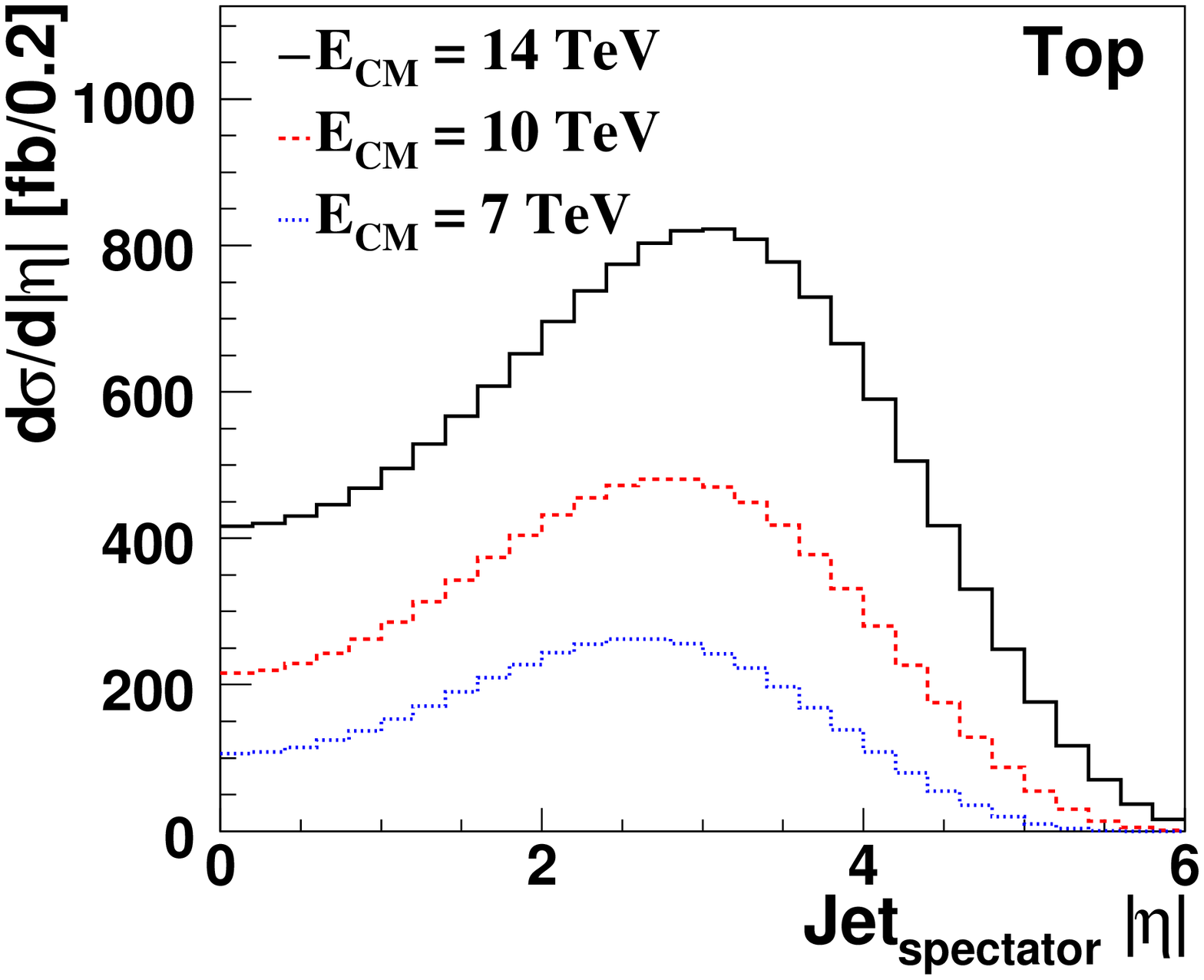}}

\caption{Absolute value of the spectator jet pseudo-rapidity at the 7~TeV
LHC, (a) comparing Born-level to NLO, (b, c) for individual contribution
of the $O(\alpha_{s})$ corrections for top and antitop quarks, respectively,
and (d) for different CM energies for top quarks.
\label{fig:spectator_eta_nlo}}
\end{figure}

The pseudo-rapidity distribution of the spectator jet is shown in
Figure~\ref{fig:spectator_eta_nlo}. The distribution is peaked in
the forward direction and only few light quark jets appear in the
central detector. This is comparable to the Tevatron case which has
an asymmetric light quark jet pseudo-rapidity distribution~\cite{Yuan:1989tc,Cao:2005pq}.
The $O(\alpha_{s})$ corrections shift the spectator jet to even more
forward pseudo-rapidities due to additional gluon radiation, though
the different contributions have opposite effects. The INIT correction
is responsible for the shift to forward pseudo-rapidities due to the
additional gluon radiation from the initial state up or down quark.
The effect of the HEAVY correction is indirect, though noticeable.
The HEAVY correction is largest for central light quark jets because
in that case the initial state heavy quark has a higher momentum fraction
$x$. The decay correction has no significant effect on the spectator
jet, as expected. Since the $O(\alpha_{s})$ corrections are small
compared to the Born-level contribution, the overall shift is small.
Comparing top to antitop quark production, Figure~\ref{fig:spectator_eta_nlo}(a)
also shows that the antitop quark production cross section is nearly
as large as the top quark production cross section in the central
region, while the top quark production cross section is significantly
larger than the antitop quark one for forward pseudo-rapidities. Central
light quark jets come from events where the $p_{z}$ of the two initial
state partons is approximately balanced. This corresponds to small
light quark momentum fraction $x$ values, where the difference between
up and down quark PDFs is small. 

Figure~\ref{fig:spectator_eta_nlo}(d) compares the spectator jet
pseudorapidity distribution for different collider CM energies. As
the collider energy goes down, the spectator jet becomes more central
and the peak in the forward region is less pronounced because the
initial state light quark energy is reduced.

\subsubsection{$b$~quark jet}

\begin{figure}[!h!tbp]
\subfigure[]{
\includegraphics[width=0.33\linewidth]{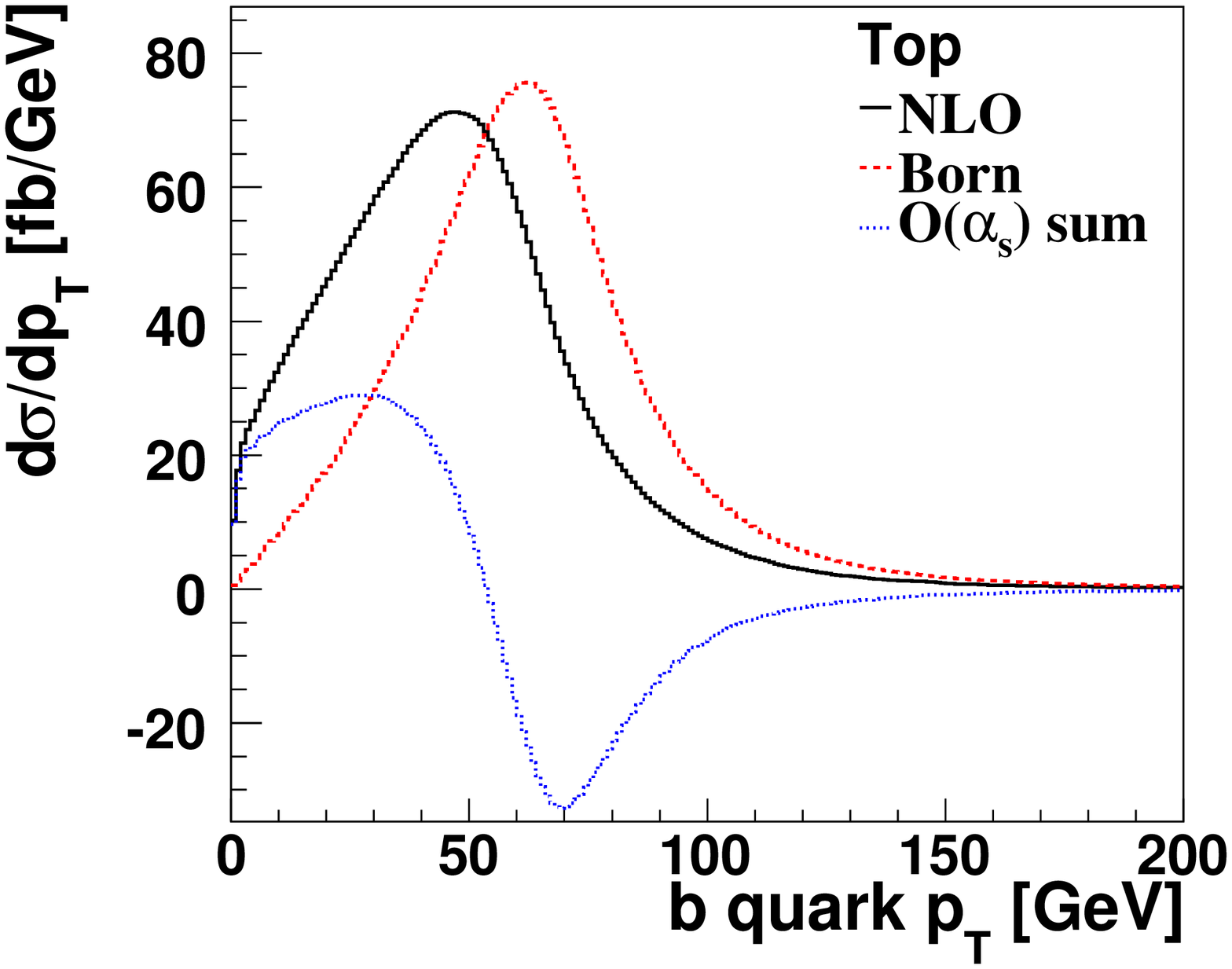}}
\subfigure[]{
\includegraphics[width=0.33\linewidth]{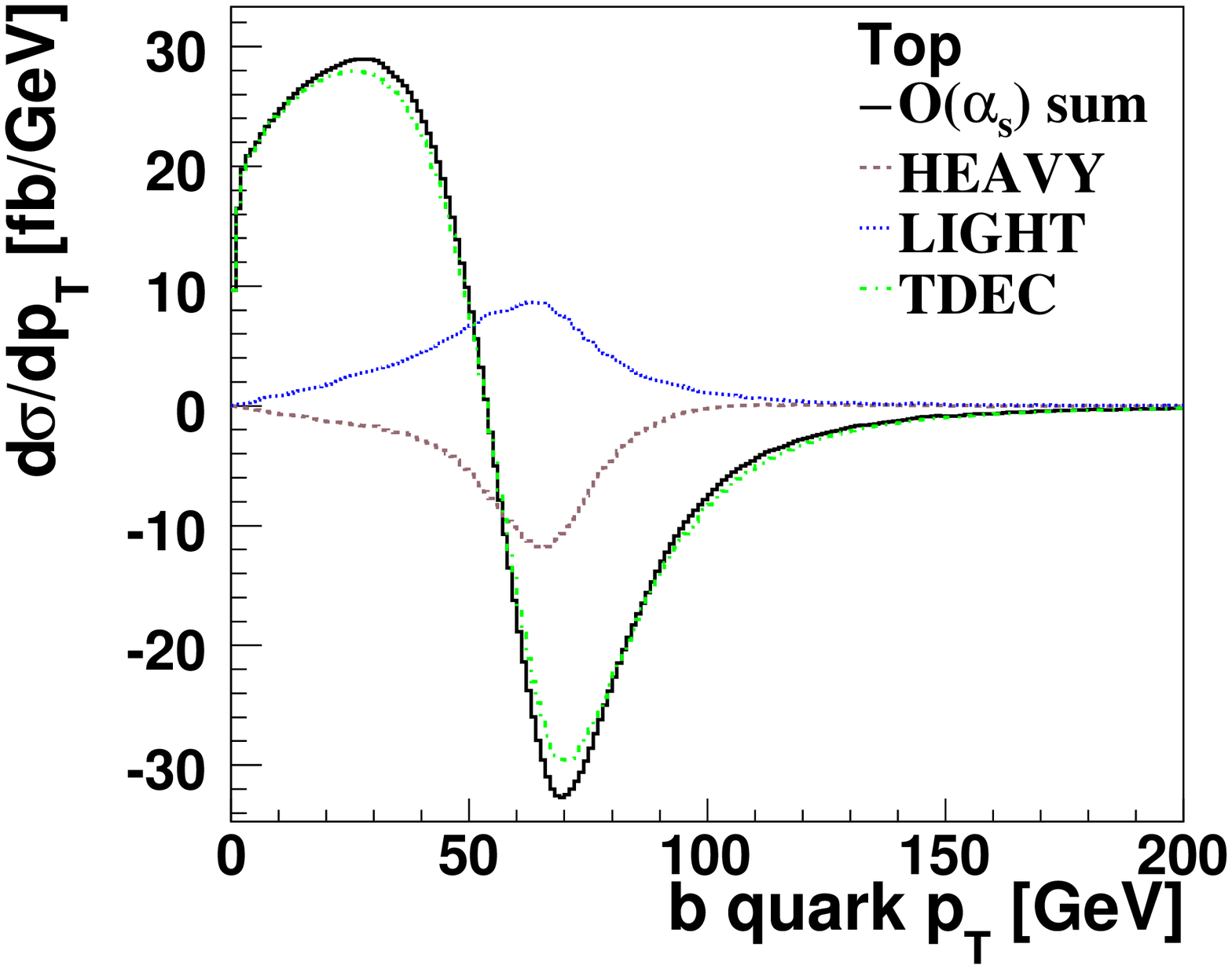}}

\subfigure[]{
\includegraphics[width=0.33\linewidth]{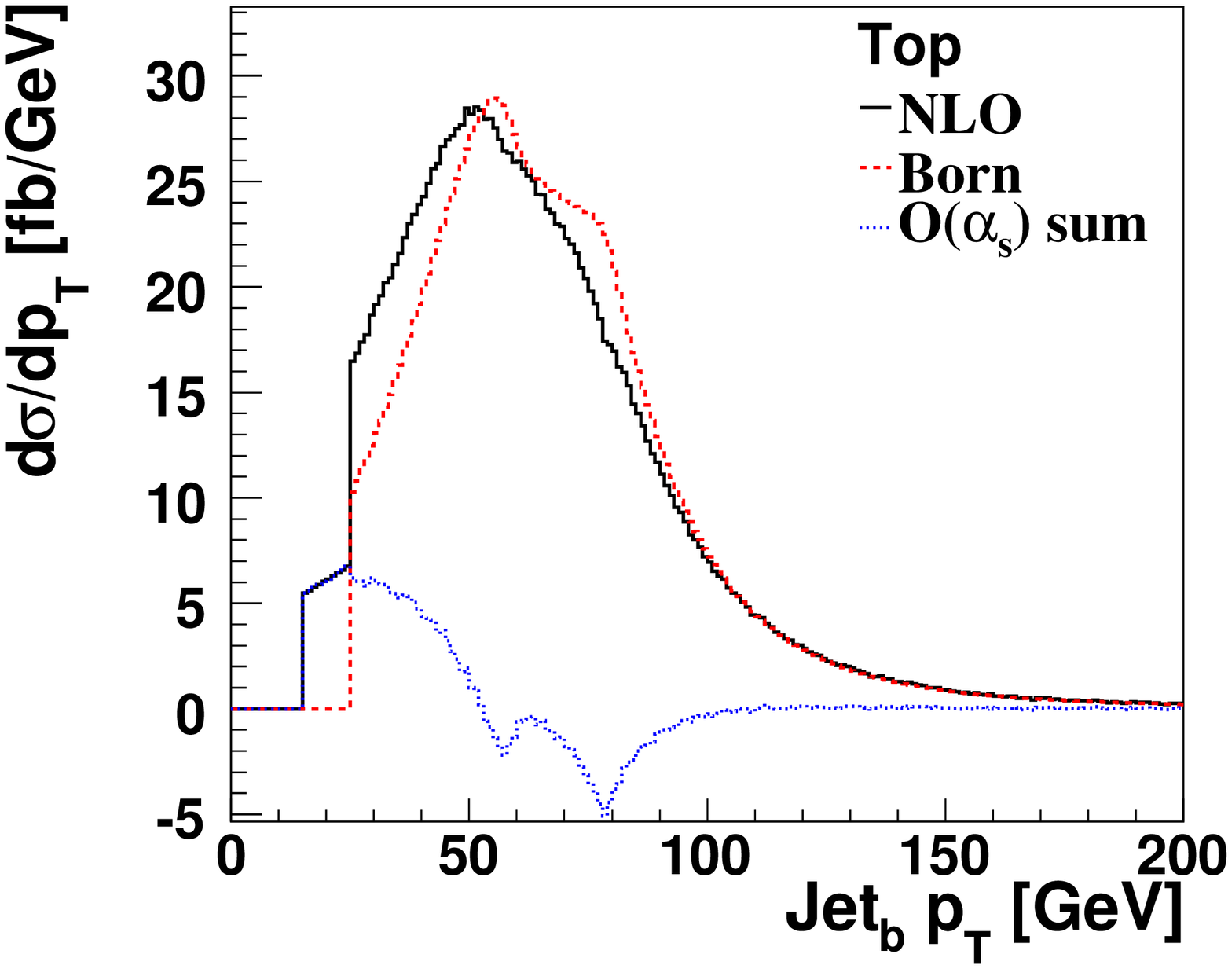}}
\subfigure[]{
\includegraphics[width=0.33\linewidth]{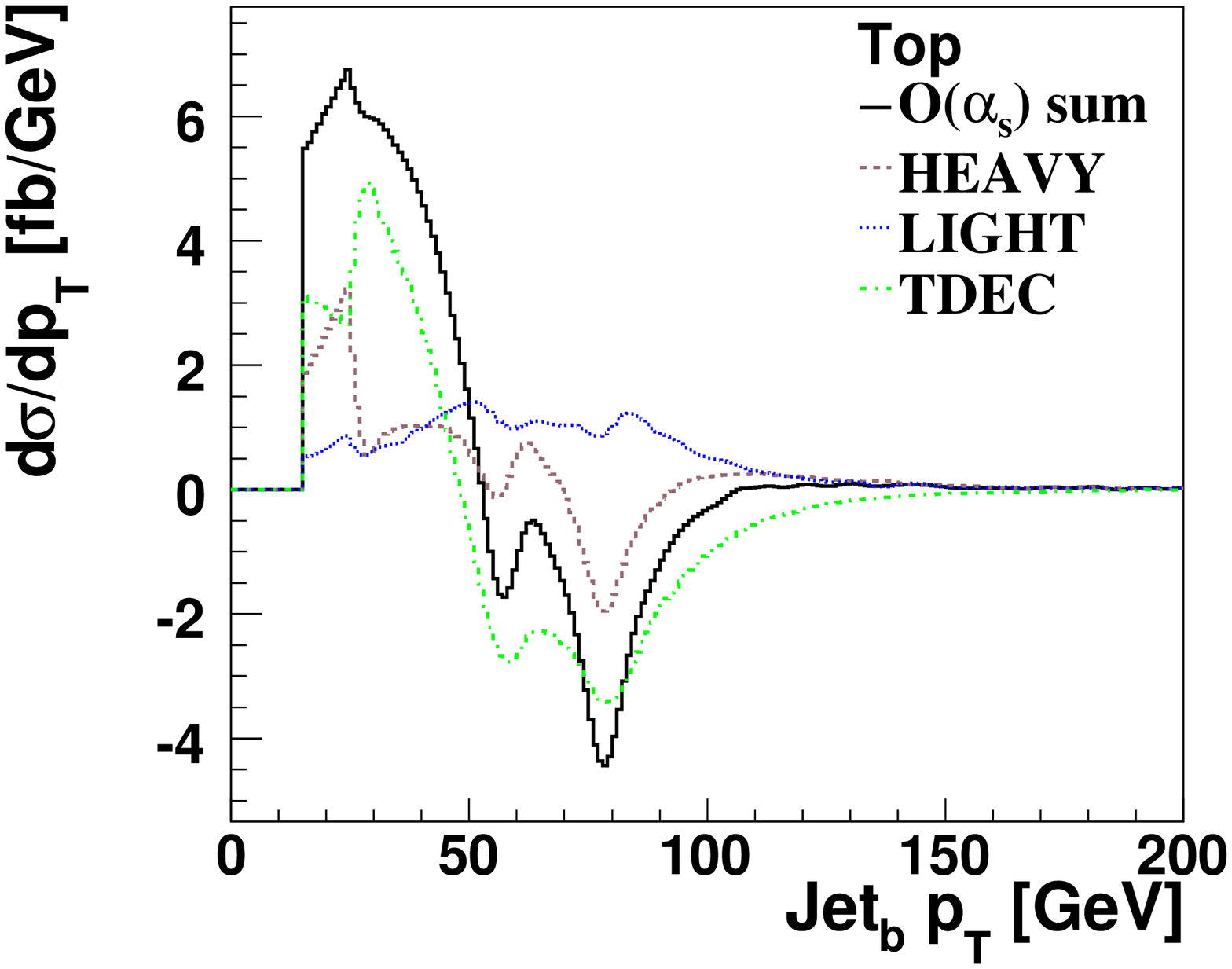}}

\caption{Transverse momentum of (a, b) the $b$~quark before selection cuts
and (c, d) the $b$~quark jet after cuts, (a, c) comparing Born-level
to $\oalphas$ corrections and (b, d) the individual $\oalphas$ contributions,
at the 7~TeV LHC.\label{fig:ptb}}

\end{figure}

Compared to the lepton and $\met$, the effects of the $\oalphas$
corrections on the reconstructed $b$~quark jet are more pronounced.
Figure~\ref{fig:ptb} shows a comparison of the $p_{T}$ of the $b$~quark
before cuts and the $b$~quark jet after cuts between Born-level
and $\oalphas$ corrections for top quark production. The antitop
quark distributions are similar. The $p_{T}$ distribution of the
$b$~quark is predominantly determined by the top quark mass and
therefore peaks at $\sim m_{t}/3$. The NLO QCD corrections shift
the peak position to lower values, both at parton level and after
selection cuts. At parton level, the only $\oalphas$ correction that
changes the shape is the TDEQ correction, which tends to shift the
distribution to lower $p_{T}$ values because a gluon radiated from
the top quark decay tends to move along the $b$~quark direction
due to collinear enhancements. The LIGHT and HEAVY corrections leave
the distribution approximately unchanged and their magnitudes cancel
each other. 

The selection cuts change the shape of the $p_{T}$~distribution
by removing events in the peak region around 60-70~GeV. For these
events the top quark is produced with low $p_{T}$, hence the light
quark also has low $p_{T}$ and the events fail the requirement of
at least two jets. The selection cuts also affect the $\oalphas$
corrections shown in Fig.~\ref{fig:ptb}(d). The TDEC correction
is the largest contribution, similar to the parton level in Fig.~\ref{fig:ptb}(b),
and it also shows the effect of the selection cuts because the top
quark and thus light quark $p_{T}$ is unchanged. The HEAVY correction 
shows a similar behavior; because it is an initial state correction
it does not change the $p_{T}$ balance between light quark and top quark. 
Real gluon emission in the LIGHT correction
does change this balance, thus the effect is not visible as clearly
anymore.

\begin{figure}[!h!tbp]
\subfigure[]{
\includegraphics[width=0.33\linewidth]{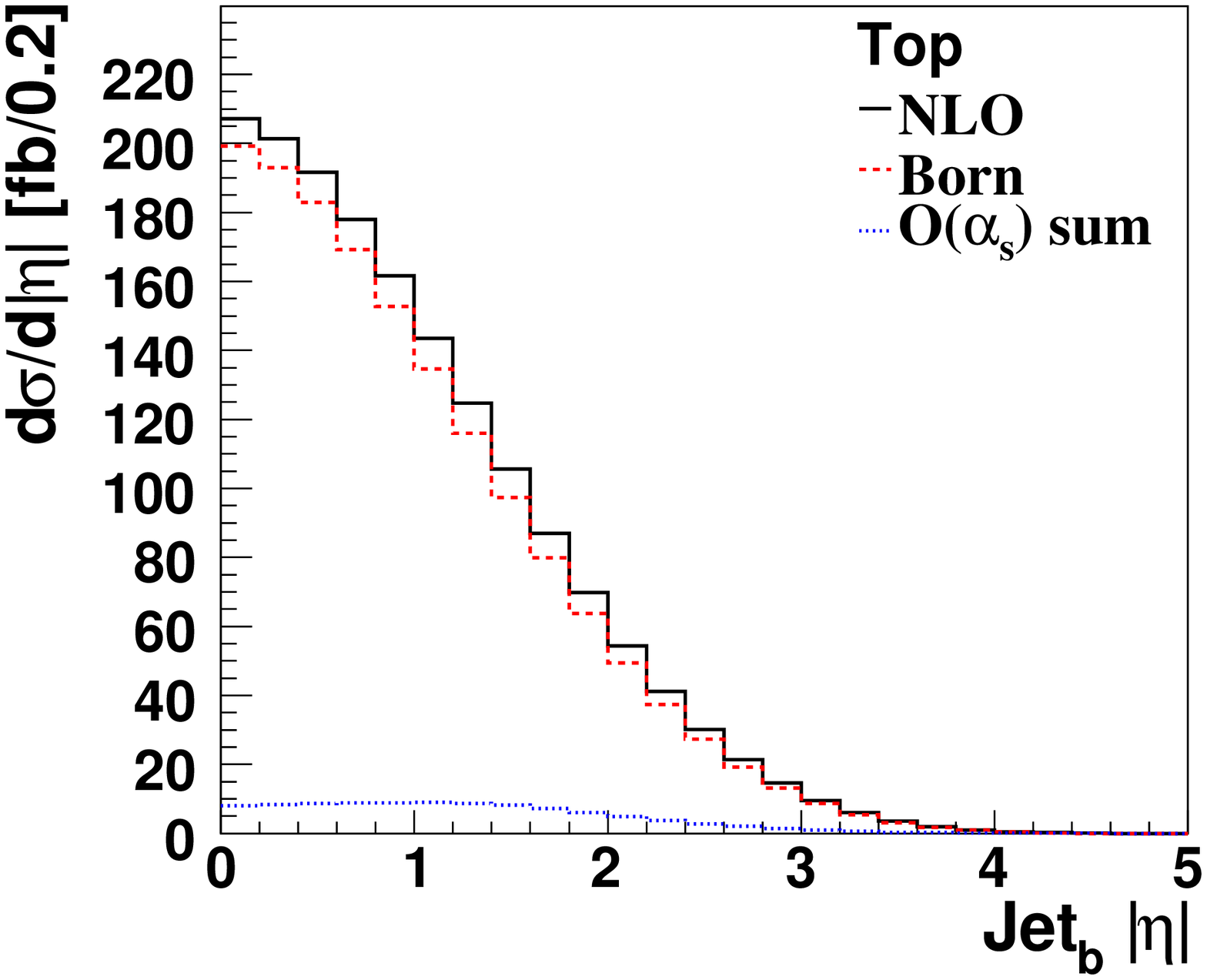}}
\subfigure[]{
\includegraphics[width=0.33\linewidth]{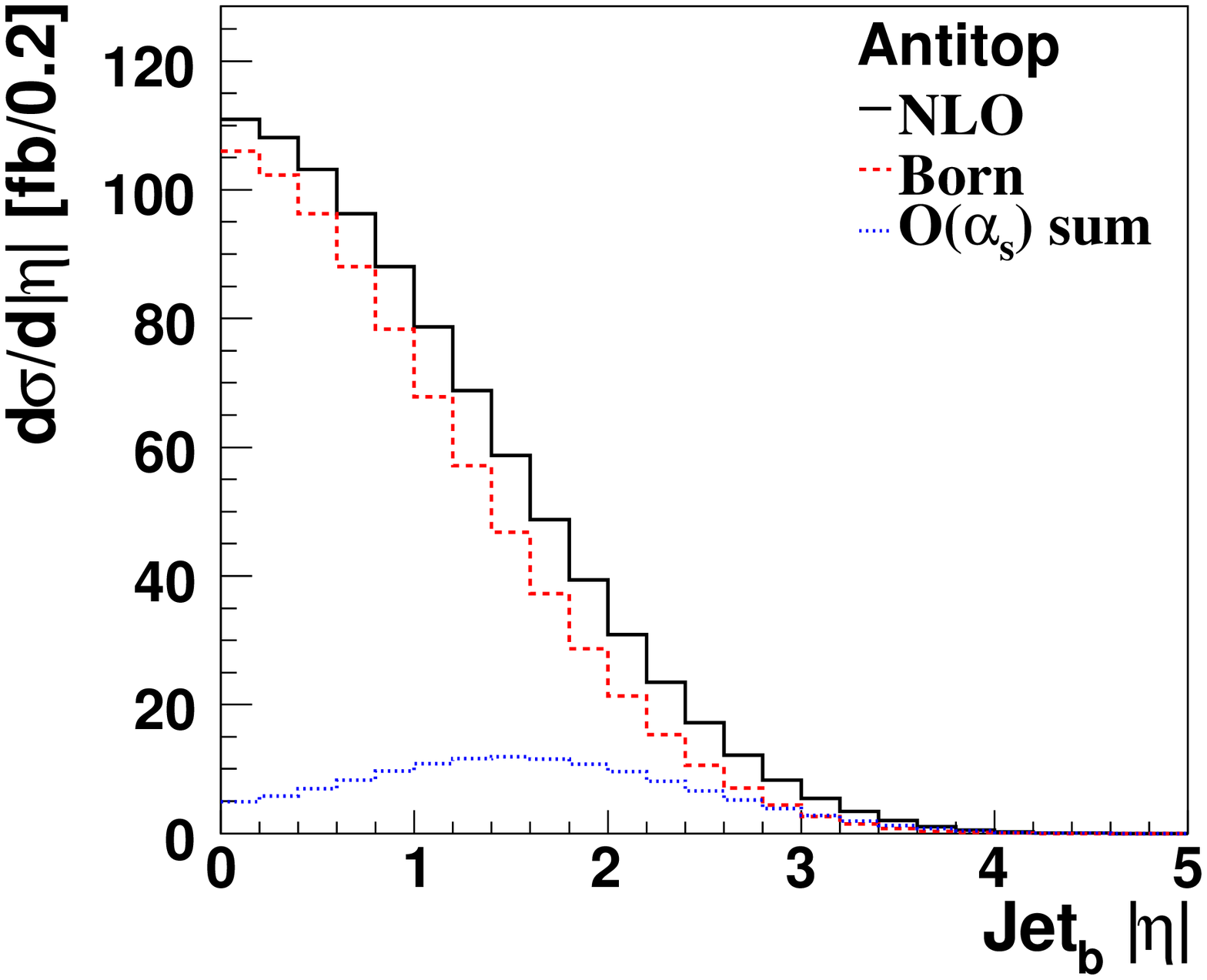}}

\subfigure[]{
\includegraphics[width=0.33\linewidth]{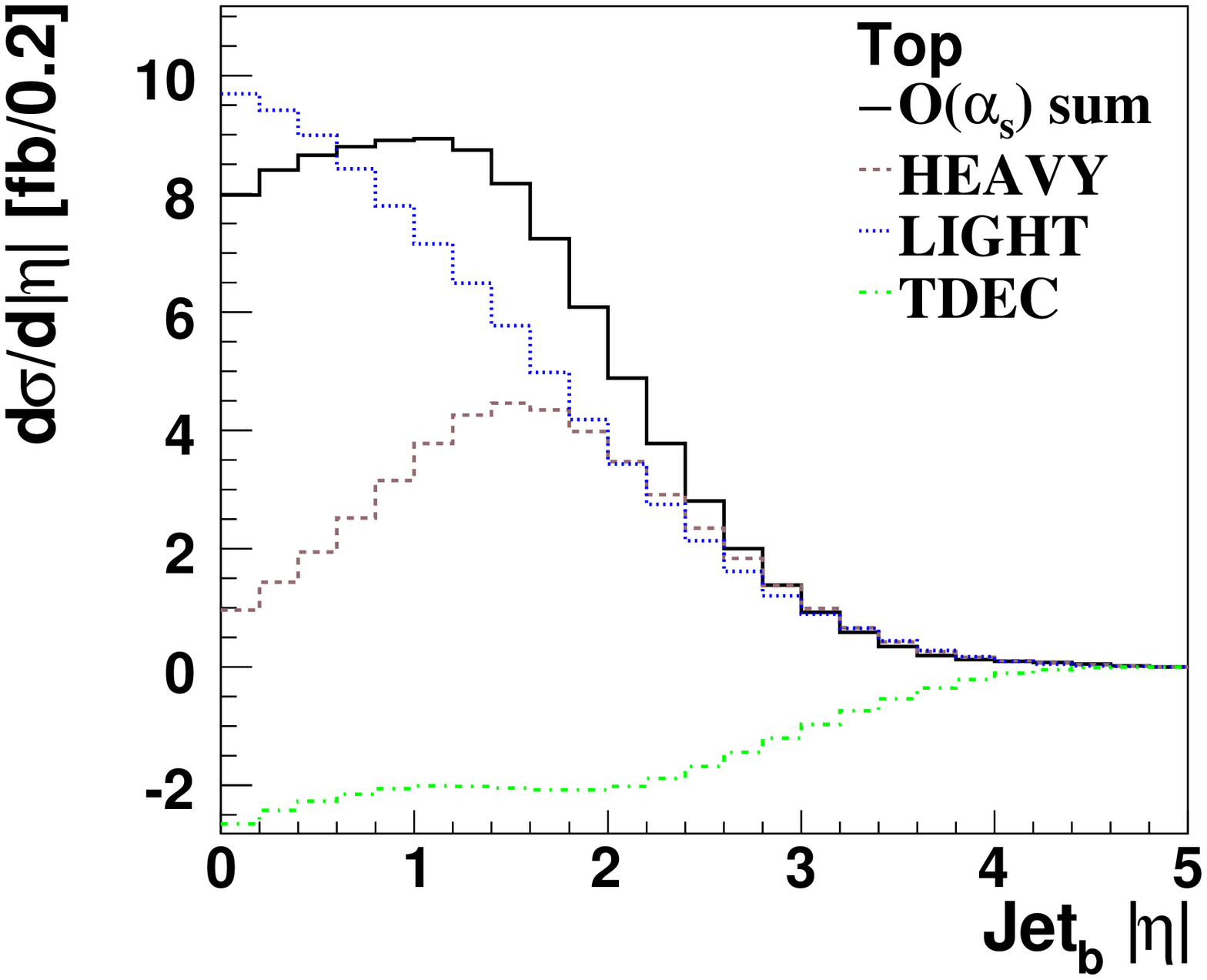}}
\subfigure[]{
\includegraphics[width=0.33\linewidth]{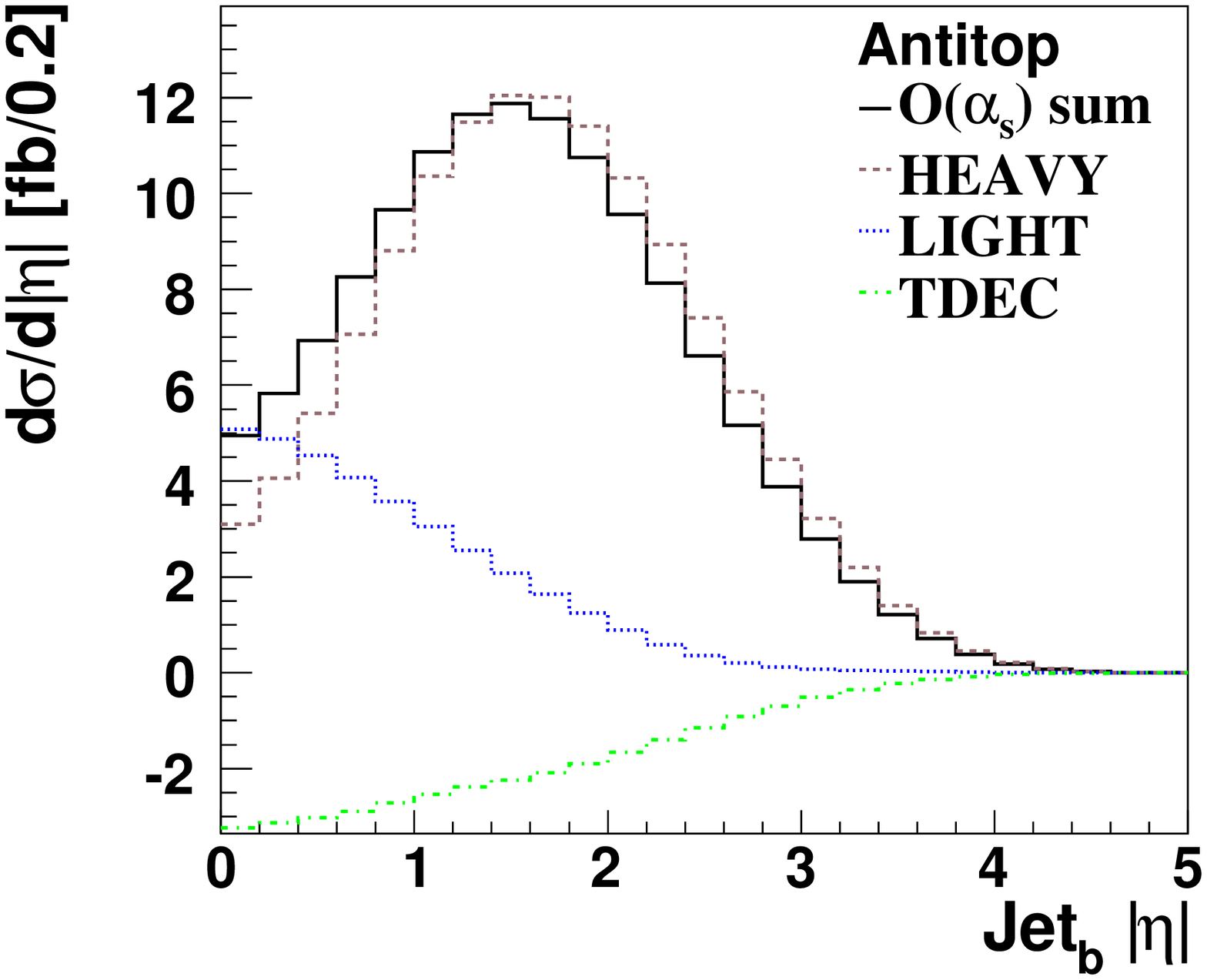}}

\caption{Pseudo-rapidity of the $b$~quark jet after selection cuts at the
7~TeV LHC, (a, b) comparing Born-level to $\oalphas$ corrections
and (c, d) individual $\oalphas$ contributions, in (a, c) for top
quarks and in (b, d) for antitop quarks. 
\label{fig:etab}}
\end{figure}

The $b$~quark jet pseudo-rapidity distribution in top quark events
is less affected by the $\oalphas$ corrections, as can be seen in
Fig.~\ref{fig:etab}. The top quark is so heavy that it is mostly
produced in the central rapidity region and thus the $b$~quark jet
from its decay also peaks around a pseudo-rapidity of zero. The shape
of the $b$~quark jet pseudo-rapidity distribution remains almost
unchanged compared to the Born-level because it comes from the top
quark decay. The difference between Born-level and $\oalphas$ is
larger for antitop quark events. While the Born-level contribution
and the LIGHT corrections are smaller for antitop quarks, the HEAVY
and TDEC corrections are similar in size for top and antitop quarks.
These distributions are similar at a CM energy of 14~TeV.

\subsubsection{Event kinematics}

The impact that different $\oalphas$ corrections have on the $p_{T}$
of the jets is also reflected in event-wide energy variables such
as the total transverse energy in the event ($H_{T}$) or the reconstructed
invariant mass of all final state objects. The total transverse energy
is defined as\begin{equation}
H_{T}=p_{T}^{lepton}+\met+\sum_{jets}p_{T}^{jet}.\label{eq:HT}\end{equation}
\begin{figure}[!h!tbp]
\subfigure[]{
\includegraphics[width=0.33\linewidth]{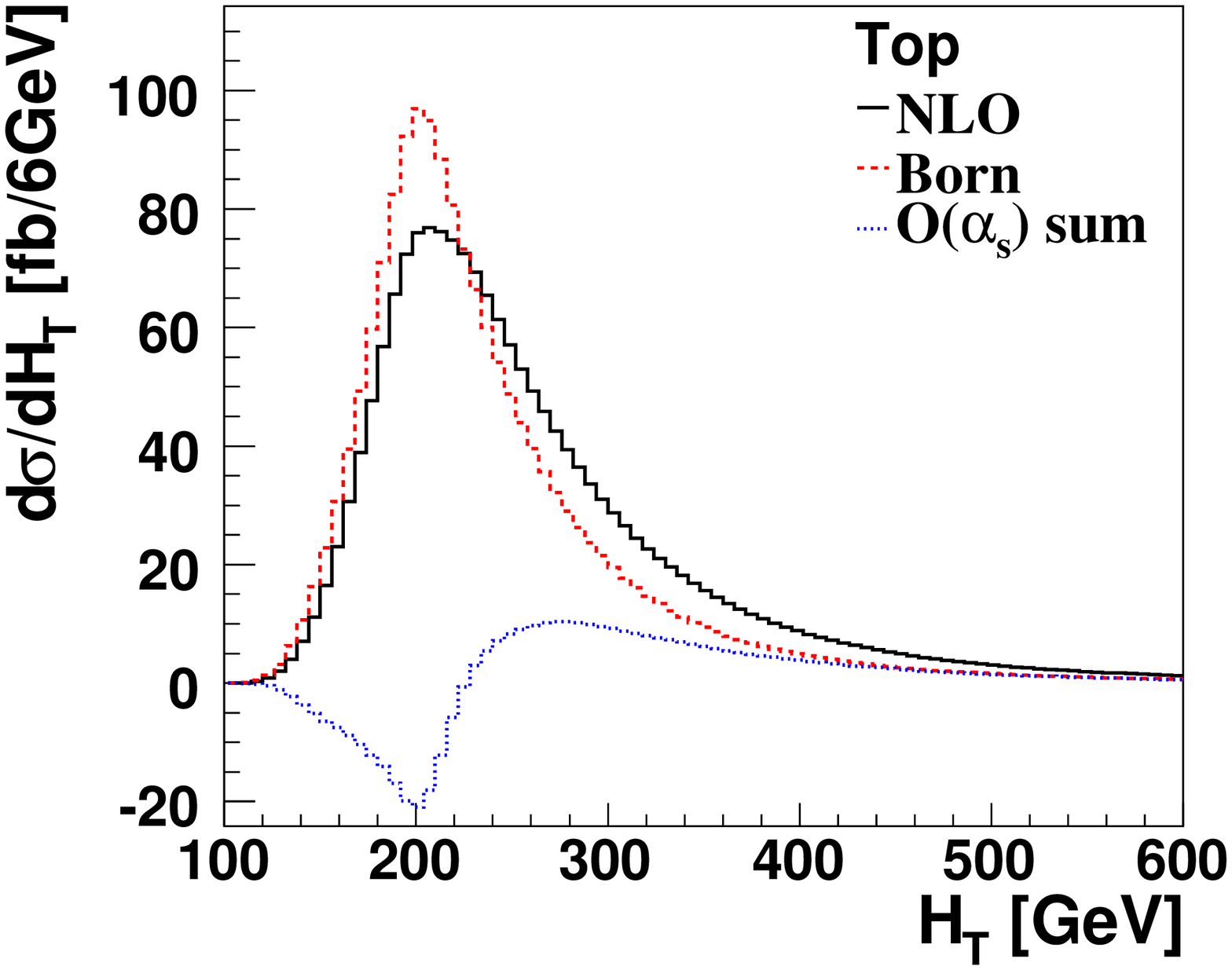}}
\subfigure[]{
\includegraphics[width=0.33\linewidth]{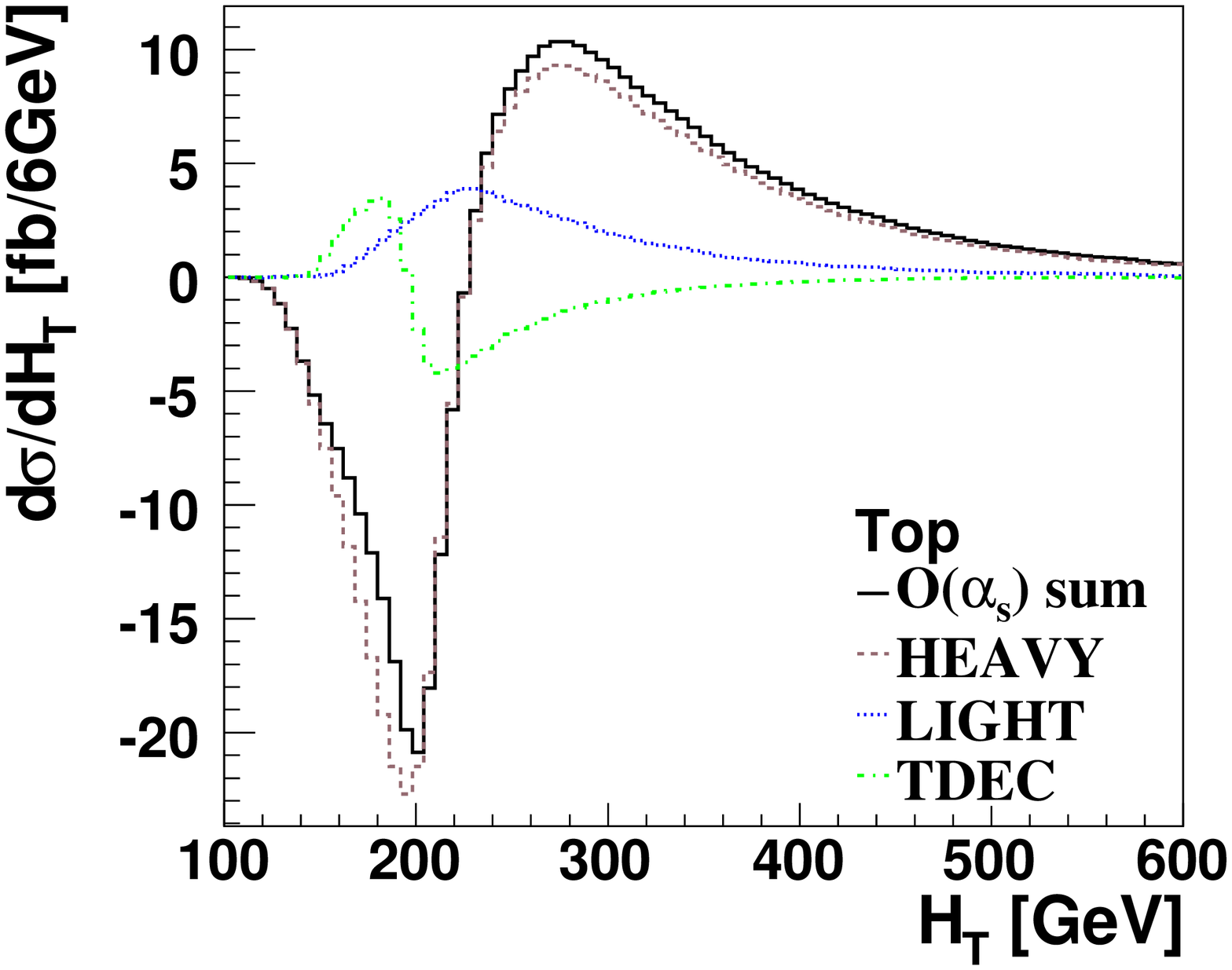}}

\subfigure[]{
\includegraphics[width=0.33\linewidth]{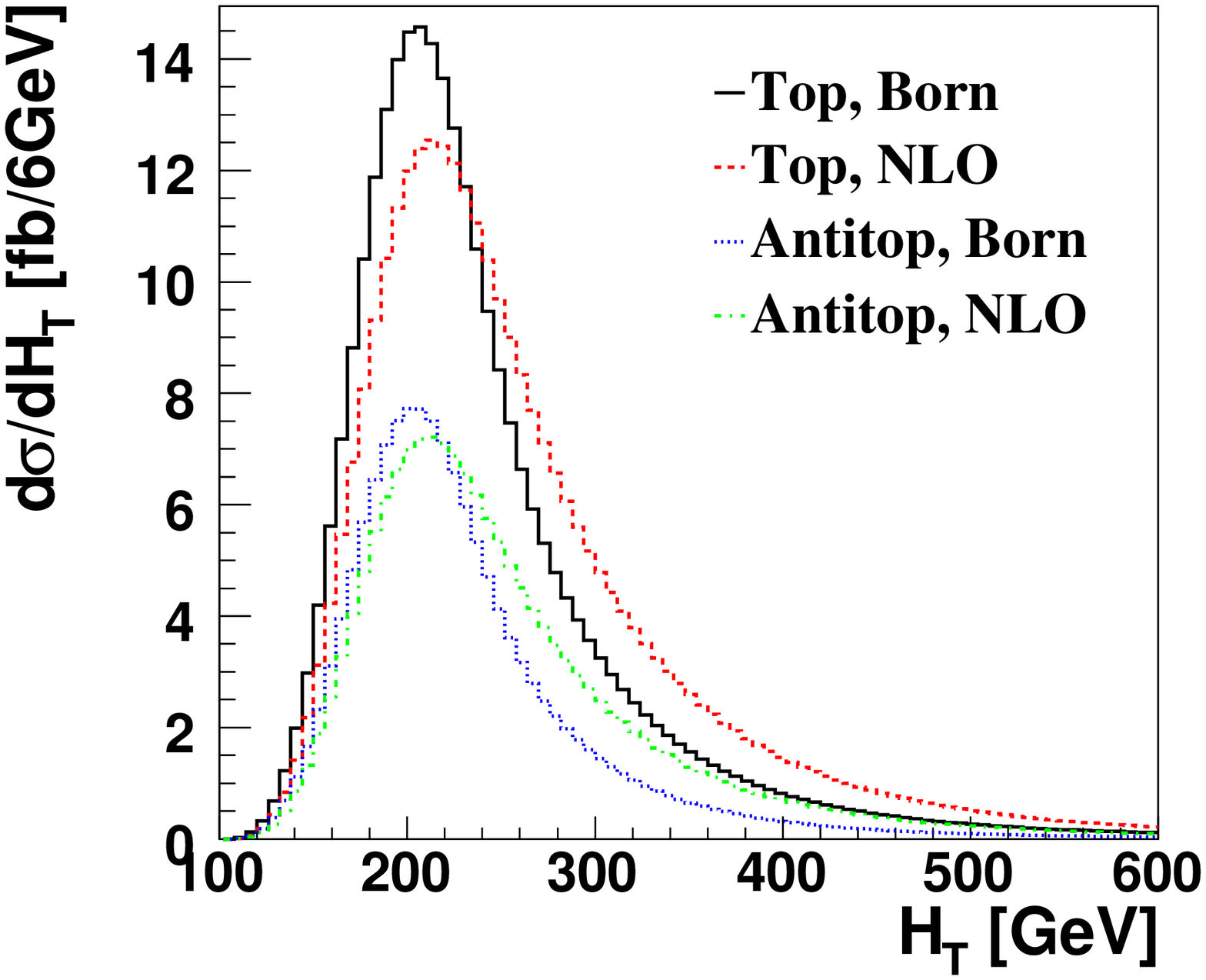}}
\subfigure[]{
\includegraphics[width=0.33\linewidth]{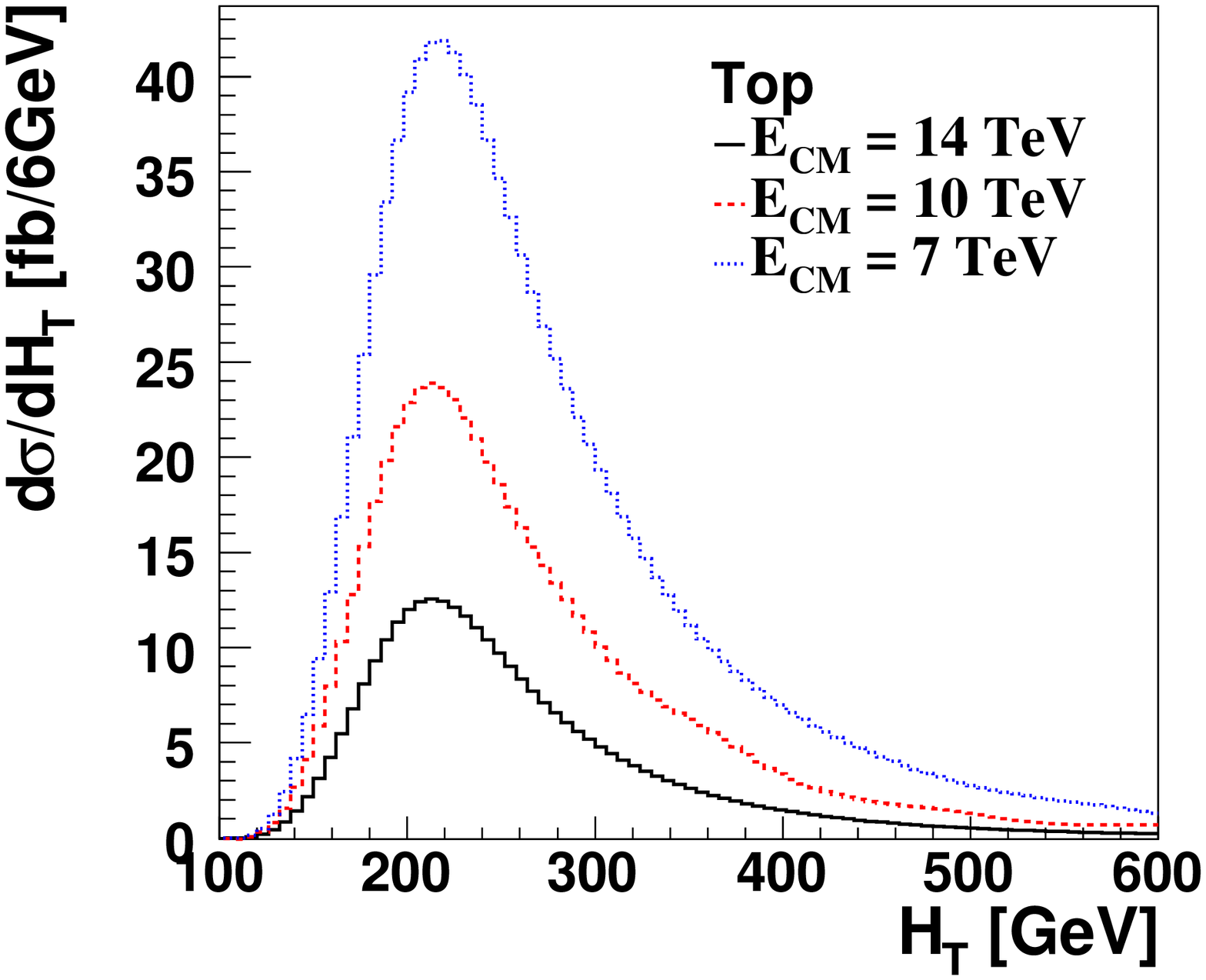}}

\caption{Total event transverse energy $H_{T}$ after selection cuts, (a) comparing
Born-level to $\oalphas$ corrections and (b) the individual $\oalphas$
contributions, (c) comparing top to antitop quark production, for
top quark production at the 7~TeV LHC, and (d) comparing different
CM energies.\label{fig:HT}}
\end{figure}

The distribution of $H_{T}$ for $t$-channel single top quark events
is shown in Fig.~\ref{fig:HT}. The Born-level $H_{T}$ distribution
peaks around 200~GeV at a CM energy of 7~TeV and shifts to slightly
higher values for larger CM energies. The HEAVY contribution decrease
the height of the peak and shifts it to higher values, while the LIGHT
and TDEC contributions are small. All three $O(\alpha_{s})$ contributions
broaden the distribution. The antitop quark production distributions
as well as those for lower CM energies are very similar because the
shape is determined mainly by the top quark mass.

\subsection{Distributions for three-jet events\label{sub:Distributions-for-Three-jet}}
As Fig.~\ref{fig:njets_jet_pt} shows, the majority of $t$-channel
events at the LHC contain three reconstructed jets rather than two.
This is the main difference between $t$-channel single top quark
production at the Tevatron and the LHC, other than that the LHC collider
is not CP symmetric. This is also true for events passing the loose
selection cuts. In this section we focus on the properties of these
three-jet events and specifically the additional jet. As expected,
the effect is not quite as large when only jets within a very small
$\eta$ range are considered because the extra jet typically has higher
$\eta$. In order to study $\oalphas$ effects it is thus important
to set the jet $\eta$ cut as high as possible and the jet $E_{T}$
cut as low as possible. 

\subsubsection{Abundance of three-jet events}
Additional light quark jets arise in the LIGHT quark
line corrections, in the TDEC corrections, and in the gluon radiation
diagrams of the HEAVY correction. The $W$-gluon fusion HEAVY process
can at least in principle also contribute when the $\bar{b}$~quark
jet from the initial state gluon splitting is not $b$-tagged. While
the typical experimental $b$-tagging efficiency is lower than 100\%,
we will assume fully efficient, perfect $b$-tagging for simplicity
and consider events with one $b$-tagged jet and two $b$-tagged jets
separately.

\begin{figure}[!h!tbp]
\subfigure[]{
\includegraphics[width=0.33\linewidth]{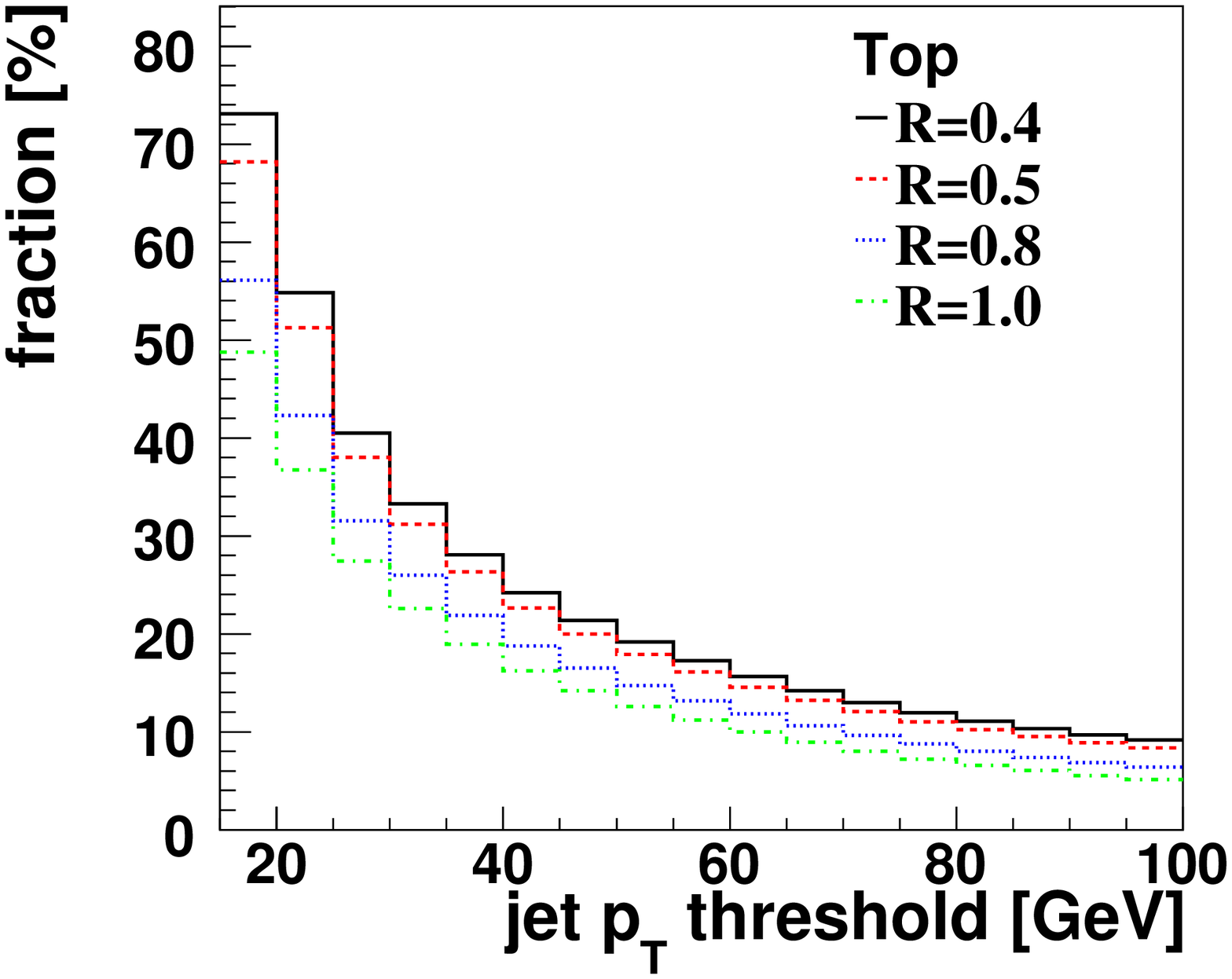}}
\subfigure[]{
\includegraphics[width=0.33\linewidth]{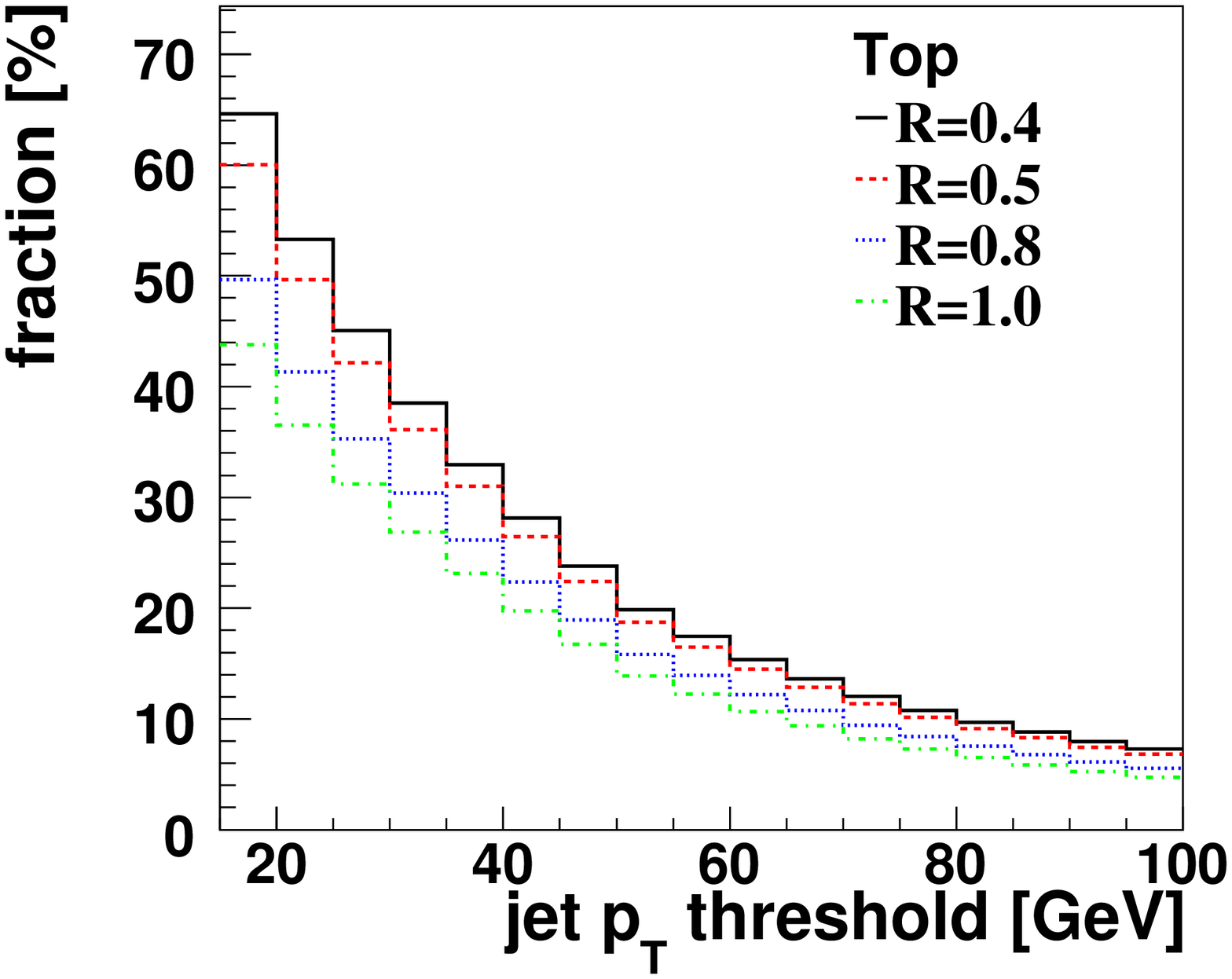}}

\caption{Fraction of 3-jet events as a function of the jet $p_{T}$ threshold
at the 7~TeV LHC for different jet cone sizes, (a) for the loose
set of cuts and (b) for the tight set of cuts. \label{fig:njets_jet_cone}}
\end{figure}

From Fig.~\ref{fig:njets_jet_pt} it is clear that the jet multiplicity
at NLO depends strongly on the jet $p_{T}$ cut. Figure~\ref{fig:njets_jet_cone}
shows that it also depends on the jet reconstruction cone size. The
dependence of the total cross section on the jet pseudo-rapidity cut
is different between Born-level and NLO, mostly as a result of the
presence of a third jet. 

\begin{figure}[!h!tbp]
\subfigure[]{
\includegraphics[width=0.33\linewidth]{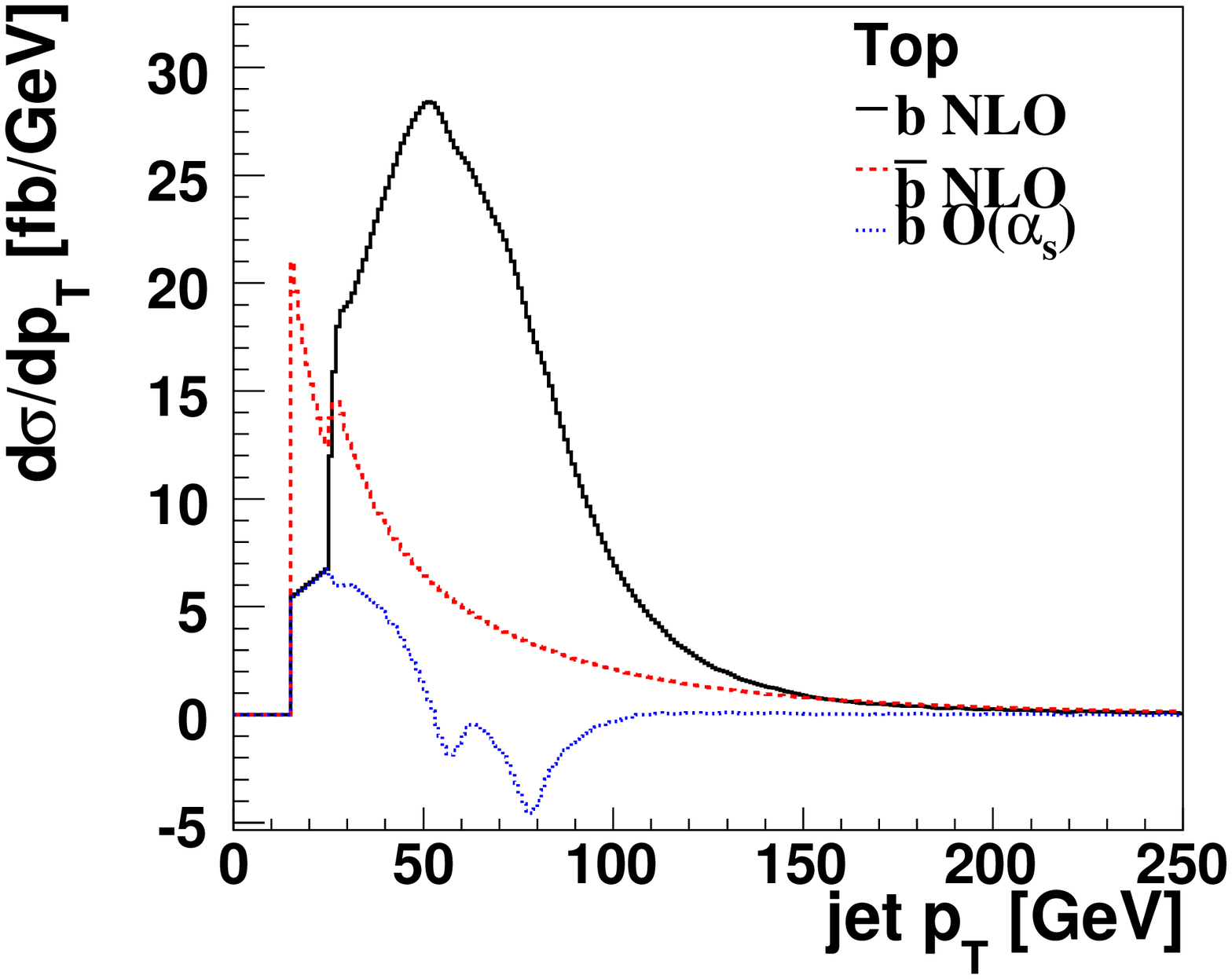}}
\subfigure[]{
\includegraphics[width=0.33\linewidth]{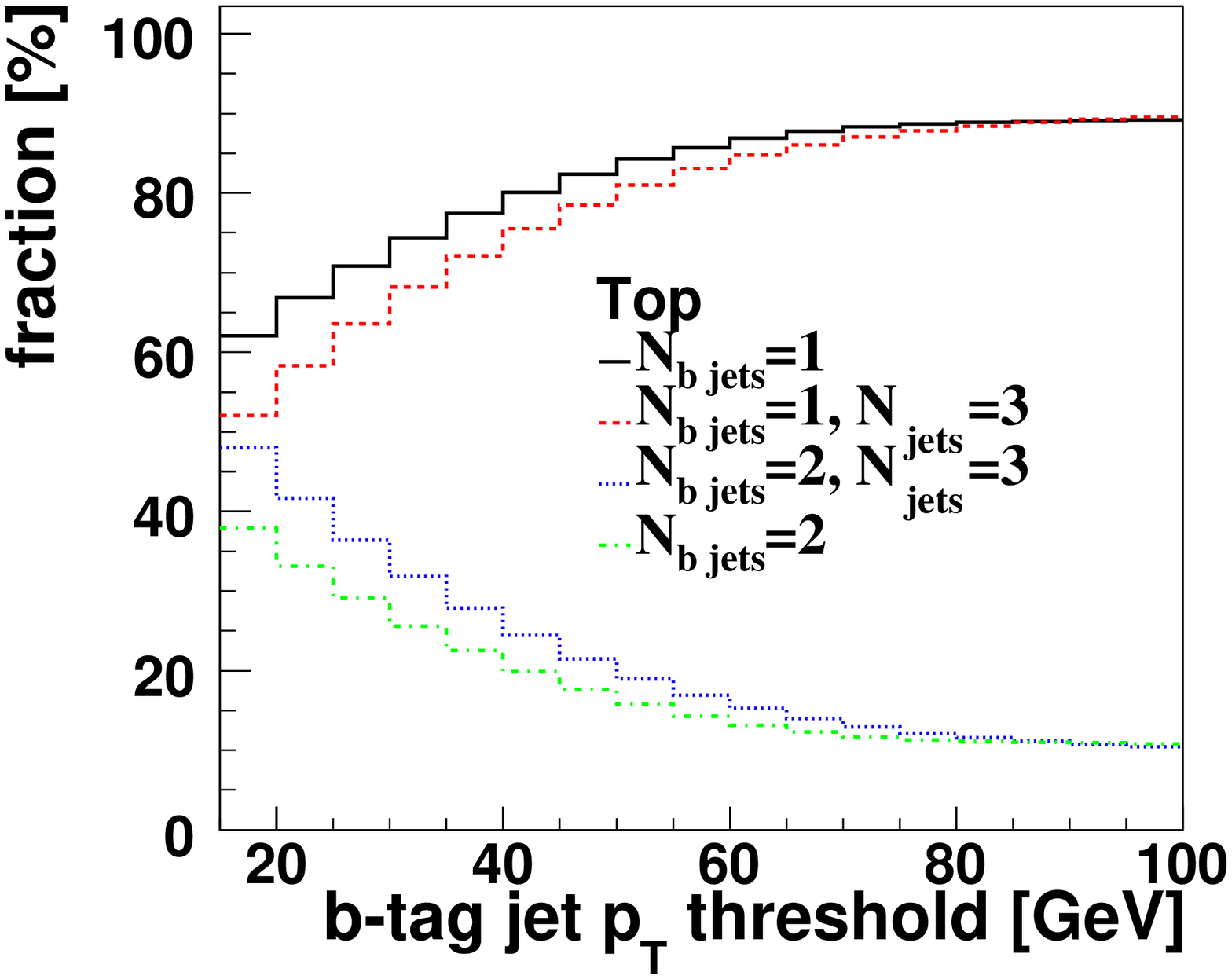}}
\subfigure[]{
\includegraphics[width=0.33\linewidth]{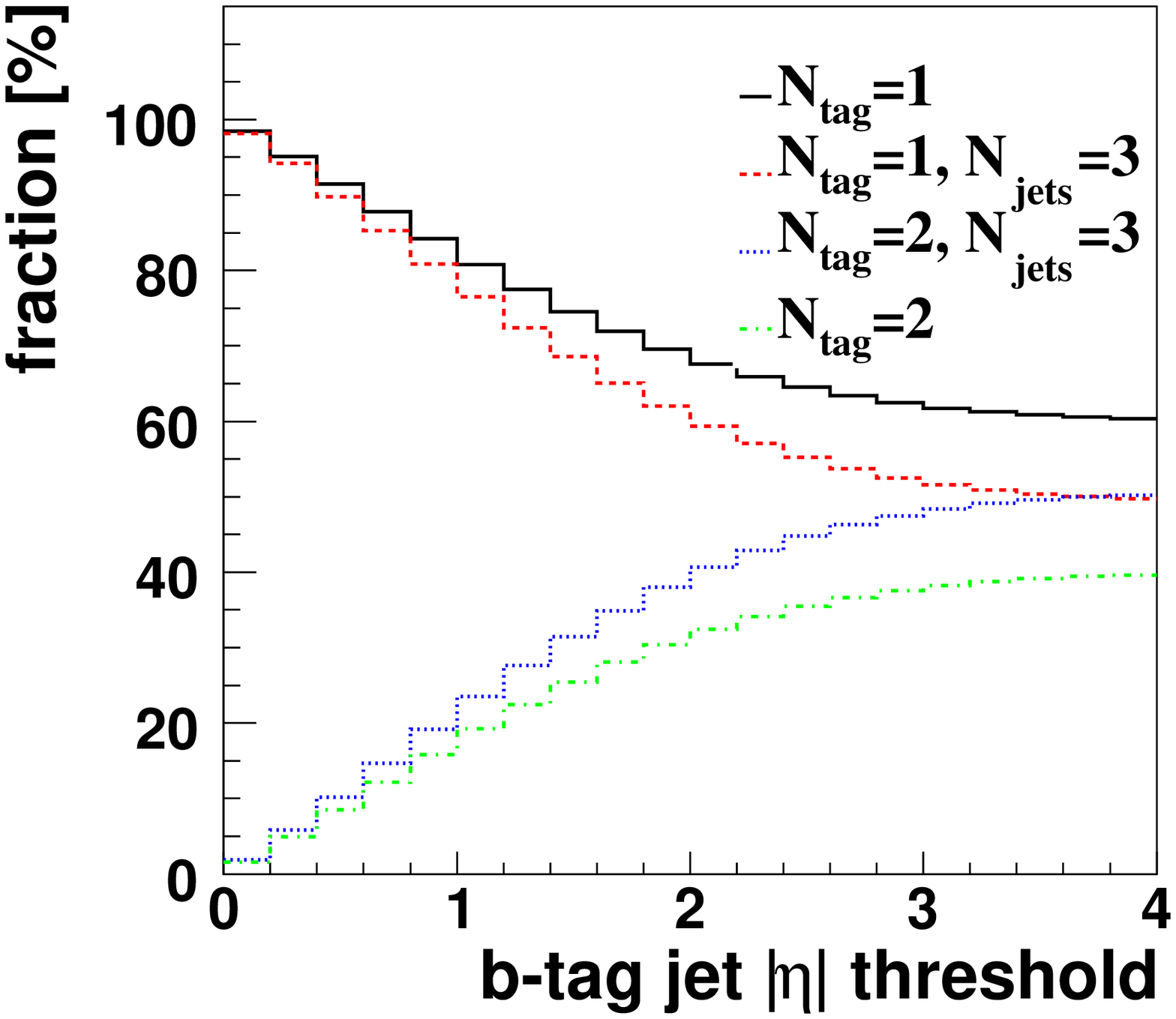}}

\caption{(a) $p_{T}$ of the $b$ and $\bar{b}$ jets in top quark production,
(b) fraction of events with one or two $b$-tagged jets as a function
of jet $p_{T}$ threshold and (c) as a function of jet $|\eta|$ threshold,
for both inclusive two-jet and exclusive three-jet events, at the
7~TeV LHC.\label{fig:ptbbbar}}
\end{figure}

Figure~\ref{fig:ptbbbar} compares the momentum of the $b$~quark
jet from the top quark decay and the $\bar{b}$~quark from the HEAVY
correction and examines the fraction of events containing one or two
$b$-tagged jets in the final state. Events with both $b$~and $\bar{b}$~jets
originate from the $W$-gluon fusion sub-process, $qg\to q^{\prime}\bar{b}t(\to bW(\to\ell^{+}\nu))$.
The fraction of events with two $b$-tagged jets is about 40\% for
low jet $p_{T}$, and remains high even for higher jet $p_{T}$. Even
for jet $p_{T}$ above 100~GeV, the fraction of events with both
a $b$~and a $\bar{b}$~jet is above 10\%. Figure~\ref{fig:ptbbbar}
also shows that for exclusive three-jet events, the fraction of events
with two $b$-tagged jets is even higher. The fraction of $b$-tagged
jets also depends on the jet $|\eta|$ threshold, as shown in Fig.~\ref{fig:ptbbbar}(c).
Since the $\bar{b}$~quark from the $W$-gluon fusion sub-process
typically moves in the forward direction, a high $|\eta|$ threshold
results in two $b$-tagged jets per event, whereas a low jet $|\eta|$
threshold results in only one.

\subsubsection{Third jet}

Here we consider the kinematic distribution of the third jet and its
flavor composition.

\begin{figure}[!h!tbp]
\subfigure[]{
\includegraphics[width=0.33\linewidth]{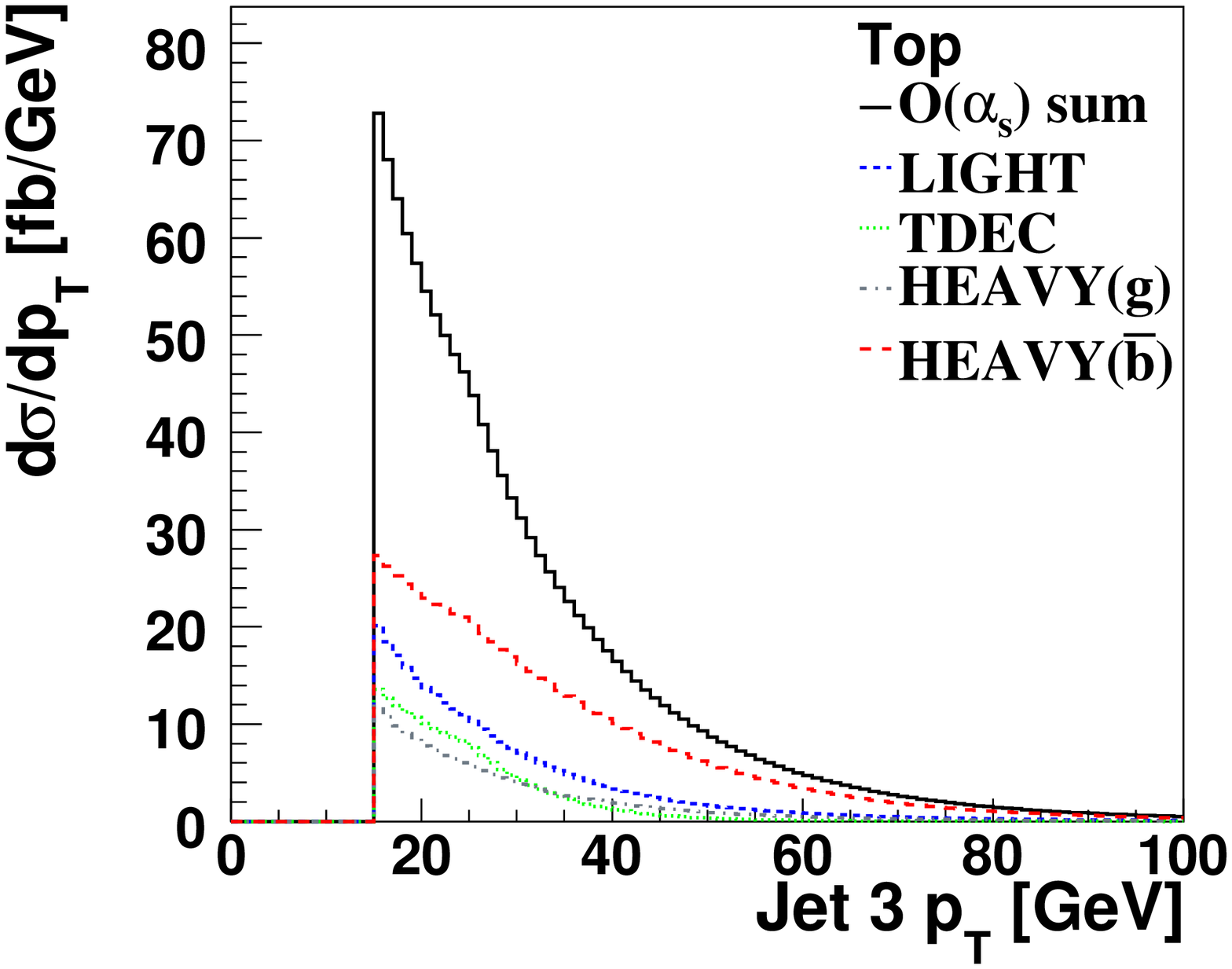}}
\subfigure[]{
\includegraphics[width=0.33\linewidth]{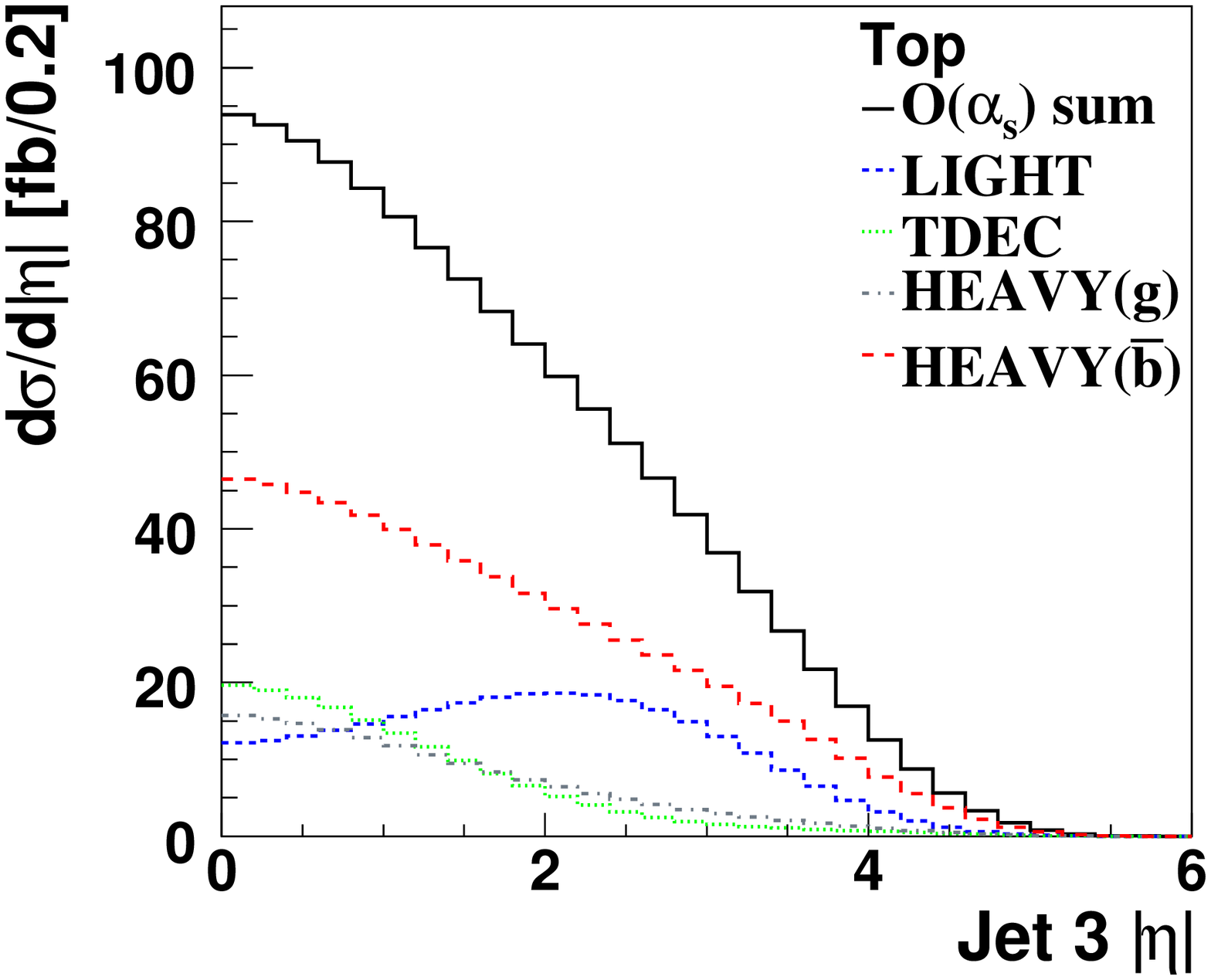}}

\caption{Third jet (a) $p_{T}$ and (b) $|\eta|$ for loose selection cuts
in top quark production at the 7~TeV LHC.\label{fig:Jet3PtEta}}
\end{figure}

Figure~\ref{fig:Jet3PtEta} shows the transverse momentum and pseudo-rapidity
of the different contributions to the third jet. The HEAVY correction
is broken up into its two components, one where the third jet originates from
a gluon (Fig.~\ref{fig:real_tchan}~(c)) and one where the third
jet originates from a $\bar{b}$-quark ($W$-gluon fusion, Fig.~\ref{fig:real_tchan}~(d)).
The $\bar{b}$-quark case accounts for about half of the total 3-jet
cross section, more at higher $p_{T}$ and higher $|\eta|$. The gluon
case accounts for about 15\% of the total 3-jet cross section, but
the gluon jet is more central and typically at lower $p_{T}$. The
LIGHT correction shows the same forward pseudo-rapidity peak as the
spectator jet (as expected) and contributes about 23\%. The TDEC correction
contributes about 12\%, mostly at low $p_{T}$ and in the central
pseudo-rapidity region. This limited range is the direct result of
the phase space available to the $b$-quark in the top quark decay.
Fig.~\ref{fig:Jet3PtEta} shows that all of the contributions are important
and none can be neglected when modeling $t$-channel events. While
the conventional wisdom that $W$-gluon fusion dominates the $t$-channel
holds true, nevertheless all diagrams contribute.

\begin{figure}[!h!tbp]
\subfigure[]{
\includegraphics[width=0.33\linewidth]{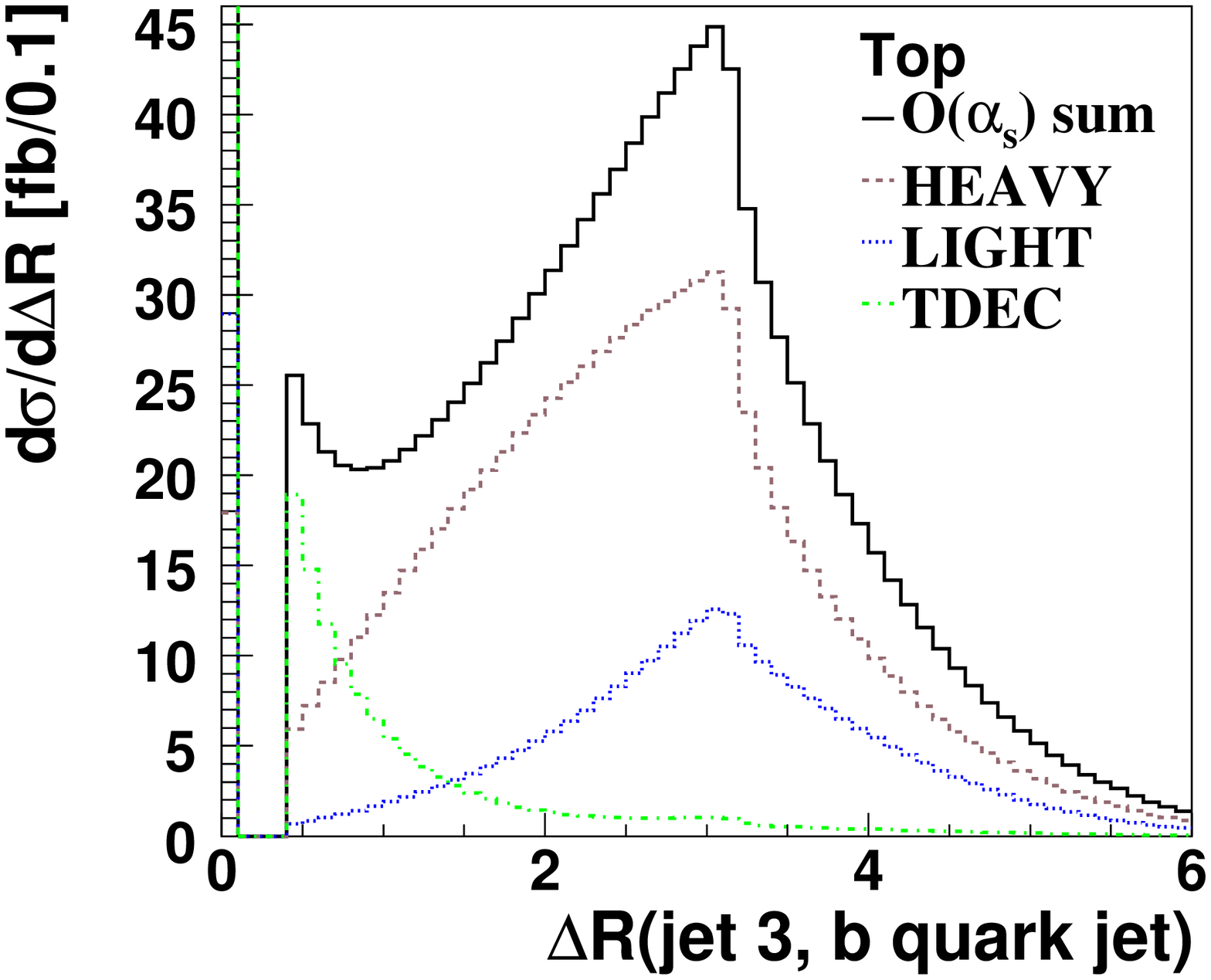}}
\subfigure[]{
\includegraphics[width=0.33\linewidth]{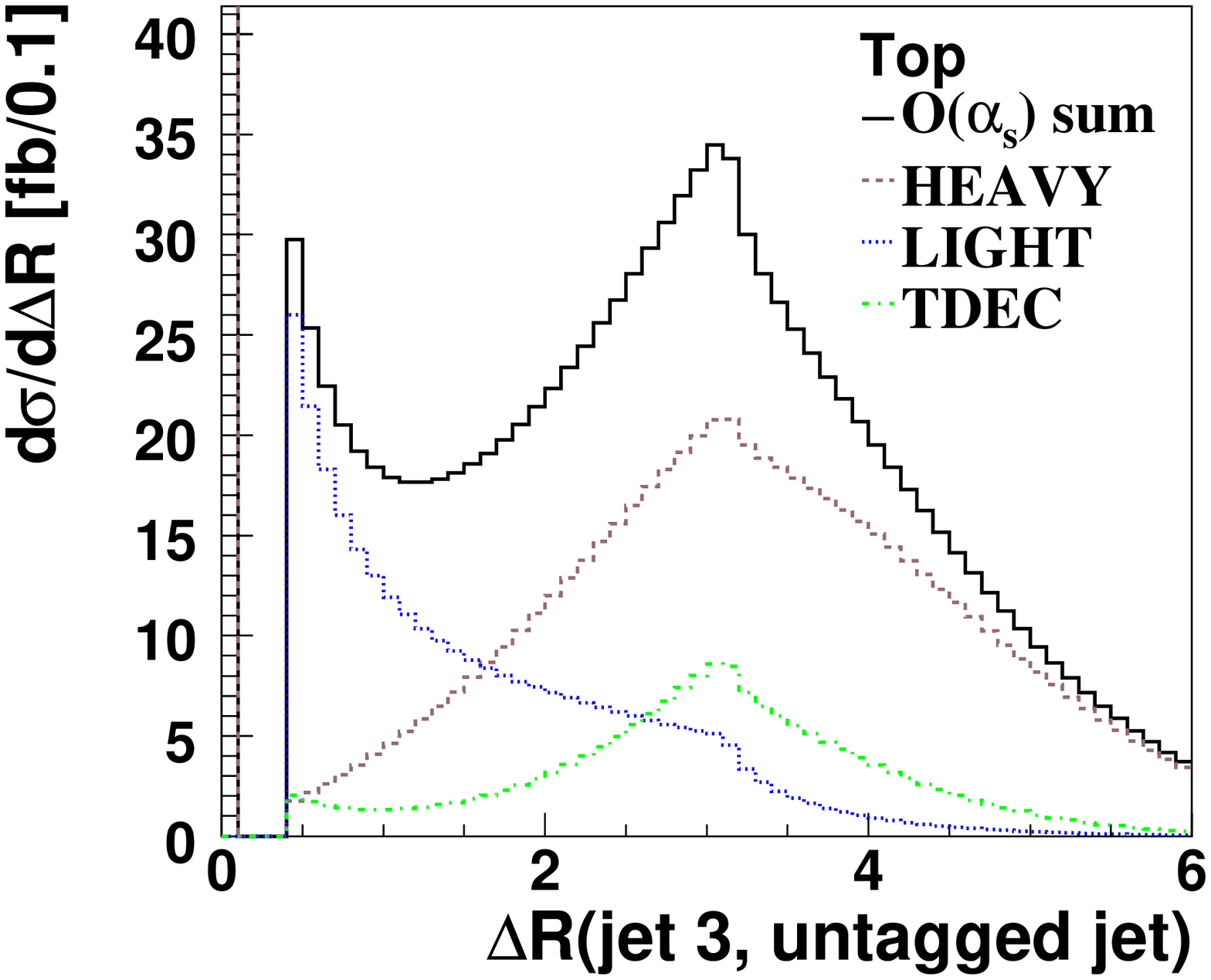}}

\caption{Separation between the third jet and (a) the tagged jet and (b) the
untagged jet after selection cuts for the various $\oalphas$ corrections
at the 7~TeV LHC. \label{fig:dRJet3bJet}}
\end{figure}

Normally an extra jet from decay-stage radiation should be included
in the top quark reconstruction in order to include all top quark
decay products. Figure~\ref{fig:dRJet3bJet}(a) shows that jets from
the TDEC contribution are close to the $b$~quark jet, thus at least
in principle easily identified. Extra jets from the production-stage
radiation (LIGHT and HEAVY corrections) are typically farther away
from the $b$-tagged jet. However, the HEAVY correction is large,
significantly larger than the TDEC correction for $\Delta R$ values
larger than about 0.8. The higher $p_{T}$ cuts for the tight selection
make this situation worse because they increase the relative size
of the production emission, and the situation is similar for antitop
quarks and for higher CM energies. Decay-stage radiation can more
easily be identified when an additional top mass constraint is used
to choose a jet pairing, see section~\ref{sub:Event-Reconstruction}. 

Figure~\ref{fig:dRJet3bJet}(b) shows the equivalent distribution
in $\Delta R$ between the extra jet and the untagged jet. In this
case the LIGHT radiation peaks close to the untagged jet as expected
from gluon radiation in the final state. There is an additional LIGHT
contribution at higher $\Delta R$ from production-stage radiation
of the light quark line. 

Note that there are entries at zero in the distributions in Fig.~\ref{fig:dRJet3bJet}.
These corresponds to events where the third jet is the leading $b$-tagged
jet (i.e. there are two higher $p_{T}$ untagged jets) or the leading
untagged jet (i.e. there are two higher $p_{T}$ tagged jets). The
third jet corresponds to the $b$-tagged jet in 13\% (12\%) of the
top (antitop) quark events. It corresponds to the untagged jet in
17\% (18\%) of the top (antitop) quark events.

\subsubsection{Identifying the spectator jet in 3-jet events}

In events containing two untagged jets, it is not
clear which of the two is the light quark jet and which is the gluon
(or $\bar{b}$-quark) jet. Two approaches are currently in use: choosing
the more forward jet in pseudo-rapidity~(as used for example by the
LHC experiments ATLAS and CMS, see section 3.4 in Ref.~\cite{Gerber:2007xk}),
and choosing the highest-$p_{T}$ jet that is not $b$-tagged (as
used for example by the Tevatron collaborations
\D0~\cite{singletopsearchPRDD0,singletopevidencePRDD0} and 
CDF~\cite{singletopobsCDFPRD}).
Here we explore the accuracy of these two methods, i.e. the fraction
of events for which the light quark jet is correctly identified in
3-jet events by each algorithm.

\begin{figure}[!h!tbp]
\subfigure[]{
\includegraphics[width=0.33\linewidth]{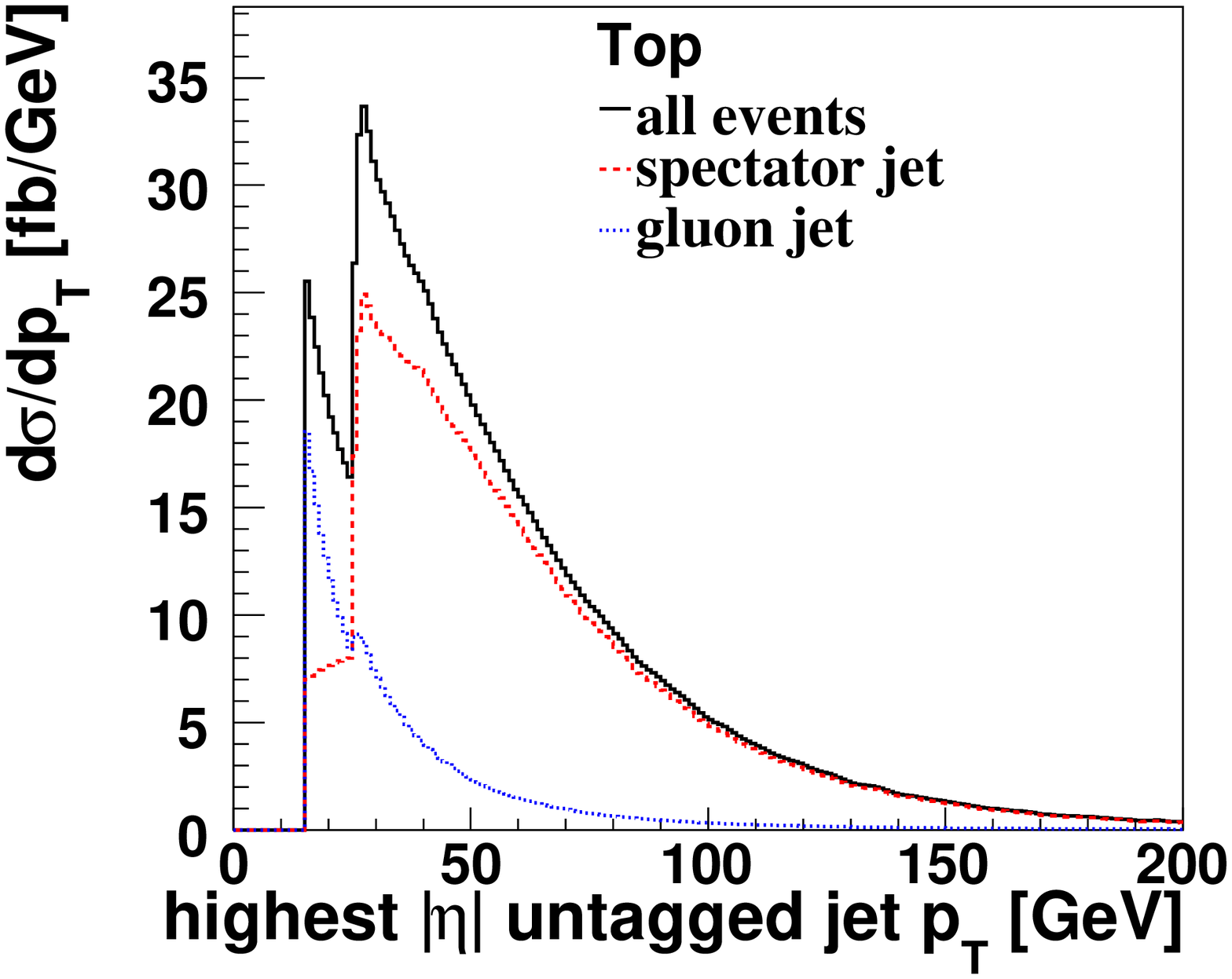}}
\subfigure[]{
\includegraphics[width=0.33\linewidth]{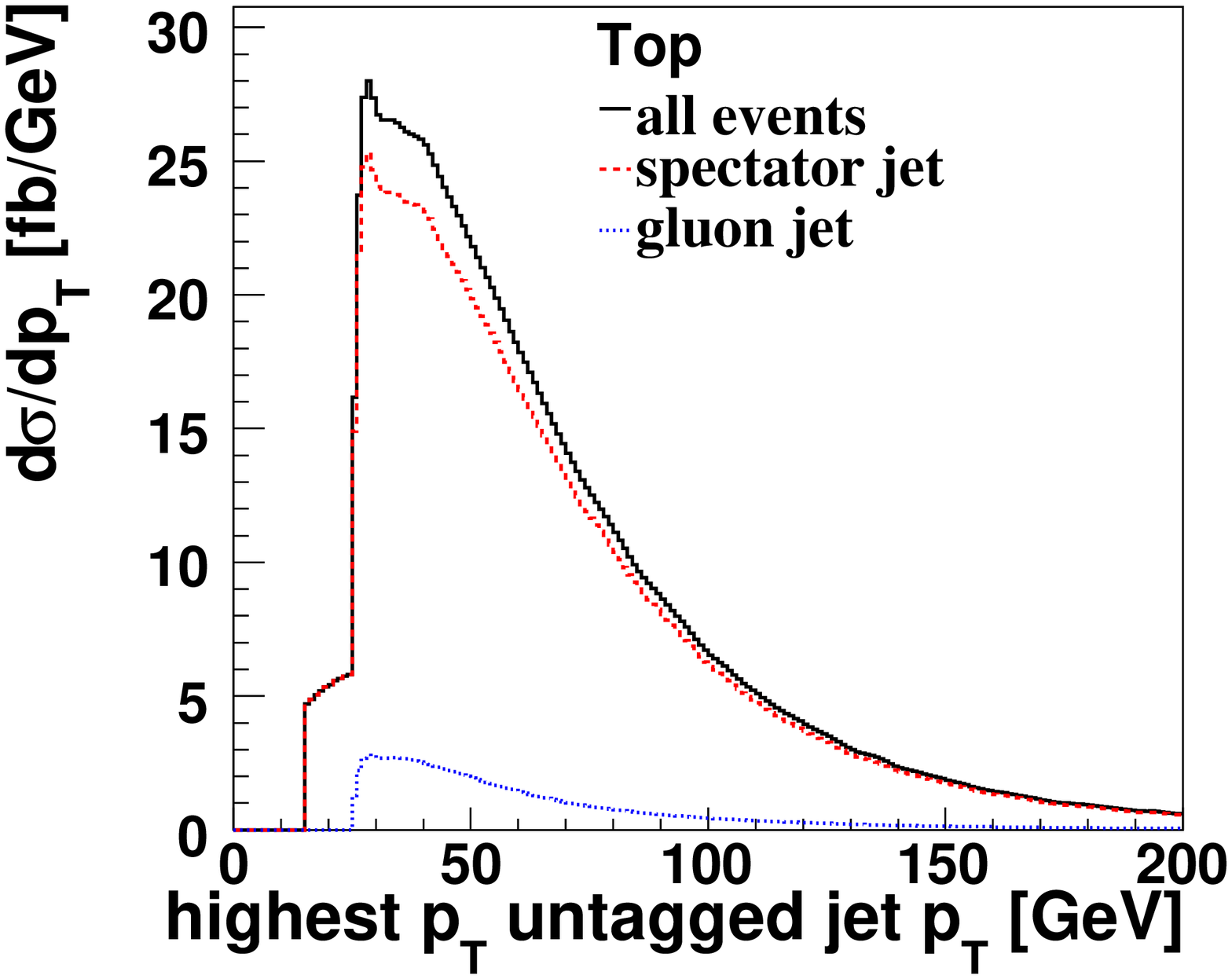}}

\caption{Transverse momentum of the untagged jet that is (a) highest in $|\eta|$
and (b) highest in $p_{T}$, for all events and separately for those
where it is the spectator jet or a gluon jet, at the 7~TeV LHC. \label{fig:UntaggedJetEff}}
\end{figure}

Figure~\ref{fig:UntaggedJetEff} shows the $p_{T}$ of the untagged
jet for these two cases, together with its composition in terms of
spectator jet or gluon jet. When the highest $|\eta|$ jet is chosen,
the spectator jet is correctly identified in only 80\% of the events.
By contrast, when the highest $p_{T}$ jet is chosen, it is correctly
identified in 92\% of the events after loose cuts. These efficiencies
are similar for antitop files and about 2\% lower at a CM energy of
14~TeV. Figure~\ref{fig:UntaggedJetEff} shows that for untagged
jet $p_{T}$ above 70~GeV, the correct jet is chosen in 95\% of the
events, approaching 100\% as the jet $p_{T}$ increases.

\subsubsection{Identifying the b quark jet in 3-jet events}

At Born-level, there is only one $b$-tagged jet,
which is identified with the $b$~quark from the top quark decay.
In events containing both a $b$~quark and a $\bar{b}$~quark, this
unambiguous association is not possible anymore. However, the additional
jet typically has lower $p_{T}$ than the $b$-quark from the top
quark decay, we therefore choose the highest $p_{T}$ $b$-tagged
jet as the $b$~quark from the top quark decay. 

\begin{figure}[!h!tbp]
\subfigure[]{
\includegraphics[width=0.33\linewidth]{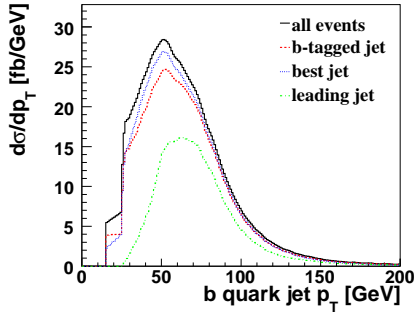}}

\caption{Transverse momentum of the $b$~quark jet, for all events and for
different $b$~quark jet reconstruction choices, after selection
cuts at the 7~TeV LHC. \label{fig:BID}}
\end{figure}

\begin{figure}[!h!tbp]
\subfigure[]{
\includegraphics[width=0.33\linewidth]{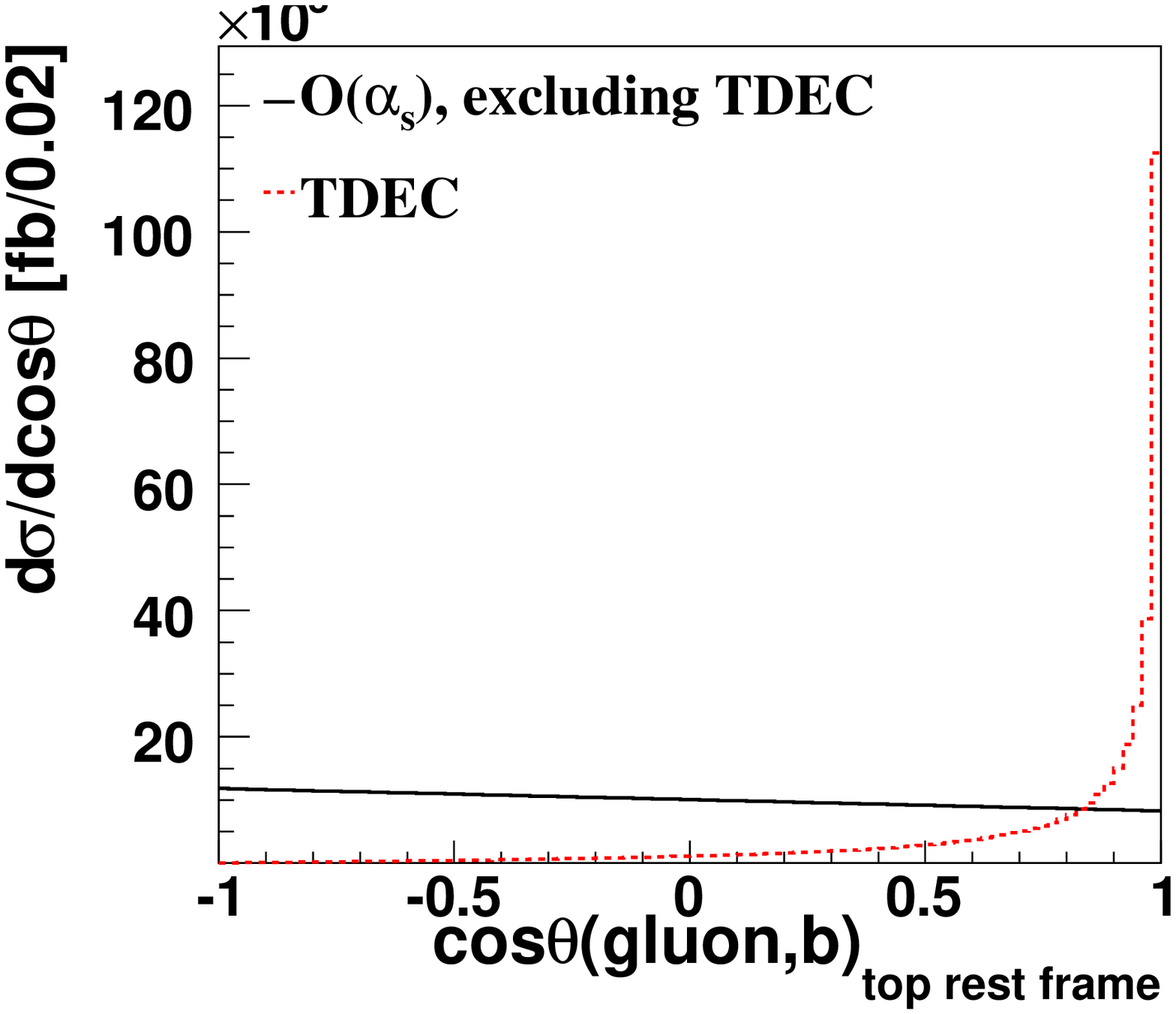}}
\subfigure[]{
\includegraphics[width=0.33\linewidth]{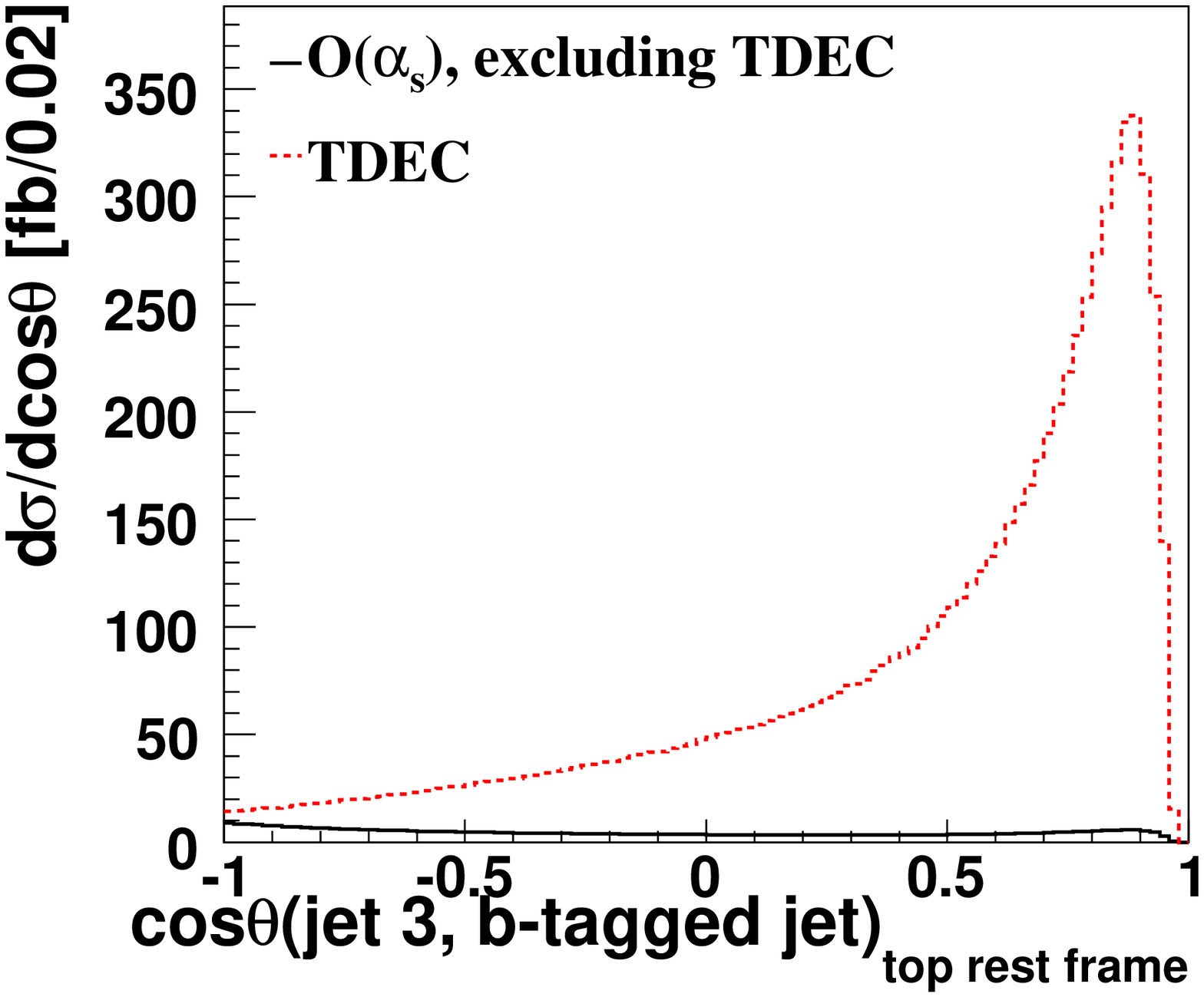}}

\caption{Cosine of the angle (a) between the gluon from the TDEC radiation
and the $b$~quark from the top quark decay and (b) between the third
jet and the $b$-tagged jet if the third jet is included in the top
quark reconstruction, both in the top quark rest frame, after selection
cuts at the 7~TeV LHC. \label{fig:CosThetaJet3bJet}}
\end{figure}

Figure~\ref{fig:BID} shows the $p_{T}$ of the $b$~quark jet from
the top quark decay, for all events and for three different algorithms
to identify the $b$~quark jet: a) choosing the leading jet in the
event (highest $p_{T}$), b) choosing the leading $b$-tagged jet,
and c) choosing the jet or jet pair that, when combined with the reconstructed
$W$~boson, gives a top quark mass closest to 173~GeV. The overall
fraction of correctly identified $b$~quark jets is 49\% for the leading
jet, 87\% for the leading $b$-tagged jet and 90\% for the best jet, 
for top quark production. These fractions for antitop quarks
are 40\%, 83\% and 91\%, respectively, slightly lower for the leading
jet and the leading $b$-tagged jet and unchanged for the best jet.
They are similar for higher CM energies and different cone sizes.
Algorithm a) is the most inefficient because there are two high-$p_{T}$
jets in each event. Simply choosing the leading $b$-tagged jet is an
obvious choice but requires b-tagging which reduces the acceptance. 
Algorithm c) improves upon b) slightly because
it also accounts for decay-stage radiation by considering a two-jet
system as the $b$~quark jet. We can investigate this decay-stage
radiation further by looking only at events where a two-jet system
is chosen. Figure~\ref{fig:CosThetaJet3bJet} shows the angle between
the two jets in this case, where both have been boosted into the top
quark rest frame.

Figure~\ref{fig:CosThetaJet3bJet} shows that in those event in which
the third jet is included in the top quark reconstruction it is close
in angle to the $b$-quark from the top quark decay as expected. At
the parton level, shown in Fig.~\ref{fig:CosThetaJet3bJet}(a), decay-stage
radiation peaks very closely to the $b$~quark from the top quark
decay, and is clearly distinguishable from the other $\oalphas$ corrections.
After event reconstruction and selection, the $cos\theta$ distribution
for decay radiation is broader, but still clearly peaks in the direction
aligned with the $b$-tagged jet.

Nevertheless, the difference between algorithm b) and c) is small,
and moreover algorithm c) will do worse when detector resolution effects
are included. Choosing the leading $b$-tagged jet does not suffer
from detector resolution problems and is almost as efficient. We will
therefore identify the leading $b$-tagged jet with the $b$~quark
jet from the top quark decay for the remainder of this paper.

\subsection{Top quark reconstruction\label{sub:Event-Reconstruction}}

The complete reconstruction of the single top quark final state, including the 
$W$~boson and the top quark itself, is necessary in order to not only take 
advantage of simple single-object kinematics but also of correlations between 
objects when separating the signal from the backgrounds. 
We use parton-level information for the $W$~boson, i.e. we reconstruct
it from the lepton and the neutrino. Experimentally, the $z$~momentum
of the neutrino ($p_{z}^{\nu}$) is not known and typically obtained
from a $W$~boson mass constraint. Here we use the parton-level $p_{z}^{\nu}$
to focus our studies on the effects of jet-related processes. We then combine
the $W$~boson with the leading $b$-tagged jet to form the top quark.

\begin{figure}[!h!tbp]
\subfigure[]{
\includegraphics[width=0.33\linewidth]{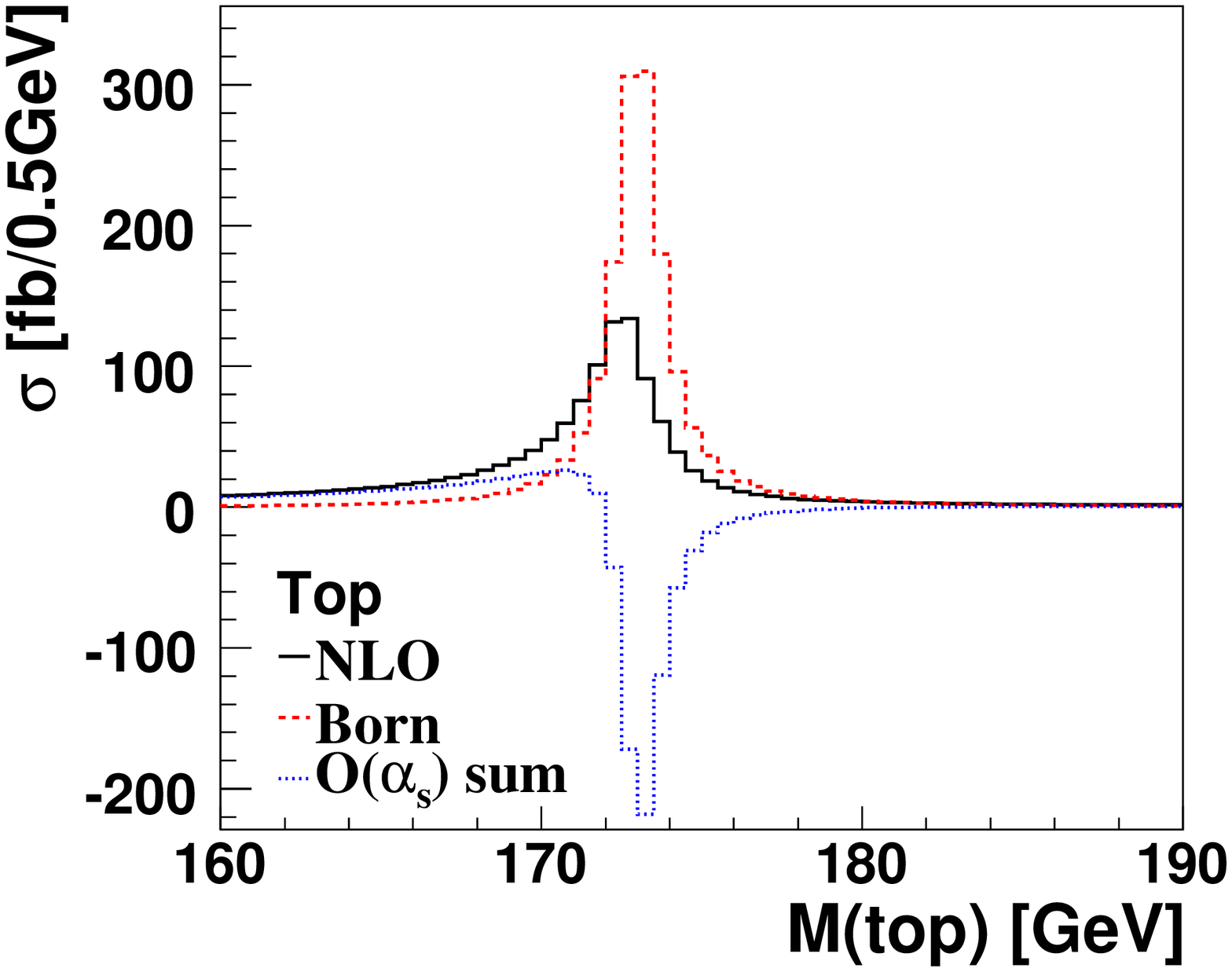}}
\subfigure[]{
\includegraphics[width=0.33\linewidth]{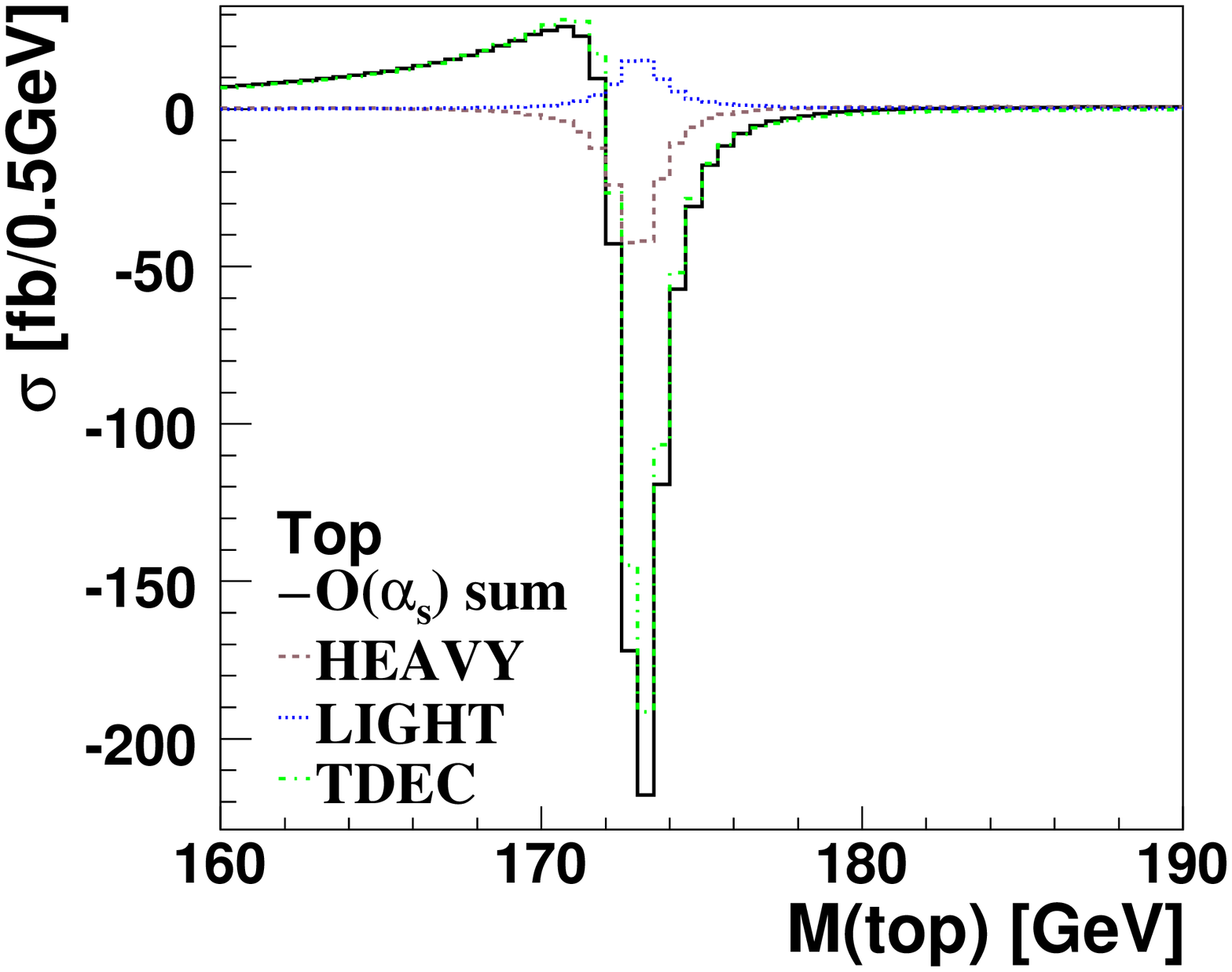}}

\caption{Invariant mass of the $W$~boson and the leading $b$-tagged jet
at the 7~TeV LHC, (a) comparing Born-level to NLO and (b) the individual
$\oalphas$ corrections. \label{fig:TopM}}
\end{figure}

Figure~\ref{fig:TopM} shows the invariant mass of the reconstructed
top quark for the different $\oalphas$ corrections. As expected,
the LIGHT and HEAVY corrections do not impact the shape of the invariant
mass distribution. The TDEC correction shifts the invariant mass to
lower values due to real-emission events where the additional jet
is not included in the top quark reconstruction.

\begin{figure}[!h!tbp]
\subfigure[]{
\includegraphics[width=0.33\linewidth]{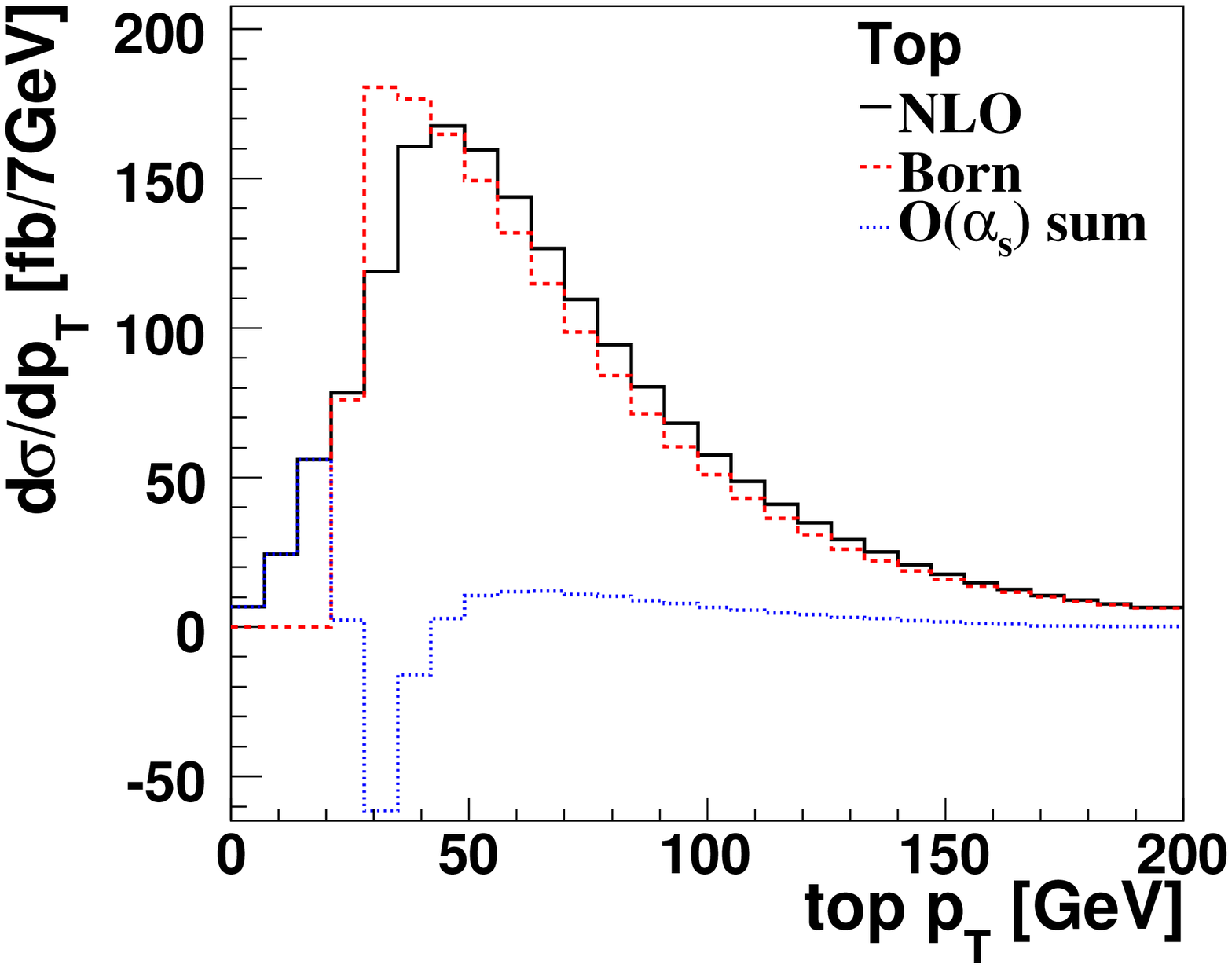}}
\subfigure[]{
\includegraphics[width=0.33\linewidth]{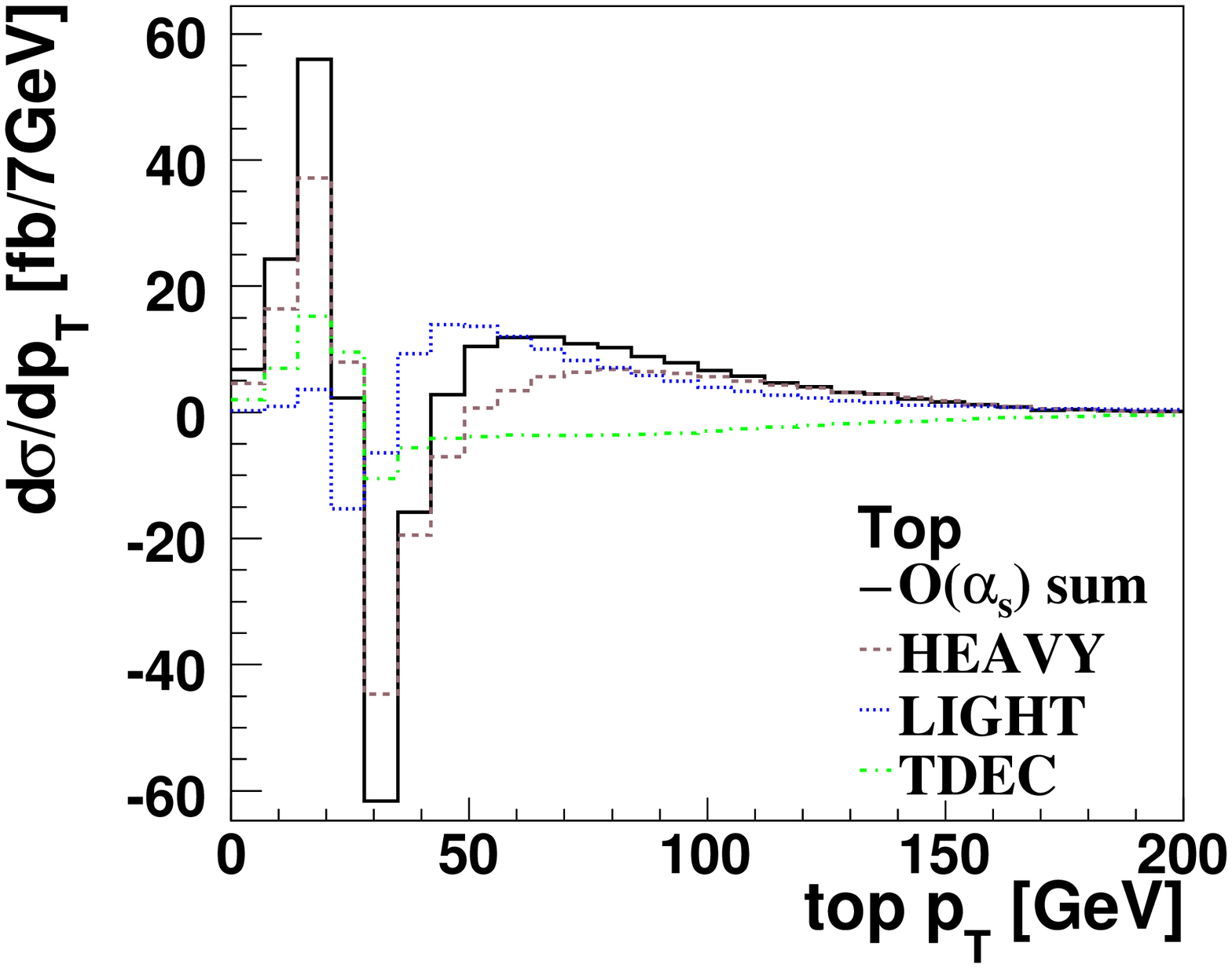}}

\caption{Transverse momentum of the reconstructed top quark, (a) comparing
Born-level to $\oalphas$ corrections and (b) the individual $\oalphas$
contributions at the 7~TeV LHC.\label{fig:TopPt}}
\end{figure}

Figure~\ref{fig:TopPt} shows the transverse momentum distribution
of the reconstructed top quark. At Born-level, this is identical to
the spectator jet $p_{T}$, but at $\oalphas$ additional real emission
changes that. TDEC emission shifts the top quark $p_{T}$ down, whereas
LIGHT and HEAVY emission tend to shift it up. The distributions for
antitop quarks and different CM energies are similar.

\begin{figure}[!h!tbp]
\subfigure[]{
\includegraphics[width=0.33\linewidth]{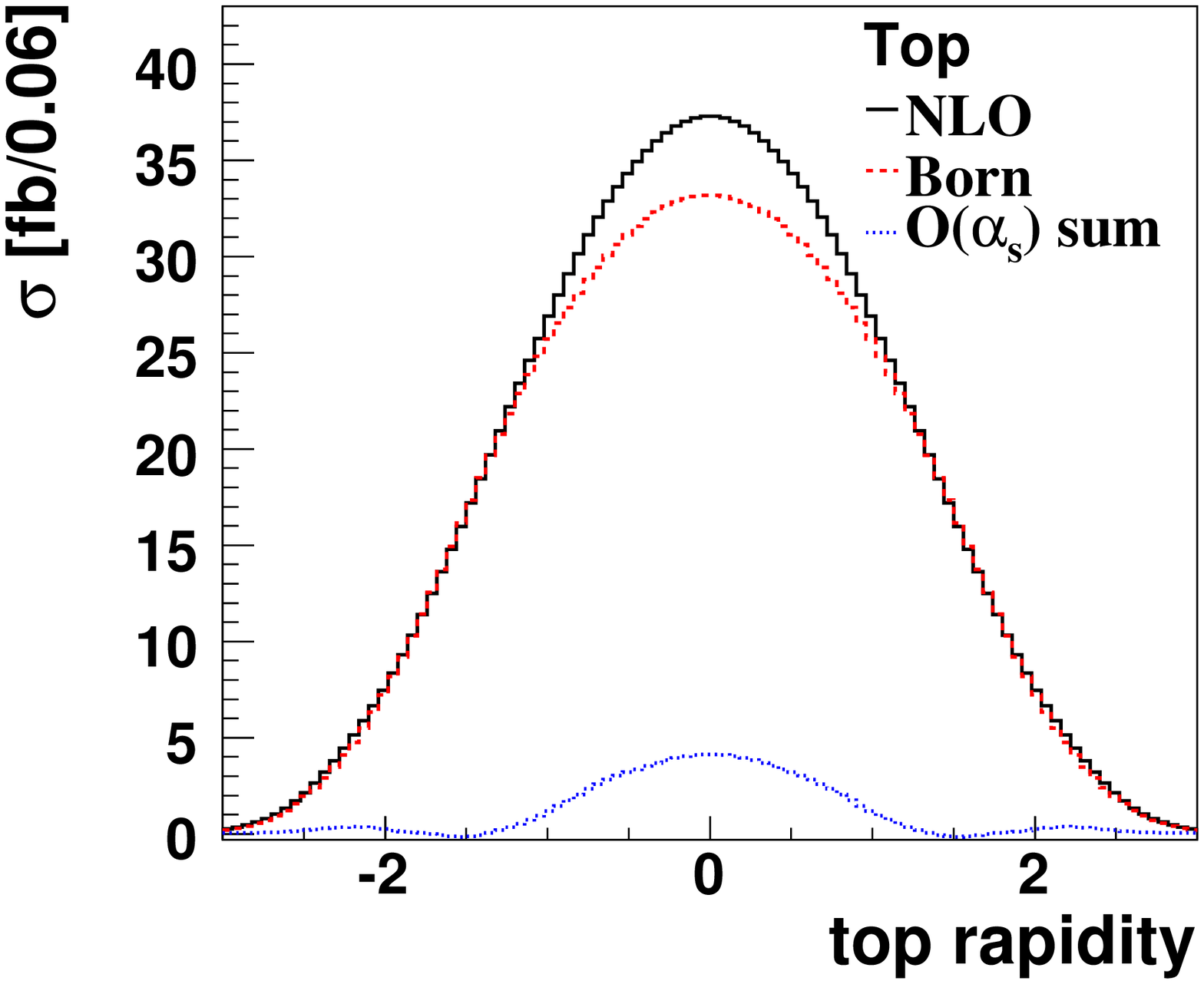}}
\subfigure[]{
\includegraphics[width=0.33\linewidth]{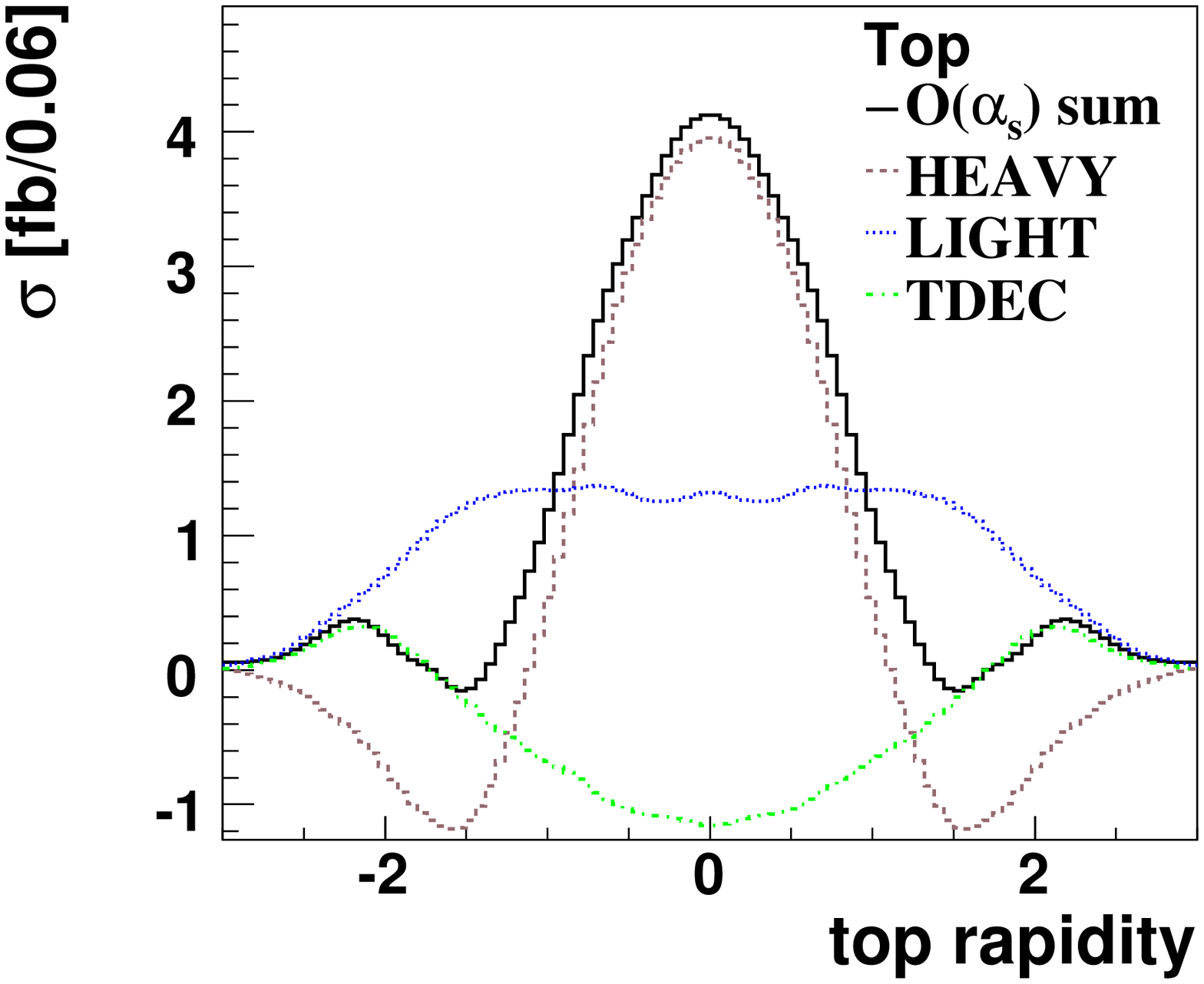}}

\subfigure[]{
\includegraphics[width=0.33\linewidth]{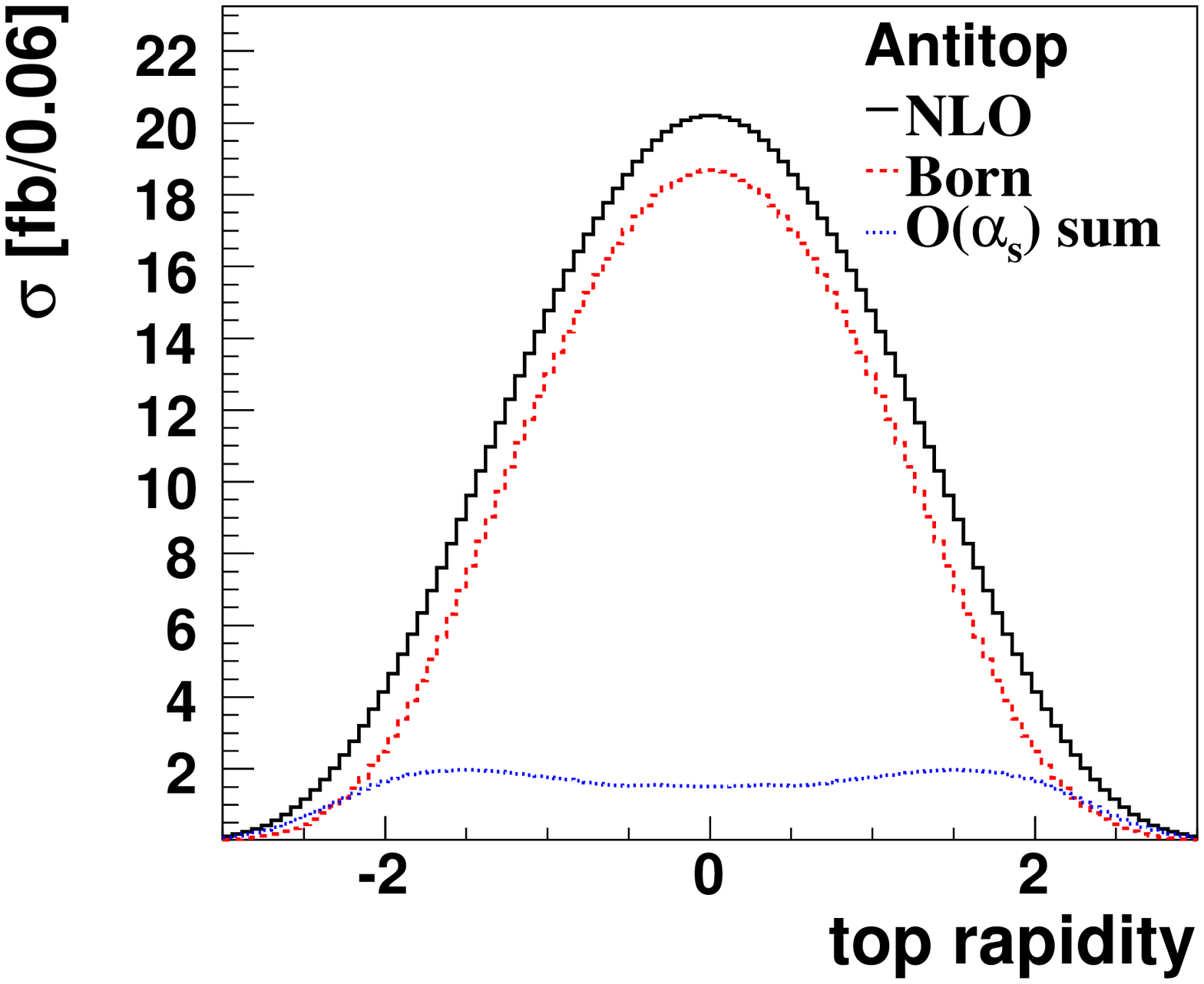}}
\subfigure[]{
\includegraphics[width=0.33\linewidth]{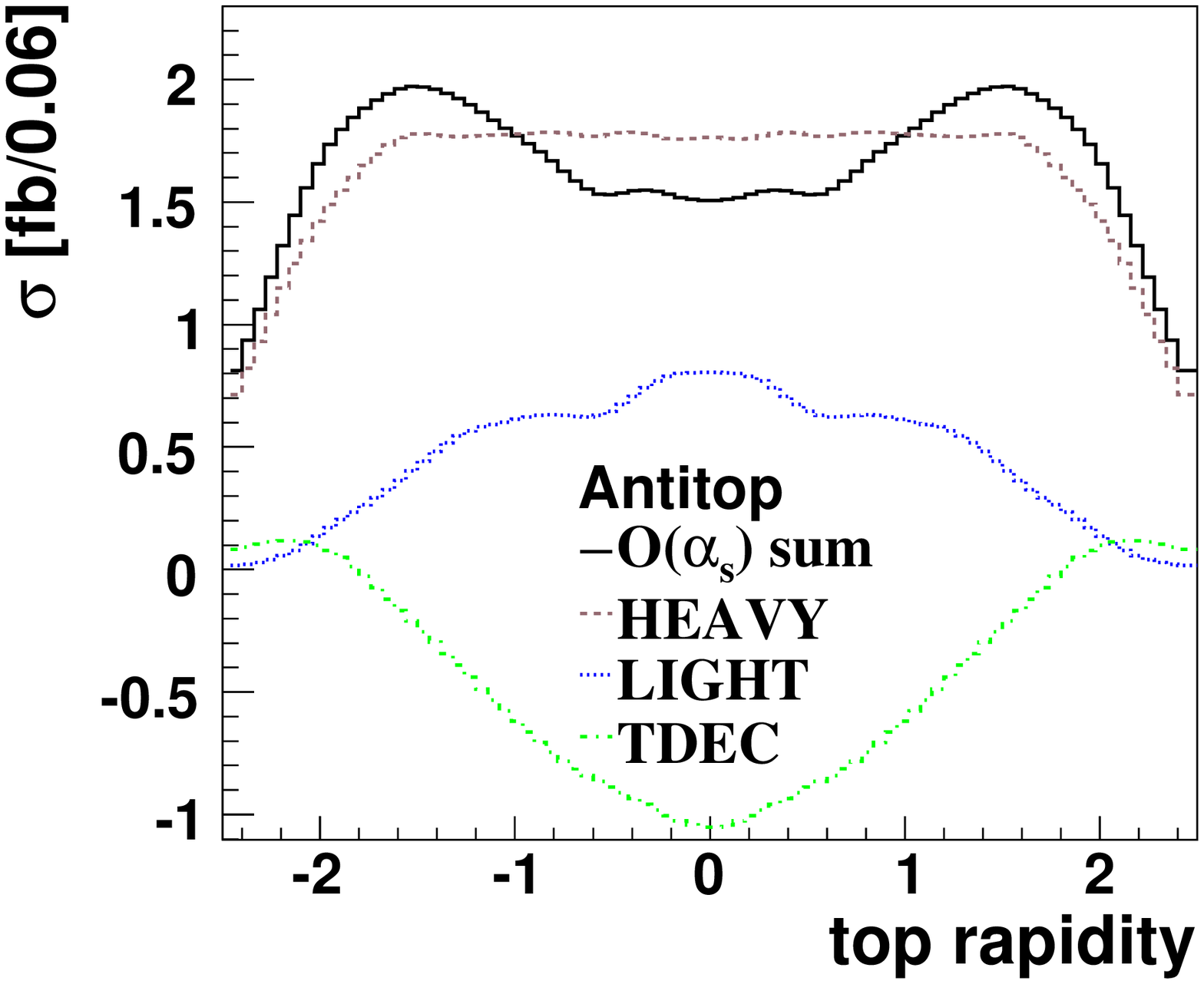}}

\caption{Rapidity of the top quark after selection cuts, (a, c) comparing Born-level
to $\oalphas$ corrections and (b, d) the individual $\oalphas$ contributions,
for (a, b) top quarks and (c, d) antitop quarks produced at the 7~TeV
LHC.\label{fig:TopRapidity}}
\end{figure}

Figure~\ref{fig:TopRapidity} compares the rapidity of the top quark
after selection cuts between top and antitop quark production. For
top quark production, shown in Figs.~\ref{fig:TopRapidity}(a) and~(b),
the HEAVY corrections shift the top quark to more central rapidities,
similar to the light quark pseudorapidity distribution in
Fig.~\ref{fig:spectator_eta_nlo}. The LIGHT and TDEC corrections have little 
to no effect as expected. The rapidity distribution is more narrow in antitop quark 
production due to the smaller parton momentum fraction of the incoming down quark compared 
to the incoming up quark that results in a top quark. At NLO, the top quark rapidity
distribution is widened, while the antitop quark rapidity distribution is only shifted
up in magnitude. This is mainly a result of the different contributions from the HEAVY
correction in the two cases. This difference between
top and antitop quark production is similar for different CM energies.

\subsection{Top quark polarization\label{sub:Object-Correlations}}

In this section we study angular correlations expected from event
kinematics, reconstructing the top quark using the leading $b$-tagged
jet and the $W$~boson. Since single top quark production in the SM is a
weak interaction process, the top quark is highly polarized in a suitable
basis, and this polarization can be measured. Detailed studies of the
top quark polarization and of angular correlations in the top quark
electroweak coupling to other particles is a sensitive probe to new 
physics~\cite{Chen:2005vr}. The spin correlations were also recently explored 
at the parton level for the 14~TeV LHC~\cite{Motylinski:2009kt}.
Here we study angular correlations between
top quark production and decay. In particular the charged lepton
is maximally correlated with the top quark spin~\cite{Mahlon:1995zn,Mahlon:1999gz}.
We can thus obtain the most distinctive distribution by plotting the
angle between the charged lepton and the spin axis of the top quark
in its rest frame.

\begin{figure}[!h!tbp]
\includegraphics[width=0.7\linewidth]{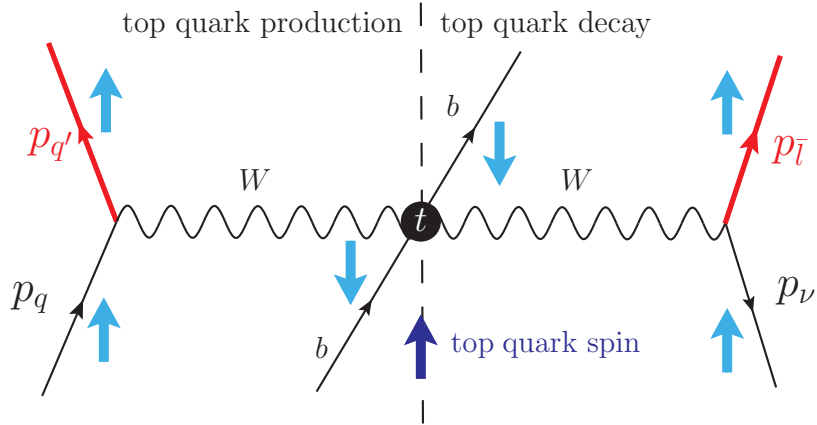}

\caption{Illustration of the spin correlation between top quark production
and top quark decay. The circle denotes the top quark rest frame and
light colored arrows indicate the spin direction.\label{fig:TopSpinBasis}}
\end{figure}

Three different axes for the spin polarization have been studied, 
each in the top quark rest frame: the helicity basis, the spectator basis, and the beamline
basis~\cite{Mahlon:1999gz}. Figure~\ref{fig:TopSpinBasis} illustrates the production and 
decay of the top quark from a spin correlation perspective, with the top
quark at rest in the center of the figure. Just as the lepton from
the top quark decay is maximally correlated with the top quark spin,
so is the down-type quark in the top quark production. This down-type
quark typically corresponds to the spectator quark in $t$-channel
top quark production. Hence the spectator basis should produce the
largest spin correlation for top quark production. In the beamline
basis, the top quark spin is measured along the direction of one of
the incoming protons. In $t$-channel antitop quark production the
down-type quark typically comes from one of the two incident protons,
but since the LHC is a proton-proton collider, a choice of direction
needs to be made. It has been proposed to simply choose the proton
that is most likely to have produced the spectator jet based on the
sign of $\eta$ of the spectator jet~\cite{Mahlon:1999gz}. Here
we also explore choosing based on the sign of $p_{z}$ of the c.m.
system. In the commonly used helicity basis, the top quark spin is
measured along the top quark direction of motion in the center of
mass frame, which is chosen as the frame of the (reconstructed top
quark, spectator jet) system after event reconstruction. 
We will examine all three bases for both top and antitop quark production here.

%

In the helicity basis, the c.m. frame needs to be reconstructed in
order to define the top quark momentum direction. This is straightforward
at Born-level but complicated by additional jet radiation at NLO.
Therefore, choosing the appropriate frame is necessary to maintain
the best spin correlation. In this study, two options for reconstructing
the c.m. frame are investigated:
\begin{enumerate}
\item $tq(j)$-frame: the c.m. frame of the incoming partons. This is the
rest frame of all the final state objects (reconstructed top quark
and all others jets). In exclusive three-jet events, this frame is
reconstructed by summing over the 4-momenta of top quark, spectator
jet and third-jet. 
\item $tq$-frame: the c.m. frame of the top quark and spectator jet. In
this case, even in exclusive three-jet events, the reference frame
is constructed by summing over only the 4-momenta of the top quark
and spectator jet.
\end{enumerate}
In exclusive two-jet events, the two frames are identical, they only
differ for exclusive three-jet events. At the Tevatron, it was found
that the $tq$-frame gives a larger degree of polarization. This is
also true at the LHC as shown in Table~\ref{tab:toppol-2jet} and
discussed below. We therefore choose the $tq$-frame when calculating
the top quark polarization in the helicity basis. 

\begin{figure}[!h!tbp]
\subfigure[]{
\includegraphics[width=0.33\linewidth]{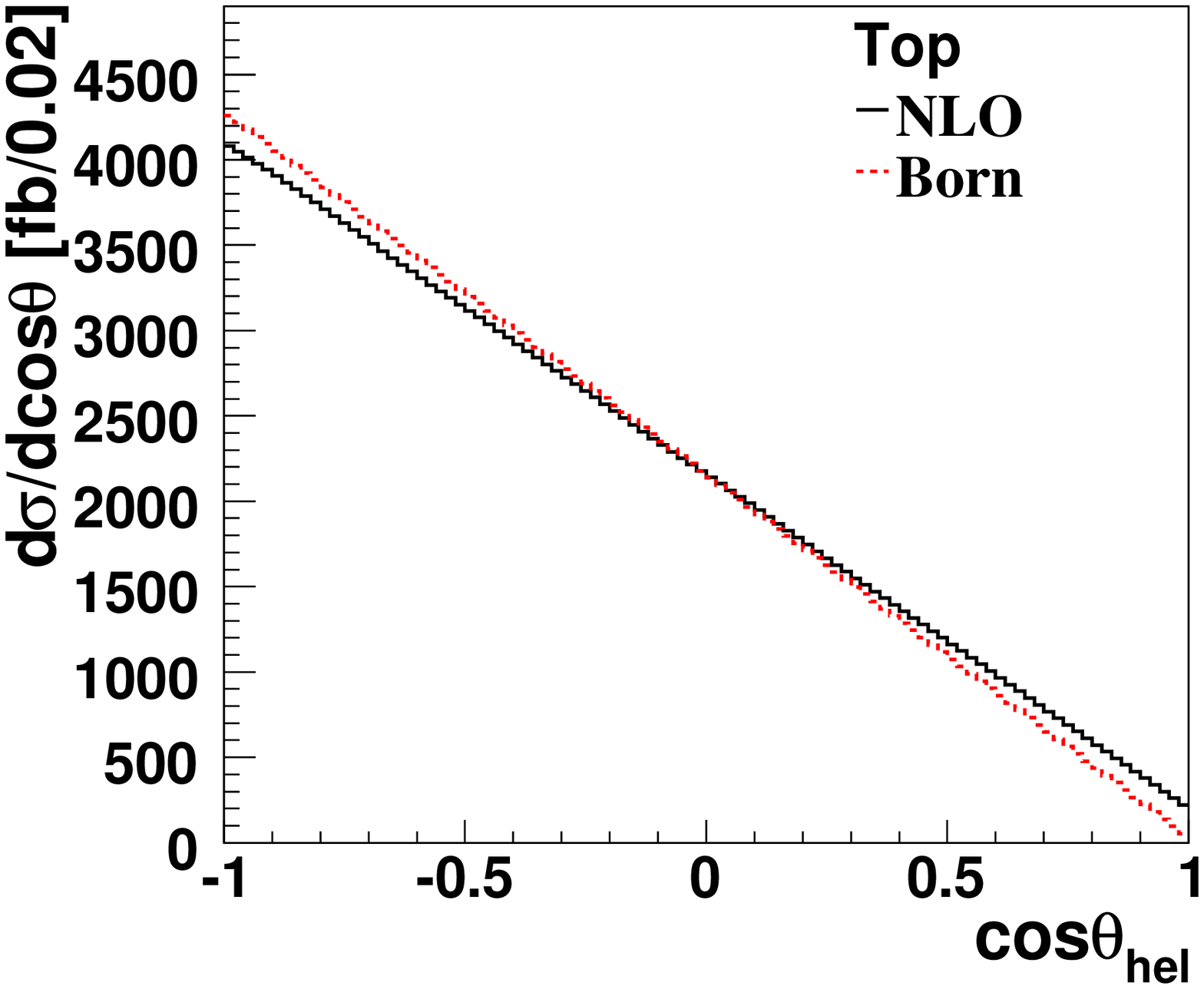}}
\subfigure[]{
\includegraphics[width=0.33\linewidth]{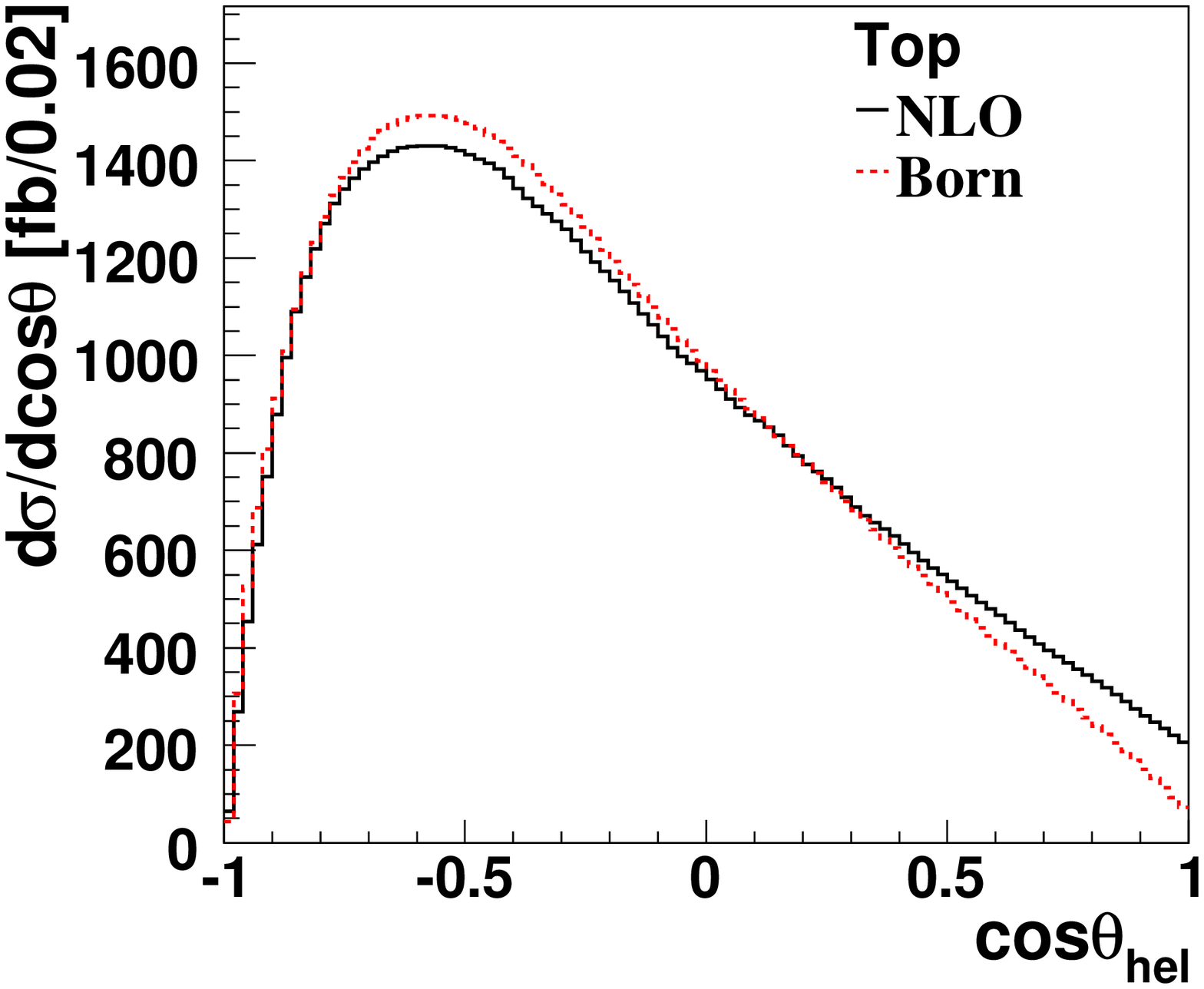}}

\caption{Top quark polarization in the helicity basis, (a) using the full parton
information and (b) after event reconstruction with selection cuts,
comparing Born-level to NLO (normalized to Born-level), at the 7~TeV
LHC.\label{fig:TopPolHel}}
\end{figure}

In the helicity basis, the polarization of the top quark is examined
as the angular distribution ($\cos\theta_{hel}$) of the lepton in
the top quark frame relative to the moving direction of the top quark
in the c.m. frame. The angular correlation in this frame is given
by \begin{eqnarray}
\cos\theta_{hel}=\frac{\vec{p}_{t}\cdot\vec{p}_{\ell}^{\;*}}{|\vec{p}_{t}||\vec{p}_{\ell}^{\;*}|},\end{eqnarray}
where $\vec{p}_{\ell}^{\;*}$ is the charged lepton three-momentum
defined in the rest frame of the top quark, whose three momentum is
denoted as $\vec{p_{t}}$, which is in turn defined in the c.m. frame.
For a left-handed top quark, the angular correlation of the lepton
$\ell^{+}$ is given by $(1-\cos\theta_{hel})/2$, and for a right-handed
top quark, it is $(1+\cos\theta_{hel})/2$. Figure~\ref{fig:TopPolHel}(a)
shows that this linear relationship for $\cos\theta_{hel}$ is indeed
a valid description for $t$-channel single top quark events at the
parton level. The figure also shows that the top quark is almost completely
polarized in the helicity basis at Born-level, and that this polarization
is weakened when including $\oalphas$ corrections. Figure~\ref{fig:TopPolHel}(b)
shows that this weakening is amplified after event reconstruction,
where the effect of the lepton-jet separation cut can also be seen,
as the drop-off of the $\cos\theta_{hel}$ distribution close to a
value of $-1$. This corresponds to the events in which the top quark
is back-to-back with the lepton, hence the spectator jet is aligned
with the lepton.

In the spectator basis, the relevant angular correlation for the $t$-channel
process is $\cos\theta_{spec}$, defined as \begin{eqnarray}
\cos\theta_{spec}=\frac{\vec{p}_{spec}^{\;*}\cdot\vec{p}_{\ell}^{\;*}}{|\vec{p}_{spec}^{\;*}||\vec{p}_{\ell}^{\;*}|}\,,\end{eqnarray}
 where $\vec{p}_{spec}^{\;*}$ is the spectator jet three-momentum
in the top quark rest frame and $\vec{p}_{\ell}^{\;*}$ is the lepton
three-momentum in the top quark rest frame. For top quark production,
this basis picks the wrong spin axis direction for the $\bar{d}b$
and $b\bar{d}$ initial states, but as pointed out in Ref.~\cite{Mahlon:1999gz},
the spectator jet is almost parallel to the initial state light quark,
thus some spin correlation is preserved even in that case. At Born-level
the polarization is identical between the helicity basis and the spectator
basis because the spin quantization axes point in opposite directions
(in the c.m. frame, the light quark and the top quark are back-to-back).
This is also true at NLO when the $tq$~frame is used as the c.m.
frame.

\begin{figure}[!h!tbp]
\subfigure[]{
\includegraphics[width=0.33\linewidth]{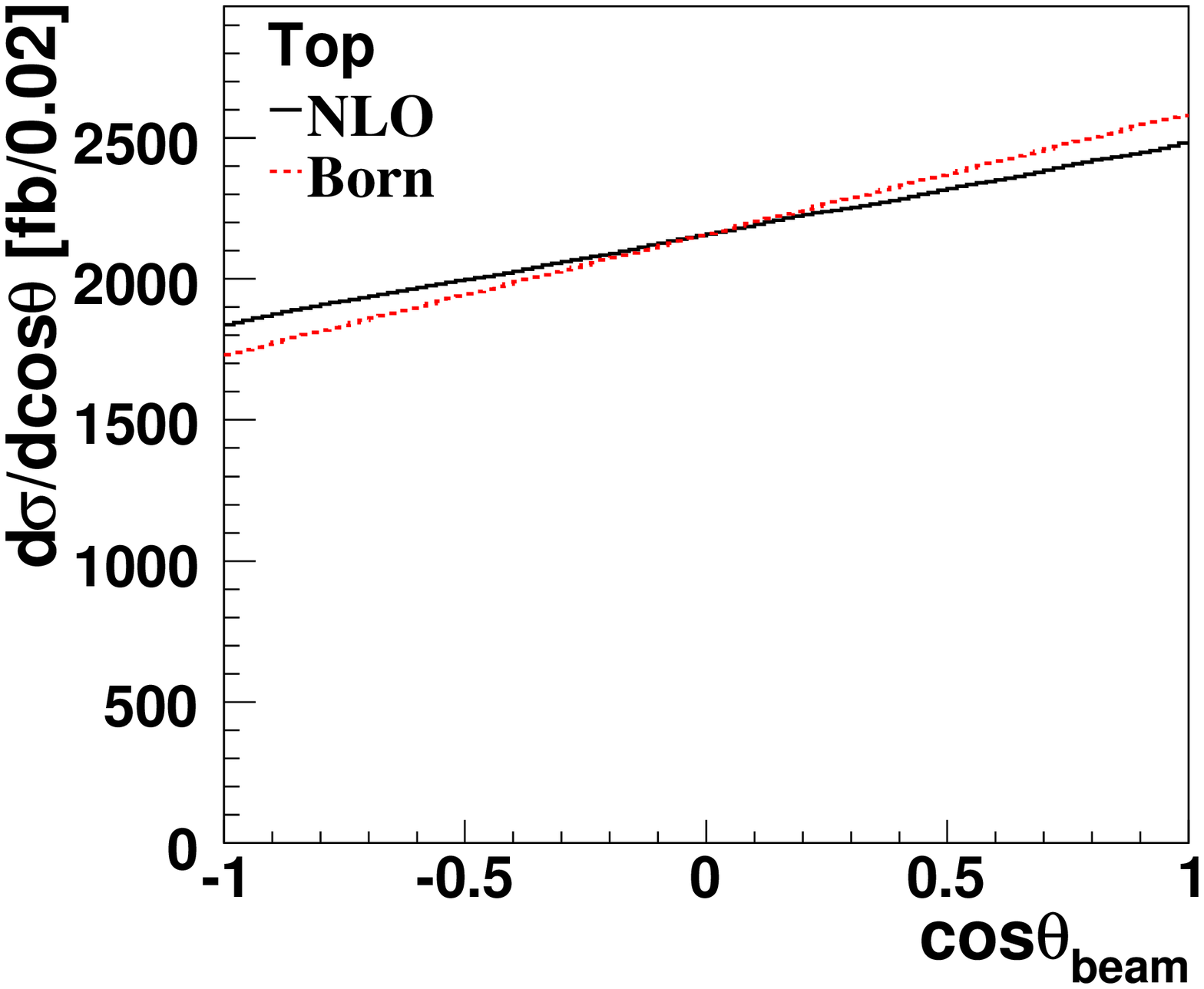}}
\subfigure[]{
\includegraphics[width=0.33\linewidth]{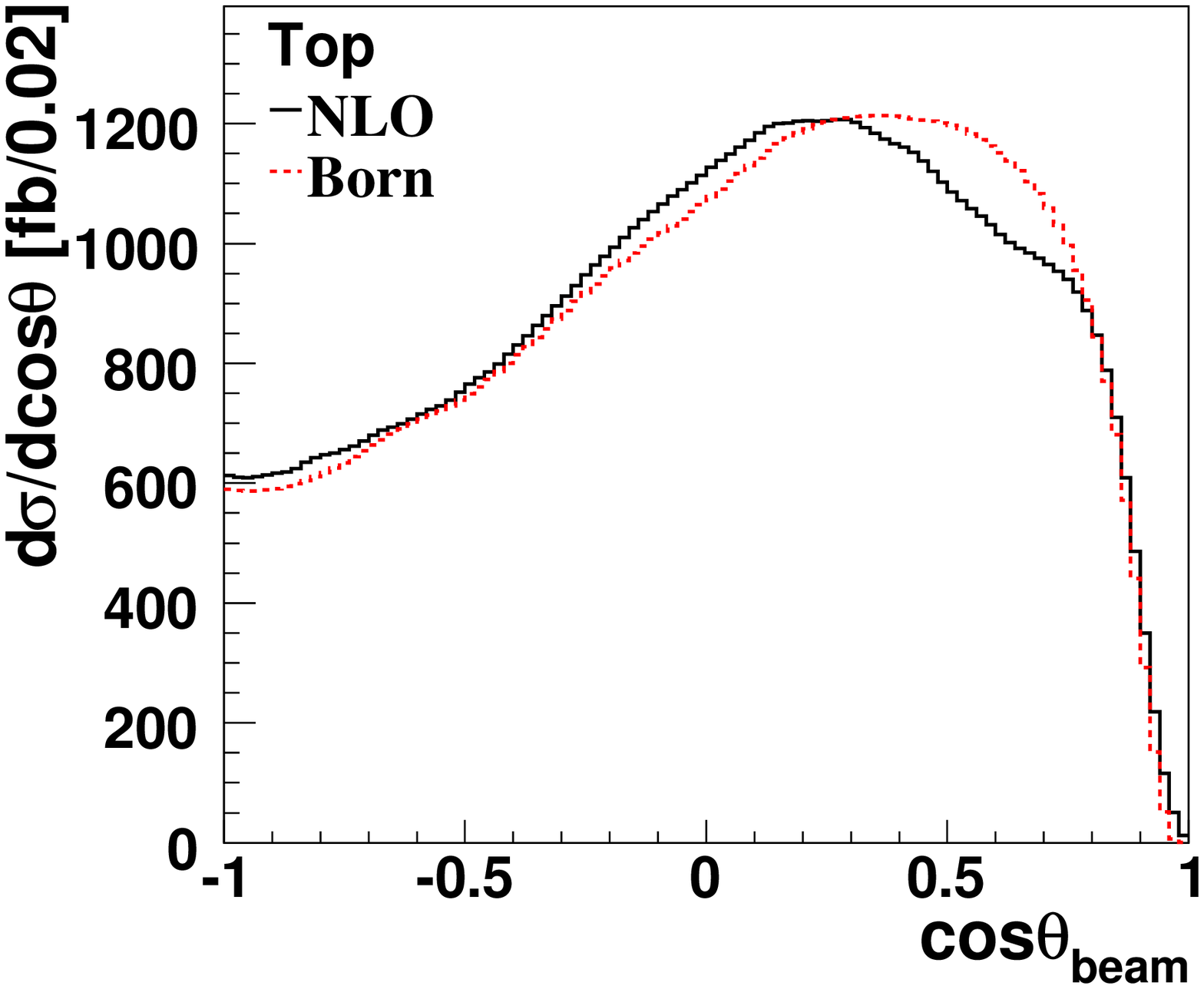}}
\subfigure[]{
\includegraphics[width=0.33\linewidth]{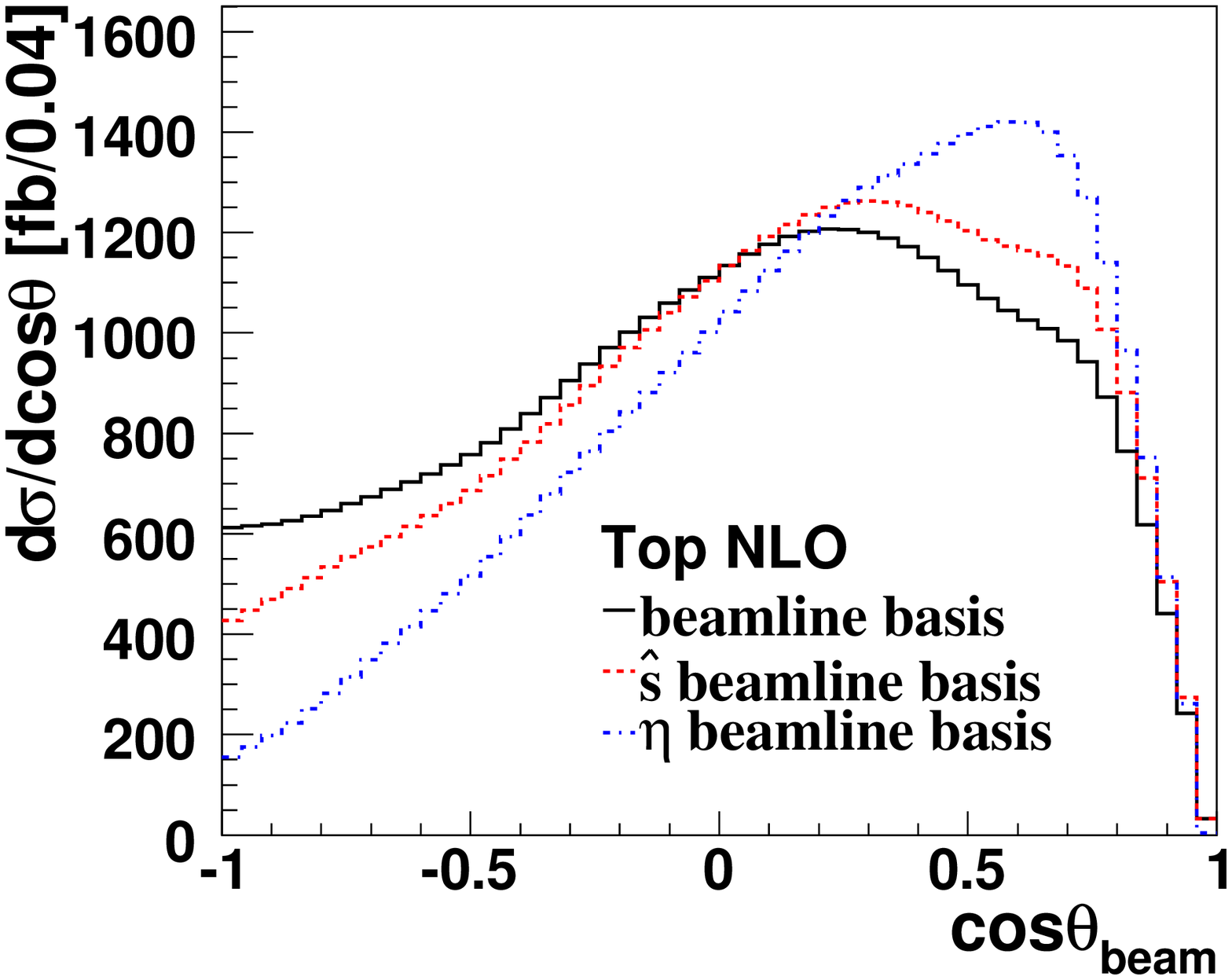}}
\caption{Top quark polarization (a, b) in the beamline basis, (a) using parton
information and (b) after event reconstruction with selection cuts,
comparing Born-level to NLO (normalized to Born-level). (c) comparison
of the beamline basis to the $\hat{s}$~beamline basis and the $\eta$~beamline
basis at the 7~TeV LHC.\label{fig:TopPolBeam}}
\end{figure}

In the beamline basis, the relevant angular correlation for the $t$-channel
process is $\cos\theta_{beam}$, defined as \begin{eqnarray}
\cos\theta_{beam}=\frac{\vec{p}_{p}^{\;*}\cdot\vec{p}_{\ell}^{\;*}}{|\vec{p}_{p}^{\;*}||\vec{p}_{\ell}^{\;*}|}\,,\end{eqnarray}
 where $\vec{p}_{p}^{\;*}$ is the three-momentum of one of the protons
in the top quark rest frame and $\vec{p}_{\ell}^{\;*}$ is the lepton
three-momentum in the top quark rest frame. For a top quark polarized
in the positive $z$-direction, the angular distribution of the lepton
$\ell^{+}$ is $(1+\cos\theta_{beam})/2$, while for a top quark polarized
in the negative $z$-direction it is $(1-\cos\theta_{beam})/2$. Since
the LHC is a proton-proton collider, it is unknown which of the two
protons provided the light quark in each event. Three different approaches
are explored here to solve this ambiguity: a) choosing the positive
$z$-direction for every event (simply called beamline basis), b)
choosing the $z$-direction of the spectator quark ($\eta$~beamline
basis)~\cite{Mahlon:1999gz}, and c) choosing the $z$-direction
of the c.m. frame of all particles in the lab frame ($\hat{s}$~beamline
basis). Figure~\ref{fig:TopPolBeam} shows the linear relationship
for $\cos\theta_{beam}$ in the beamline basis. The distribution is
much more flat than that in the helicity basis, and Fig.~\ref{fig:TopPolBeam}
demonstrates that the $\hat{s}$~beamline basis and the $\eta$~beamline
basis both significantly improve this situation because the correct
down-quark direction is picked more often. After event reconstruction
the situation is similar: the spin correlation is further
reduced and shows a drop-off close to one due to the limited lepton
$\eta$~range.

To better quantify the change in polarization, it is useful to define
the degree of polarization $\mathcal{D}$ of the top quark. This is
given as the ratio\[
\mathcal{D}=\frac{N_{-}-N_{+}}{N_{-}+N_{+}},\]
 where $N_{-}$ ($N_{+}$) is the number of left-handed (right-handed)
polarized top quarks in the helicity basis. Similarly, in the spectator
(beamline) basis, $N_{-}$ ($N_{+}$) is the number of top quarks
with polarization against (along) the direction of the spectator jet
(proton) three momentum in the top quark rest frame. The angular distribution
is then given by~\cite{Mahlon:1998uv}\textbf{\begin{eqnarray*}
\frac{1}{\sigma}\frac{d\sigma}{d(\cos\theta)} & = & \frac{N_{-}}{N_{-}+N_{+}}\frac{1+\cos\theta}{2}+\frac{N_{+}}{N_{-}+N_{+}}\frac{1-\cos\theta}{2}\\
 & = & \frac{1}{2}\left(1+D\cos\theta_{i}\right).\end{eqnarray*}
}Simple algebra leads to the following identity: \begin{eqnarray}
\mathcal{D} & = & -3\int_{-1}^{1}x\,{\frac{d\sigma}{\sigma dx}}\, dx\,,\label{dpola}\end{eqnarray}
 where ${\displaystyle \frac{d\sigma}{\sigma dx}}$ is the normalized
differential cross section as a function of the polar angle $x$.
Here, $x$ denotes $\cos\theta_{hel}$ in the helicity basis, etc.
The polarization fraction and asymmetry are not given here but can
be calculated easily from the numbers provided~\cite{Cao:2005pq}.

\begin{table}
\begin{centering}
\begin{tabular}{|l|c|c|c|c|}
\hline 
 & \multicolumn{2}{|c|}{Top} & \multicolumn{2}{|c|}{Antitop}\\
 & LO & NLO & LO & NLO\\
\hline
Helicity basis:  & & & & \\
~~Parton level ($tq$-frame) & 0.99 & 0.91 & 0.93 & 0.86\\
~~Parton level ($tq(j)$-frame) &   & 0.66 &      & 0.62\\
~~Loose selection ($tq$-frame) & 0.70 & 0.60 & 0.63 & 0.56\\
~~Loose selection ($tq(j)$-frame) &   & 0.58 &      & 0.54\\
~~Tight selection ($tq$-frame) & 0.71 & 0.72 & 0.69 & 0.73\\
~~Tight selection (($tq(j)$-frame) &  & 0.76 &      & 0.76\\
\hline
Spectator basis: & & & & \\
~~Parton level & -0.99 & -0.91 & 0.93 & 0.86\\
~~Reconstructed events & -0.70 & -0.60 & 0.63 & 0.56\\
\hline
Beamline basis: & & & & \\
~~Parton level & -0.20 & -0.15 & 0.14 & 0.11\\
~~Loose selection & -0.14 & -0.09 & 0.07 & 0.06\\
~~Tight selection & -0.28 & -0.26 & -0.20 & -0.21\\
\hline
$\eta$~beamline basis: & & & & \\ 
~~Parton level & -0.64 & -0.64 & -0.77 & -0.67\\
~~Loose selection & -0.56 & -0.49 & -0.62 & -0.49\\
~~Tight selection & -0.61 & -0.50 & -0.63 & -0.53\\
\hline
$\hat{s}$~beamline basis: & & & & \\
~~Loose selection & -0.41 & -0.26 & -0.38 & -0.14\\
\hline
\end{tabular}
\par\end{centering}

\caption{Degree of polarization $\mathcal{D}$ for inclusive two-jet single
top quark events, at parton level before cuts and after selection
cuts in the $t$-channel process at the 7~TeV LHC. The $tq(j)$-frame
in the helicity basis denotes the c.m. frame of the incoming partons,
while the $tq$-frame denotes the rest frame of the top quark and
spectator jet. \label{tab:toppol-2jet}}
\end{table}

Table~\ref{tab:toppol-2jet} shows $\mathcal{D}$ for inclusive two-jet
events at parton level and after the loose set of selection cuts.
The result for exclusive three-jet events is shown in Table~\ref{tab:toppol-3jet}.
As explained above, the degree of polarization is the same in the
helicity basis using the $tq$-frame and the spectator basis as a
result of the event kinematics. This is true even after reconstruction.
The top quark is almost completely polarized in the helicity and spectator
bases, and the $O(\alpha_{s})$ corrections only degrade that picture
slightly. At the parton level, the degree of top quark polarization in the
helicity basis is larger in the $tq$-frame than in the $tq(j)$-frame. After
event reconstruction there is a dependence on the selection cuts. For the 
loose selection cuts, the $tq$-frame is better than the $tq(j)$-frame, whereas
for the tight selection cuts the $tq(j)$-frame is better. The tight selection
cuts require a forward untagged jet and a central, large $b$-tagged jet $p_T$, 
thus modifying the event kinematics, supressing the HEAVY correction and 
enhancing the LIGHT corrections, see Figs.~\ref{fig:etab} 
and~\ref{fig:spectator_eta_nlo}. As a result the degree of polarization in the
$tq(j)$-frame is increased. The beamline basis produces only
small degrees of polarization as expected. This gets better when choosing
a direction in the $\eta$~beamline basis. Especially for tight cuts
which require the spectator jet to be forward, the polarization in
the $\eta$~beamline basis improves, but even then it is still smaller
than in the helicity or spectator basis. $\mathcal{D}$ is also consistently
smaller for antitop quarks than for top quarks for all methods. In
the exclusive three-jet sample, the degree of polarization is further
reduced because the third jet affects the kinematics of either the
spectator jet or the top quark. The spin polarization measurements
at higher CM energies show the same result.

Our study shows that the helicity basis (using the $tq$-frame) and
the spectator basis are equally good to study the top quark polarization.
In the $s$-channel process, the measured polarization could be enhanced
significantly by choosing a direction for the incoming down-type quark
based on the boost of the virtual $W$~boson~\cite{Heim:2009ku}.
In the $t$-channel this is also useful for the beamline basis. However,
even after such a choice that basis still produces a smaller degree of
polarization than the helicity or spectator bases.

\begin{table}
\begin{centering}
\begin{tabular}{|l|c|c|}
\hline 
 & Top & Antitop \\
\hline
Helicity basis:  & & \\
~~Loose cuts ($tq$-frame) & 0.60 & 0.53\\
~~Loose cuts ($tq(j)$-frame) & 0.58 & 0.50\\
~~Tight cuts ($tq$-frame) & 0.67 & 0.66\\
~~Tight cuts ($tq(j)$-frame) & 0.73 & 0.71\\
\hline
Spectator basis:  & & \\
~~Loose cuts & -0.60 & -0.53\\
\hline
Beamline basis: & & \\
~~Loose cuts & -0.14 & -0.08\\
~~Tight cuts & -0.28 & -0.35\\
\hline
$\eta$~beamline basis: & & \\
~~Loose cuts & -0.47 & -0.48\\
~~Tight cuts & -0.60 & -0.59\\
\hline
$\hat{s}$~beamline basis: & & \\
~~Loose cuts & -0.29 & -0.22\\
\hline
\end{tabular}
\par\end{centering}

\caption{Degree of polarization $\mathcal{D}$ at for exclusive three-jet $t$-channel
events after selection cuts at the 7~TeV LHC. The $tq(j)$-frame
in the helicity basis denotes the c.m. frame of the incoming partons,
while the $tq$-frame denotes the rest frame of the top quark and
spectator jet.\label{tab:toppol-3jet}}
\end{table}

\section{Conclusions\label{sec:Conclusions}}

We have presented a next-to-leading order study of $t$-channel single
top and antitop quark production at the LHC proton-proton collider for
several CM energies, including $\oalphas$ QCD corrections to both the production and 
decay of the top quark. The $\oalphas$ corrections affect top and antitop quark
production differently and also show a dependence on the CM energy.
The inclusive $t$-channel cross section for the
production of a single top (antitop) quark with a mass of 173~GeV
at 7~TeV is $40.0\pm4.6$~pb ($21.9\pm4.1$~pb), where the uncertainty
includes scale, top quark mass, and PDF components. For top quark production
this is only an increase of 4\% compared to Born-level. For antitop quark
production the increase compared to Born-level is larger at 15\%. The behavior
is similar for higher collider CM energies. The impact of
kinematical cuts on the acceptances has been studied for a loose and
a tight set of cuts, corresponding to typical event selections used
by the ATLAS and CMS collaborations at the LHC. We find that the acceptances
are sensitive to the $\Delta R_{{\rm cut}}$ we imposed on the jet
cone size and the lepton isolation. With the choice of $\Delta R_{{\rm cut}}=0.4$,
the difference between the Born-level and NLO acceptances
is about $5\%$ for a loose cut set and slightly larger when
changing $\Delta R_{{\rm cut}}$ from 0.4 to 0.5. For antitop quark production
the difference between the Born-level and NLO acceptances is about 15\%,
and again slightly larger when increasing $\Delta R_{{\rm cut}}$ from 0.4 to 0.5.

We have categorize the $\oalphas$ contributions to the $t$-channel single
top process into three gauge invariant sets: the light quark line
corrections, the heavy quark line corrections and the top quark decay
corrections. This allows us to analyze the $\oalphas$ corrections in detail
and facilitates comparisons with event generators. The corrections affect the 
shape of some of the important kinematic distributions and result in a large 
fraction of events containing three reconstructed jets in the final state for
the loose set of kinematic cuts. The acceptance for $t$-channel single
top quark events and the fraction of 3-jet events depend strongly
on the jet $p_T$ cut. 
The kinematic distributions affected by the radiative corrections
include those that separate the $t$-channel single
top signal from the various backgrounds, such as the pseudo-rapidity
distribution of the spectator jet. We find that the $\oalphas$ LIGHT
and HEAVY corrections have almost opposite effects on various
pseudo-rapidity distributions, due to the difference in the parton
distribution functions between the valence quarks and sea quarks.
The former shifts the spectator jet to even higher pseudo-rapidities,
while the later shifts it to more central pseudo-rapidity regions.
The summed contributions cause the spectator jet to be even more forward
which will change the prediction of the acceptance for $t$-channel
single top quark events. Also, a large fraction of three jet events
contain two $b$-jets due to the collinear enhancement in the $W+g$
fusion process. This also impacts the experimental choice of the light quark jet
in 3-jet events. Choosing the most forward jet (highest $|\eta|$) is correct
in only 80\% of the events, while choosing the highest $p_T$ jet is correct
in over 90\% of the events. 

In order to study top quark properties such as the top quark polarization,
induced from the effective $t$-$b$-$W$ couplings, we reconstruct
the top quark by combining the $W$~boson with the leading $b$-tagged jet. 
We use the top quark thusly reconstructed to explore spin correlations
in three different bases: the helicity basis, the spectator basis
and the beamline basis. The degree of polarization is very large in the helicity 
and spectator bases. Its reduced after event reconstruction, especially for
antitop quark production. The radiative corrections reduce the degree of polarization 
further, both at parton level and after event reconstruction.

\begin{acknowledgments}
S. H. and R. S. are supported in part by the U.S. National Science Foundation under 
Grant No. PHY-0757741. Q. H. C. is supported in part by the Argonne National Laboratory 
and University of Chicago Joint Theory Institute (JTI) Grant No. 03921-07-137, and by the 
U.S. Department of Energy under Grants No. DE-AC02-06CH11357 and No. DE-FG02- 90ER40560. 
C. P. Y. acknowledges the support of the U.S. National Science Foundation under Grants No. 
PHY-0555545 and PHY-0855561. Q.H.C thanks Shanghai Jiaotong University for hospitality 
where part of this work was done. C. P. Y. would also like to thank the hospitality of 
National Center for Theoretical Sciences in Taiwan and Center for High Energy Physics, 
Peking University, in China, where part of this work was done.
\end{acknowledgments}

\addcontentsline{toc}{section}{\refname}\bibliography{phen-tchan}

\end{document}